\documentclass[fleqn,usenatbib]{mnras}

\usepackage{newtxtext,newtxmath}

\usepackage[T1]{fontenc}
\usepackage{ae,aecompl}

\usepackage{graphicx}	
\usepackage{amsmath}	
\usepackage{amssymb}	
\usepackage{times}

\usepackage{bm}
\usepackage{wasysym}
\usepackage[dvipsnames]{xcolor}
\usepackage{ulem}

\usepackage{etoolbox}

\defcitealias{laibe08}{L08}
\defcitealias{phantom18}{P18}

\title{Dust growth, fragmentation and self-induced dust traps in \textsc{Phantom}}

\author[Vericel et al.]{%
Arnaud Vericel$^{1}$\thanks{E-mail: arnaud.vericel@gmail.com}, 
Jean-Fran\c{c}ois Gonzalez$^{1}$,
Daniel J. Price$^{2}$,
Guillaume Laibe$^{1}$ and
\newauthor
Christophe Pinte$^{2,3}$
\\
$^{1}$Univ Lyon, Univ Claude Bernard Lyon 1, Ens de Lyon, CNRS, Centre de Recherche Astrophysique de Lyon, UMR5574, F-69230, Saint-Genis-Laval, France\\
$^{2}$School of Physics \& Astronomy, Monash University, Clayton VIC 3800, Australia\\
$^{3}$Univ. Grenoble Alpes, CNRS, IPAG, F-38000, Grenoble, France
}

\date{Accepted 2021 July 30. Received 2021 July 30; in original form 2020 September 17}

\pubyear{2021}

\begin{document}
\label{firstpage}
\pagerange{\pageref{firstpage}--\pageref{lastpage}}
\maketitle

\begin{abstract}
We present the implementation of a dust growth and fragmentation module in the public Smoothed Particle Hydrodynamics (SPH) code \textsc{Phantom}. This module is made available for public use with this paper. The coagulation model considers locally monodisperse dust size distributions around single values that are carried by the SPH particles. Along with the presentation of the model, implementation and tests, we showcase growth and fragmentation in a few typical circumstellar disc simulations and revisit previous results. The module is also interfaced with the radiative transfer code \textsc{mcfost}, which facilitates the comparison between simulations and ALMA observations by generating synthetic maps. Circumstellar disc simulations with growth and fragmentation reproduce the `self-induced dust trap' mechanism first proposed by Gonzalez et al., which supports its existence. Synthetic images of discs featuring this mechanism suggest it would be detectable by ALMA as a bright axisymmetric ring at several tens of au from the star. With this paper, our aim is to provide a public tool to be able to study and explore dust growth in a variety of applications related to planet formation.

\end{abstract}




\section{Introduction}
\label{sec:intro}

High resolution imaging of protoplanetary discs with the Atacama Large Millimetre/Submillimetre Array (ALMA) has revealed rich substructure in mm continuum emission that includes rings, gaps and spiral arms \citep{andrews18,huang18}. The current state of the art in modelling these discs is to simulate gas plus one, or more, dust populations of constant size \citep[e.g.][]{dipierro15,price18,mentiplay19,perez19,calcino19,benni20}. 
Gas and dust are coupled via an aerodynamic drag force \citep{whipple73}. Whereas gas orbits at a sub-Keplerian speed because of its own pressure support, the pressureless dust phase feels an azimuthal headwind and loses angular momentum. This transfer leads to the dust drifting radially towards the star and settling vertically towards the mid-plane \citep{safro69,adachi76,weiden77,dubrulle95,haghi05}. These processes are most efficient for mm to cm grain sizes, i.e. for dust populations detected by instruments such as ALMA.
While simulating gas, dust and their mutual effects, calculations indicate that interaction with embedded planets is the likely explanation for the observed rings and gaps \citep{parmel04,dipierrolaibe17,mentiplay19,perez19,toci20}. Stellar companions are also invoked to explain spiral arms, cavities or misalignments \citep{kley12,zhu15,munoz16,dong16,price18,dong18,calcino19,cuello19,cuello20,nealon20,gonzalez20}. While this general simulation approach has proven successful, particularly for planet-induced kinematic signatures \citep{pinte18,pinte19,pinte20}, it fails to explain how planets were formed in these discs in the first place, or how the dust grows and evolves on longer time scales.

From a theoretical point of view, dust evolution has been studied extensively in 3D without dust growth to understand dust radial drift, settling or concentrations in pressure bumps \citep[e.g.][]{weiden77,dubrulle95,haghi05,fromang06}. Efficient dust concentrations have been shown to lead to the Streaming Instability --- a local hydrodynamic instability leading to direct planetesimal formation \citep{streaming05,joha07,yang17,schafer16,auffinger18,abod19,listreaming19}. This mechanism could be the cornerstone of planet formation. Understanding how particles can reach such high concentration levels in discs is thus of utmost importance. 
One mechanism that could provide the required initial conditions is the `self-induced dust trap' proposed by \citet{gonzalez17} and \citet{vericelgonzalez20}, where dust growth/fragmentation, back-reaction and large scale gradients leads to dust pile-ups and subsequent formation of a local pressure maximum in the disc.

Dust growth is a complex and computationally challenging process in which coagulation of small dust particles forms larger bodies by a `snowball effect' \citep{lisstev93,dominiktielens97,blumwurm08}. Since the pioneering work of \citet{weiden97}, the numerical challenges of dust growth have led the community to follow different approaches to tackle this problem. Grid-based codes have considered the resolution of the Smoluchowski equation \citep{smolu16} in vertically and azimuthally averaged discs \citep{brauer08, birnstiel09, pinilla12, birnstiel12, drazk16, drazk17} where gas undergoes Keplerian shear and the vertical equilibrium is reached instantaneously. In most of these models the gas is considered as a static background and back-reaction, i.e. the drag force exerted by dust onto the gas, is neglected, although it can be of significant importance during the disc evolution \citep{gonzalez17,2018MNRAS.479.4187D}. However, a few 2D grid-based codes compute the Smoluchowski equation at the same time as the evolution of both gas and dust, more particularly for the study of planet hosting discs \citep{li05,li19,draz19,laune20}. 
Alternatively, dust growth has also been modelled with Monte Carlo methods in grid-based codes \citep[e.g.][]{zsom08,ormel08,draz14}{, although these need a large number of size bins to avoid overdiffusion.}

While the Eulerian nature of a grid-based code handles dust size distributions naturally in each cell, particle tracking and complex geometries present challenges.
As a result, alternative methods have also been developed with a static gas background on top of which dust super-particles are set \citep{kri16,kriciesla16,schoo18,misener19}. In these studies, dust growth is modelled using the `single-size' approximation, i.e. considering that every dust super-particle carries a single size that can evolve because of the local conditions.

This method was pioneered by \citet[][hereafter L08]{laibe08} following \citet{stepvala97}, using a 3D SPH code that solves the equations of motion self-consistently, including gas drag and back-reaction. This code, which we will refer to as \textsc{LyonSPH}, initially developed by \citet{barriere05} has been further developed over the years to include fragmentation \citep{gonzalez15,gonzalez17}, porosity \citep{garciathesis18,anto20} and snow line \citep{vericelgonzalez20} effects.
In this paper, we extend and adapt this growth algorithm to \textsc{Phantom}: a public, fast, modular and optimised SPH code widely used by the community \citep[][hereafter P18]{phantom17ascl,phantom18}. By doing so, our aim is to provide a tool to study dust growth in a range of planet-forming environments, including discs around multiple stars, flybys, and in misaligned and warped discs.

The paper is organised as follows. Section~\ref{sec:model} describes the growth and fragmentation model while Section~\ref{sec:implementation} treats its numerical implementation. We present two tests in Section~\ref{sec:tests} as well as circumstellar disc simulations in Section~\ref{sec:simus}. Finally, we discuss our findings and conclude in Sections~\ref{sec:discu} and~\ref{sec:conclu}.

\section{Model}
\label{sec:model}

\begin{figure}
\centering
\resizebox{\hsize}{!}{
\includegraphics[width=\columnwidth]{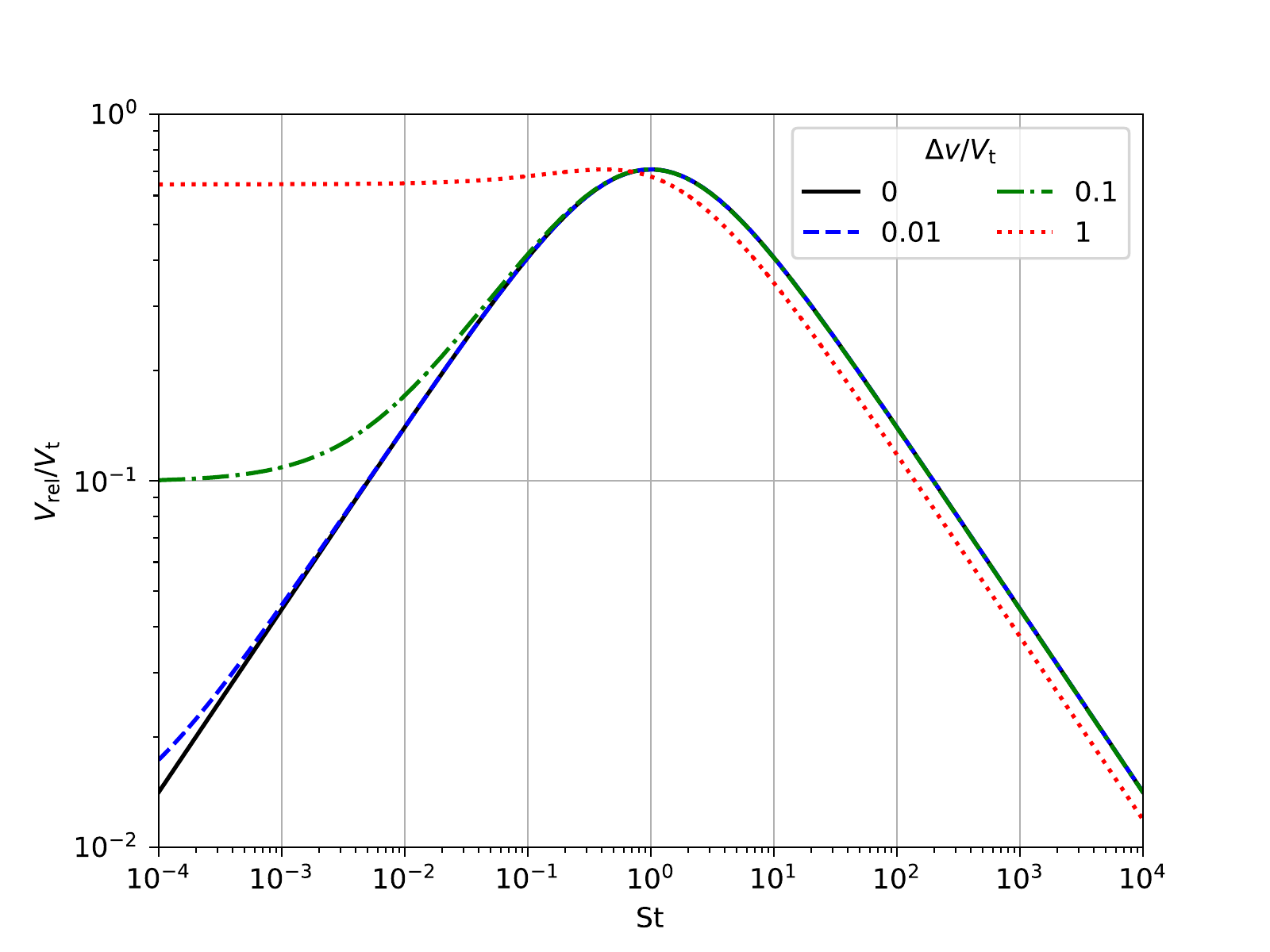}
}
\caption{Evolution of the relative velocity (Equation~\ref{eq:vrel}) as a function of the Stokes number for 4 different values of the ratio $\Delta v / V_\mathrm{t}$. In the simulations performed for this paper, the differential velocity is always small compared to the turbulent velocity (and by extension compared to the sound speed), thus we only show values in the range $0 \leq \Delta v/V_{\rm t} \leq 1$.}
\label{vrelonvt}
\end{figure}

Dust coagulation is the result of the collision between particles, the outcome of which is determined by their relative velocity, noted $V_{\mathrm{rel}}$.
The relative velocity can come from multiple sources including disc turbulence, Brownian motion, dust radial and azimuthal drift and dust settling. Since we use the `single size' approximation locally, particles in that vicinity feel the same drag force, which cancels the radial, azimuthal and vertical component of their relative velocity \citep[see][for a discussion]{vericelgonzalez20}. Moreover, the Brownian motion only plays a significant role for very small grains \citep[micrometre and smaller,][]{birn10}, which is a regime that is in practice never met in our simulations.
For these reasons, we only consider the turbulence-driven component of the relative velocity, which is transmitted to the dust from the gas via drag. We use the form proposed by \citet{stepvala97}, that is
\begin{equation}
V_{\mathrm{rel}} = \sqrt{2^{3/2}\mathrm{Ro}\alpha} \dfrac{\sqrt{\mathrm{Sc}-1}}{\mathrm{Sc}}c_\mathrm{s} = \sqrt{2}V_\mathrm{t}\dfrac{\sqrt{\mathrm{Sc}-1}}{\mathrm{Sc}},
\label{eq:vrel}
\end{equation}
where $\mathrm{Ro}$ is the Rossby number that we consider constant and equal to 3, $\alpha$ is the viscosity parameter defined by \citet{shaksuny73} and $c_\mathrm{s}$ the gas sound speed. Various turbulent relative velocity formulations have been proposed, we refer the reader to \citetalias{laibe08} and \citet{laibe14drift} for a discussion about their differences.
In Equation~(\ref{eq:vrel}), $V{_\mathrm{t}}$ is called the turbulent velocity and $\mathrm{Sc}$ is the Schmidt number, expressed as
\begin{equation}
\mathrm{Sc} = \left(1 + \mathrm{St}\right) \sqrt{1 + \dfrac{\Delta \bm {v}^2}{V_\mathrm{t}^2}},
\label{eq:schmidt}
\end{equation}
where $\mathrm{St}$ is the Stokes number and $\Delta \bm v$ is the differential velocity between the gas and dust phases. The dust relative velocity (Equation~\ref{eq:vrel}) is plotted for different values of $\Delta v$ in Fig.~\ref{vrelonvt}.
For small differential velocities with respect to $V_\mathrm{t}$, the Schmidt number approximates as $1+\mathrm{St}$, which reduces the relative velocity to
\begin{equation}
V_\mathrm{rel} \simeq \frac{\sqrt{2 \mathrm{St}}}{1+\mathrm{St}} V_\mathrm{t}.
\end{equation}
The Stokes number is defined as the product of the stopping time $t_\mathrm{s}$ set by the gas drag and the Keplerian frequency $\Omega_\mathrm{k}$ set by the star's gravitational field
%
\begin{equation}
\mathrm{St} = t_\mathrm{s} \Omega_\mathrm{k},
\label{eq:stokesdef}
\end{equation}
where $t_\mathrm{s}$ depends on the drag regime considered \citep[see][]{laibeprice2012b}. In the usual Epstein regime \citep{epstein24}, this gives
\begin{equation}
\mathrm{St} = \sqrt{\dfrac{\pi \gamma}{8}} \dfrac{\rho_\mathrm{s}s}{f\left(\rho_\mathrm{g}+\rho_\mathrm{d}\right) c_\mathrm{s}}\Omega_\mathrm{k},
\label{eq:Stepstein}
\end{equation}
where $\gamma$ is the adiabatic index, $\rho_{\rm s}$ is the dust intrinsic density, and $\rho_{\rm g}$ and $\rho_{\rm d}$ are the volume densities of the gas and dust phases. The definition of the Stokes number corresponds to the mixture \citep{dipierrolaibe17,kanagawa17}. $f$ is the Kwok parameter \citep{kwok75} correcting drag for supersonic differential velocities and expressed as
\begin{equation}
f = \sqrt{1 + \dfrac{9\pi}{128} \dfrac{\Delta \bm v^2}{c_\mathrm{s}^2}}.
\label{eq:kwok}
\end{equation}

The radial drift process is fastest when $\mathrm{St} \sim 1$. Considering this, we rewrite the Stokes number as the ratio between the dust size and its optimal drift size for which $\mathrm{St} = 1$, called $s_{\rm opt}$, which can be expressed as
\begin{align}
s_{\rm opt} & = \sqrt{\dfrac{8}{\pi \gamma}}
\dfrac{f(\rho_\mathrm{g}+\rho_\mathrm{d})c_{\mathrm{s}}}{\rho_\mathrm{s}\Omega_\mathrm{k}}, \\
& = \sqrt{\dfrac{8}{\pi \gamma}}
f(1+\varepsilon) \dfrac{\rho_\mathrm{g}}{\rho_\mathrm{s}} H,
\label{eq:sopt}
\end{align}
where $\varepsilon = \rho_\mathrm{d}/\rho_\mathrm{g}$ is the dust-to-gas ratio and $H = c_{\mathrm{s}}/\Omega_{\mathrm{k}}$ is the scale height of the gas disc.

\subsection{Growth}
\label{subsec:growth}

In the model, we consider a single dust SPH particle to represent a swarm of equal-sized physical grains. This means that at the location of the SPH particle, the size distribution is considered highly peaked around a certain value: the one that the particle carries.
We represent grain growth inside each SPH particle by assuming perfect coagulation of two grains during a collision time. Since the distribution is locally monodisperse (i.e. the size is locally uniform),
only same-size coagulation events are modelled and the grain mass $m_\mathrm{d}$ doubles during a collision
\begin{equation}
\dfrac{\mathrm{d}m_{\rm d}}{\mathrm{d}t} = m_{\rm d}f_\mathrm{col},
\label{eq:dmdtintro}
\end{equation}
where $f_\mathrm{col}=\sigma_\mathrm{d}n_\mathrm{d}V_\mathrm{rel}$ is the collision frequency, $\sigma_\mathrm{d}$ is the dust cross section and $n_\mathrm{d}$ is the dust numerical density. We consider compact {spherical} grains, for which the growth rate is more conveniently expressed using the size
\begin{equation}
\dfrac{\mathrm{d}s}{\mathrm{d}t} = \dfrac{s}{3}f_\mathrm{col}.
\label{eq:dsdtintro}
\end{equation}
The growth rate is associated with the physical grains within the SPH particle and not to the SPH particle itself. This is fundamental, because we wish to conserve the mass of each SPH particle to preserve numerical robustness. This constraint is satisfied considering a larger size {corresponds to} {fewer} physical grains {represented by} a given SPH particle, without affecting its total mass.
This representation holds as long as the mass of the SPH particle itself is much larger than the mass of a single physical grain it `contains'. This translates as the condition
\begin{equation}
s \ll \left(\dfrac{M_\mathrm{d}}{n^\mathrm{SPH}_\mathrm{d}}\dfrac{3}{4\pi\rho_\mathrm{s}}\right)^\frac{1}{3},
\end{equation}
where $M_\mathrm{d}$ is the total dust mass, {and} $n^\mathrm{SPH}_\mathrm{d}$ is the number of dust SPH particles.
Quantitatively speaking, this gives
\begin{equation}
s \ll 362 \left(\dfrac{M_\mathrm{d}}{10^{-4}~M_\odot}\right)^{\frac{1}{3}} \left(\frac{n^\mathrm{SPH}_\mathrm{d}}{10^6}\right)^{-\frac{1}{3}} \left(\dfrac{\rho_{\mathrm{s}}}{1000~\mathrm{kg.m}^{-3}}\right)^{-\frac{1}{3}}\mathrm{km},
\end{equation}
which is entirely satisfied for simulations with sizes between a few tens of micrometres to a few tens of kilometres at most. {E}quation~(\ref{eq:dsdtintro}) finally gives the growth rate
\begin{equation}
\dfrac{\mathrm{d}s}{\mathrm{d}t} = \dfrac{\rho_\mathrm{d}}{\rho_\mathrm{s}}V_{\mathrm{rel}}.
\label{eq:dsdt}
\end{equation}
\subsection{Fragmentation}
\label{subsec:frag}

\begin{figure}
\centering
\resizebox{\hsize}{!}{
\includegraphics[width=\columnwidth]{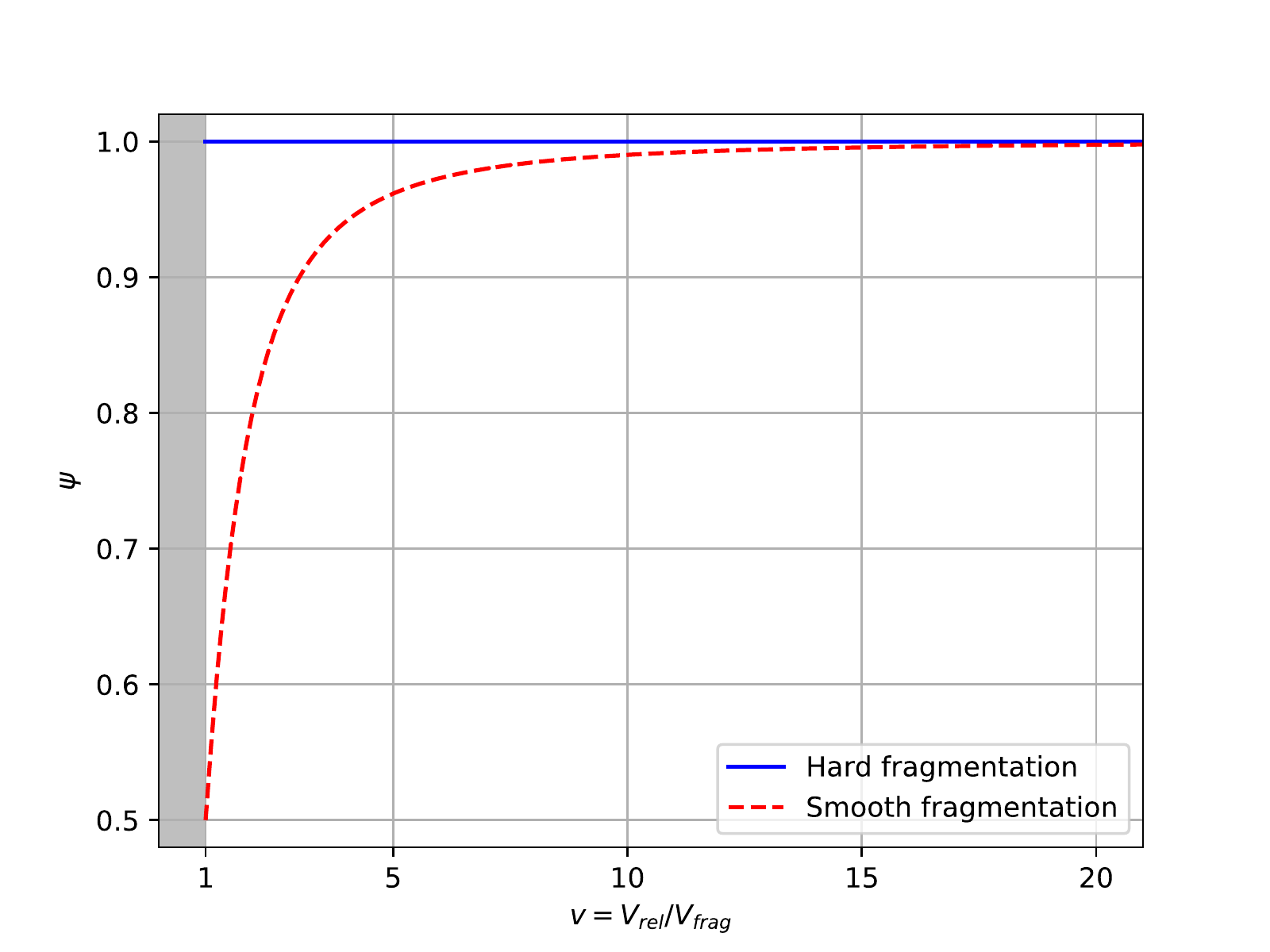}
}
\caption{Evolution of the fragmentation parameter $\psi$ as a function of the ratio $v = V_\mathrm{rel}/V_\mathrm{frag}$ for both Hard and Smooth fragmentation models. The grey zone corresponds to growth ($V_\mathrm{rel} < V_\mathrm{frag}$), where $\psi$ is not defined.}
\label{psi}
\end{figure}

While Equation~(\ref{eq:dsdt}) describes the growth process, fragmentation can also be considered in the model. More specifically, as the grains relative velocity increase, the kinetic energy of the collision also increases. At sufficiently high relative velocities, the kinetic energy can destroy the grains chemical bonds and make them fragment to smaller sizes. A threshold relative velocity above which grains are fragmenting upon collisions can then be considered. This threshold {is often called} the `fragmentation velocity', noted $V_\mathrm{frag}$. In case of fragmentation, the resulting growth rate is therefore negative and expressed as
\begin{equation}
\dfrac{\mathrm{d}s}{\mathrm{d}t} = - \dfrac{\rho_\mathrm{d}}{\rho_\mathrm{s}}V_{\mathrm{rel}}\psi,
\label{eq:-dsdt}
\end{equation}
where $\psi$ refers to two models of fragmentation:
{\begin{enumerate}
\item{`Hard', that is symmetrical to the growth case \citep{gonzalez15};}
\item{`Smooth', which considers a more progressive loss of mass with increasing relative velocities \citep{garciathesis18}.}
\end{enumerate}}
Defining the ratio of the {relative velocity} to the fragmentation velocity as $v = V_\mathrm{rel}/V_\mathrm{frag}$, the parameter $\psi$ is given by
\begin{equation}
\psi = \left\{\begin{array}{cl}
1 & \mbox{ : Hard fragmentation},\\
v^2/(1+v^2) & \mbox{ : Smooth fragmentation},\\
\end{array}\right.
\end{equation}
The evolution of $\psi$ with the relative to fragmentation velocity ratio is shown in Fig. \ref{psi}. The Hard model considers that most of the mass is lost during fragmentation, whatever the relative velocity. Conversely, the Smooth model only considers large loss of mass for very high relative velocities. Both models are implemented in \textsc{Phantom} and can be selected by the user at runtime.
%
%
\subsubsection{Fragmentation sizes and distance to the star}

When considering fragmentation, we can estimate the equilibrium size that the dust particles can reach when their relative velocity is in equilibrium with the fragmentation velocity. Following \citet{gonzalez17}, by writing the equality $V_{\rm rel} = V_{\rm frag}$, we find more precisely
\begin{equation}
\sqrt{2\tilde{\alpha}} c_{\rm s} \dfrac{\sqrt{\mathrm{St}}}{1+\mathrm{St}} = V_{\rm frag},
\end{equation}
where $\tilde{\alpha} = 2^{1/2}\mathrm{Ro}\alpha$ and Sc is considered equal to $1+\mathrm{St}$.
{Using} this leads to the following quadratic equation
\begin{equation}
\mathrm{St}^2 + 2 \left(1-\dfrac{\tilde{\alpha}c_{\rm s}^2}{V_{\rm frag}^2}\right)\mathrm{St} + 1 = 0,
\end{equation}
which has two solutions, that we call $s_{\rm frag}^{\pm}$, expressed as \citep[see Appendix~A of][]{gonzalez17}
\begin{equation}
s_{\rm frag}^{\pm} = s_{\rm opt} \left( \dfrac{\tilde{\alpha}c_{\rm s}^2}{V_{\rm frag}^2} - 1 \pm \dfrac{\sqrt{\tilde{\alpha}}c_{\rm s}}{V_{\rm frag}} \sqrt{\dfrac{\tilde{\alpha} c_{\rm s}^2}{V_{\rm frag}^2} - 2} \right),
\label{eq:sfrag}
\end{equation}
where `+' and `-' refer to the cases where $\mathrm{St} > 1$ and $\mathrm{St} < 1$ respectively.
One should note that Equation~(\ref{eq:sfrag}) is not defined when $\tilde{\alpha}c_{\rm s}^2 < 2V_{\rm frag}^2$. This corresponds to the case where the maximum relative velocity is smaller than the fragmentation velocity, which means that fragmentation never occurs. {Using} this inequality with a power-law prescription for the temperature, we find a corresponding fragmentation radius $r_{\rm frag}$
\begin{equation}
r_{\rm frag} = r_0 \left(\dfrac{\tilde{\alpha}}{2} \dfrac{c_{\rm s,0}^2}{V_{\rm frag}^2}\right)^{\frac{1}{q}},
\label{eq:rfrag}
\end{equation}
where $q$ is the temperature power-law index and $r_0$ is a reference radius.
Beyond $r_{\rm frag}$, fragmentation is not possible.

\section{Implementation}
\label{sec:implementation}

The dust growth model is introduced in both \textsc{Phantom}'s two-fluid and one-fluid algorithms \citep{laibeprice2012a,laibeprice2012b,laibeprice14a,laibeprice14b,laibeprice14c,price15,ballabio18,price20}.
Firstly, we specify the implementation with the two-fluid algorithm. In that configuration, the accuracy of the implementation relies on the evaluation of a few gas-related quantities on each dust particle. This is needed in particular for $\Delta v$, $\mathrm{St}$ and $c_\mathrm{s}$, which are not defined for a given dust particle but rather for a gas-dust pair or for gas only.

\subsection{Two fluid means two smoothing lengths}
\label{subsec:twofluid}

The SPH method allows one to evaluate any given quantity by means of interpolations over the particle neighbours. Modelling dust and gas as two separate fluids comes at the price of increasing the total number of particles and having to handle two sets of resolution lengths. 
For a gas or dust quantity (e.g the density), the typical length over which the neighbours are used is the gas or dust smoothing length (more precisely a few smoothing lengths, depending on the kernel), which are expressed as
\begin{equation}
h_\mathrm{gas} = h_\mathrm{0} \left(\dfrac{m_a}{\rho_a}\right)^{1/3},
\end{equation}
where $h_0$ is a constant setting the average number of neighbours, $m_a$ is the mass of a gas particle and $\rho_a$ its volume density, and
\begin{equation}
h_\mathrm{dust} = h_\mathrm{0} \left(\dfrac{m_i}{\rho_i}\right)^{1/3},
\end{equation} 
where $m_i$ is the mass of a dust particle and $\rho_i$ its volume density.
We used the {index} $i$ for dust particles and $a$ for gas particles, as in e.g. \citet[][]{mona97,laibeprice2012a,laibeprice2012b}.

As gas and dust do not necessarily have the same behaviour, one needs to be careful when dealing with two sets of smoothing lengths, especially if an interpolated quantity for a particle type requires a loop over the other one. This is typically the case for the drag between gas and dust, and in our case additionally for $\Delta v$, $\mathrm{St}$ and $c_\mathrm{s}$.
The risk with such a method is that if the dust-to-gas ratio increases significantly, the dust smoothing length can become much smaller than that of the gas. This can jeopardise further interpolations due to a lack of neighbours found inside the corresponding smoothing length. Secondly, as gas is pressure-supported but dust is not, reducing the dust smoothing length under that of the gas can also locally clump close dust particles together as they would not feel any repulsive force from the gas.
To limit both of these risks, we follow \citet{laibeprice2012a} and use in our cross-species interpolations the maximum of the gas and dust smoothing lengths, which is slightly more computationally expensive but safer. We denote this maximum smoothing length $h_{ia} = \mathrm{max}(h_i,h_a)$ for future use.

\subsection{Sound speed and gas density}
\label{subsec:cs-rho}

The gas density is interpolated using the kernel-weighted sum over the neighbours
\begin{equation}
\rho_{\mathrm{g},i} = \sum_a m_a W(\bm{r}_{ia},h_{ia}),
\label{eq:rhoginterp}
\end{equation}
where $W$ is the kernel function (quintic M$_6$ in our case). 
The other quantities are then interpolated using $\rho_{\mathrm{g},i}$, that is for a function $g$
\begin{align}
g_i & = \dfrac{\sum_a m_a g_a W(\bm{r}_{ia},h_{ia})}{\sum_a m_a W(\bm{r}_{ia}, h_{ia})}, \\
    & = \dfrac{1}{\rho_{\mathrm{g},i}}\sum_a m_a g_a W(\bm{r}_{ia},h_{ia}).
\end{align}
We found that this form, slightly different than the usual interpolation ({weighted by $m$ instead of $m/\rho$}), was generally more accurate than its counterpart, especially in regions where the dust concentrates. This choice of implementation does not affect the {speed} of execution of the code.

The sound speed on each dust particle is interpolated using
\begin{equation}
c_{\mathrm{s},i} = \dfrac{1}{\rho_{\mathrm{g},i}} \sum_a m_a c_\mathrm{s,a}W(\bm{r}_{ia},h_{ia}).
\label{eq:csinterp}
\end{equation}
\subsection{Differential velocity}
\label{subsec:dv}

We evaluate each component of the differential velocity vector before computing the norm
\begin{equation}
\Delta v_{x,i} = \dfrac{1}{\rho_{\mathrm{g},i}} \sum_a m_a\Delta {v}_{x,ia}W(\bm{r}_{ia},h_{ia}),
\label{eq:dvx}
\end{equation}
\begin{equation}
\Delta v_{y,i} = \dfrac{1}{\rho_{\mathrm{g},i}} \sum_a m_a\Delta {v}_{y,ia}W(\bm{r}_{ia},h_{ia}),
\label{eq:dvy}
\end{equation}
\begin{equation}
\Delta v_{z,i} = \dfrac{1}{\rho_{\mathrm{g},i}} \sum_a m_a\Delta {v}_{z,ia}W(\bm{r}_{ia},h_{ia}),
\label{eq:dvz}
\end{equation}
\begin{equation}
||\Delta \bm v_{i}|| = \Delta v_i =\sqrt{\Delta v_{x,i}^2+\Delta v_{y,i}^2 + \Delta v_{z,i}^2}.
\label{eq:dvinterp}
\end{equation}

One should notice that $\Delta v$ is not evaluated by the double-humped kernel $D$ used in \citet{laibeprice2012a} but rather by the usual density kernel $W$. Indeed, here we do not need to evaluate $\Delta \bm v$ along the line of sight of a gas-dust pair, thus the scalar product $\Delta \bm v_{ia} \cdot \bm r_{ia}$ is avoided and the evaluation is more accurate with the usual bell-shaped kernel $W$.

\subsection{Stokes number}
\label{subsec:stokes}

The evaluation of the Stokes number depends on the drag regime that the grains experience. In \textsc{Phantom}, the Epstein and Stokes regimes are automatically selected depending on the Knudsen number $K_\mathrm{n} = 9 \lambda_\mathrm{g}/4s$ \citep{weiden77,stepvala96}. To follow this and remain as general as possible in our implementation, we evaluate the local Stokes number (which corresponds to the mixture) by computing the stopping time using $\rho_{\mathrm{g},i}$, $c_{\mathrm{s},i}$ and $\Delta v_{i}$ rather than interpolating the stopping time between each gas-dust pair
\begin{equation}
\mathrm{St}_i = t_\mathrm{s}\left(\rho_{\mathrm{g},i}, c_{\mathrm{s},i}, \Delta v_{i} \right) \Omega_{\mathrm{k},i}.
\label{eq:Stinterp}
\end{equation}
The stopping time $t_\mathrm{s}$ is expressed as \citep{laibeprice2012a}
\begin{equation}
t_\mathrm{s} = \dfrac{\rho_\mathrm{g} \rho_\mathrm{d}}{K(\rho_\mathrm{g} + \rho_\mathrm{d})},
\label{eq:tsglobal}
\end{equation}
where $K$ is the drag coefficient. Considering the Epstein drag regime again, this gives for example \citepalias{phantom18}
\begin{equation}
\mathrm{St}_i = \sqrt{\dfrac{\pi \gamma}{8}} \dfrac{\rho_\mathrm{s}s_i}{f_i\left(\rho_{\mathrm{g},i}+\rho_i\right) c_{\mathrm{s},i}}\Omega_{\mathrm{k},i}.
\end{equation}

\subsection{One-fluid implementation}
\label{subsec:onefluid}

To simulate the most strongly coupled populations of grains, \textsc{Phantom} also has a `one-fluid' algorithm that represents the gas-dust mixture with only one set of particles and thus one resolution \citep{laibeprice14a,laibeprice14b,laibeprice14c,price15,ballabio18}
{We also implemented t}he growth model with the one-fluid formalism, although its validity is restricted to relatively small Stokes numbers in order to respect the terminal velocity approximation \citep{streaming05}. With that in mind, the one-fluid model with dust growth can be used for a short amount of time and mainly to accelerate the evolution of tightly coupled grains. We also added to the code a conversion tool that allows the user to convert to the two-fluid method when the one-fluid regime reaches its limit.

The implementation itself is straightforward in one-fluid since the cross species interpolations vanish. To compute the values of interest, we simply use the values stored onto the mixture particles and transform them out of the barycentric frame, that is applying the changes of variables
\begin{align}
\rho_{\rm g} & = (1 - \epsilon) \rho, \\
\rho_{\rm d} & = \epsilon \rho, \\
{\bm v}_{\rm g} & = \bm {v} - \epsilon \Delta \bm{v}, \\
{\bm v}_{\rm d} & = \bm {v} + (1 - \epsilon) \Delta \bm{v},
\end{align}
where $\rho = \rho_{\rm g} + \rho_{\rm d}$ and $\epsilon = \rho_{\rm d}/\rho$ is the dust fraction.

\subsection{Timestepping}
\label{subsec:timestep}

\subsubsection{Numerical scheme}

The size evaluation by the integrator follows the same predictor-corrector scheme that is described in \citetalias{phantom18}.
\begin{align}
s^{n+\frac{1}{2}} & = s^n + \dfrac{1}{2}\Delta t \left(\dfrac{\mathrm{d}s}{\mathrm{d}t}\right)^n, \\
s^* & = s^{n+\frac{1}{2}} + \dfrac{1}{2}\Delta t \left(\dfrac{\mathrm{d}s}{\mathrm{d}t}\right)^n, \\
\left(\dfrac{\mathrm{d}s}{\mathrm{d}t}\right)^{n+1} & = \dfrac{\mathrm{d}s}{\mathrm{d}t}\left(s^*\right), \\
s^{n+1} & = s^{n+\frac{1}{2}} + \dfrac{1}{2}\Delta t \left(\dfrac{\mathrm{d}s}{\mathrm{d}t}\right)^{n+1}. 
\end{align}
The size integration is also implemented with individual timestepping, which can be significantly faster if dust grain sizes span a wide range.

\subsubsection{Constraints}
The timestepping for growth and fragmentation is constrained by the Courant-Friedrichs-Lewy \citep[CFL]{courant28} condition, which requires the timestep to be smaller than the typical growth timescale
\begin{equation}
\Delta t \leq C_\mathrm{cour} \tau_\mathrm{g},
\end{equation}
where $C_\mathrm{cour}$ is a constant typically of order unity and $\tau_\mathrm{g}$ is the growth time scale
\begin{equation}
\tau_\mathrm{g} = \dfrac{s}{|{\mathrm{d}s}/{\mathrm{d}t}|}.
\end{equation}
{C}onsidering that $\Delta v \ll V_\mathrm{t}$ and $c_{\rm s}$ (such that $\mathrm{Sc} \simeq 1 + \mathrm{St}$ and $f \simeq 1$), this gives
\begin{equation}
\dfrac{\tau_\mathrm{g}}{t_\mathrm{s}} = \dfrac{1+\varepsilon}{\varepsilon} \sqrt{\dfrac{8}{\pi\gamma}}\dfrac{\left(1+\mathrm{St}\right)}{\sqrt{2^{3/2}\mathrm{Ro}\alpha\mathrm{St}}}.
\label{eq:tgts}
\end{equation}
%
\begin{figure}
\centering
\resizebox{\hsize}{!}{
\includegraphics[width=\columnwidth]{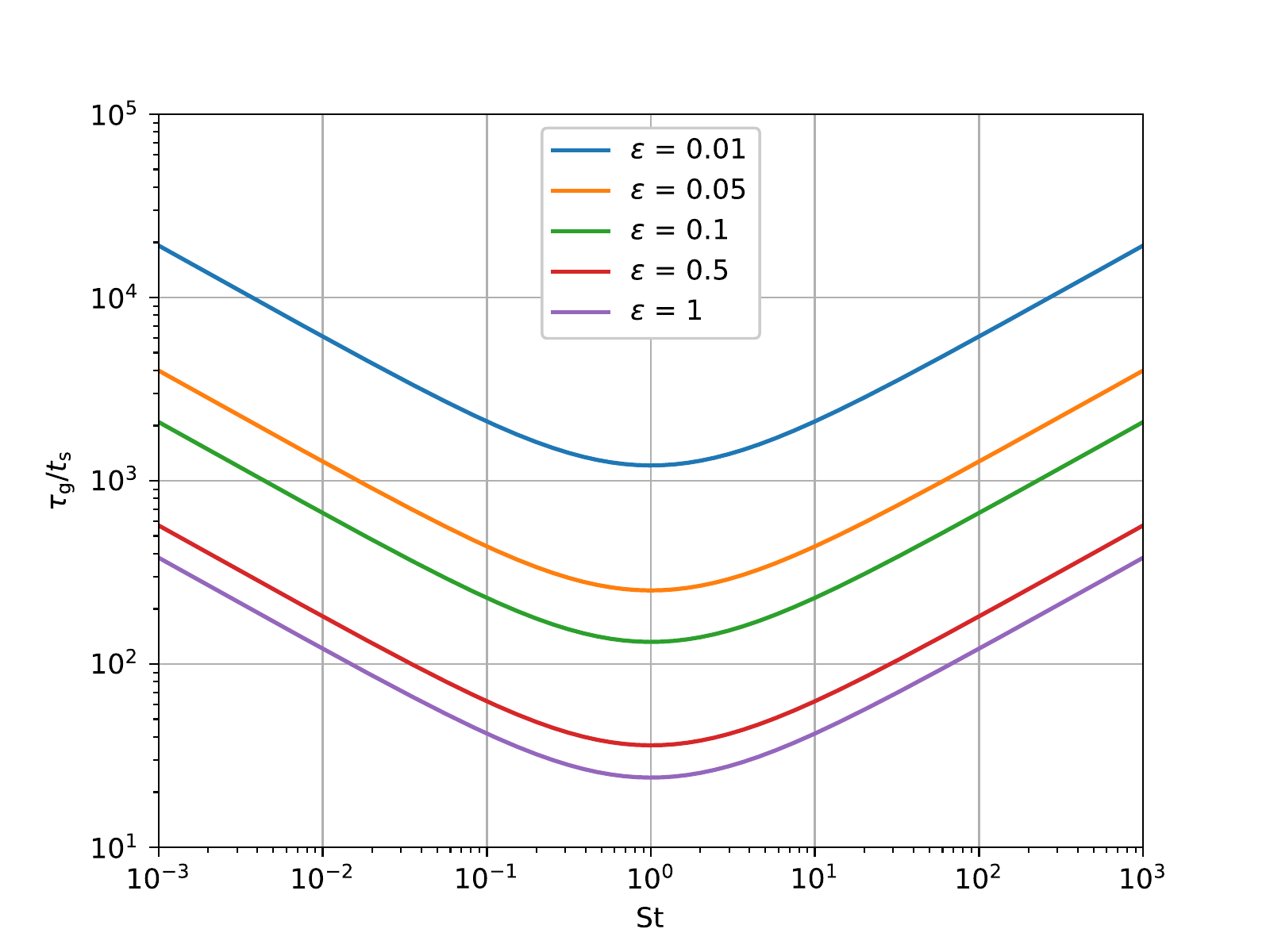}
}
\caption{Ratio $\tau_{\rm g}/t_{\rm s}$ (Equation~\ref{eq:tgts}) as a function of the Stokes number for different values of the dust-to-gas ratio.}
\label{cfl}
\end{figure}
As seen in Fig.~\ref{cfl}, the minimum timestep required to respect the CFL condition is much larger than the stopping time. This condition is already satisfied in \textsc{Phantom}, since the timestep is always at most equal to the stopping time. This also results in very small timesteps for tightly coupled grains, i.e.\ the smallest ones, which can drastically slow down the simulations.
To limit this effect when fragmentation is involved, we incorporate a minimum allowed grain size, under which dust can not fragment. We check and adjust the grain size both during the predictor and corrector steps. 

\section{Benchmark tests}
\label{sec:tests}

\textsc{Phantom} is equipped with a set of tests that are performed frequently to hunt down potential bugs. Implementing a dust growth algorithm comes with a few more tests that can be added to these checkups. In this section, we (re-)present two dust growth-related tests: the existing \textsc{Dustybox} and the new \textsc{Farmingbox}.

\subsection{The return of the \textsc{Dustybox}}
\label{subsec:dustybox}

\begin{figure}
\centering
\resizebox{\hsize}{!}{
\includegraphics[width=\columnwidth]{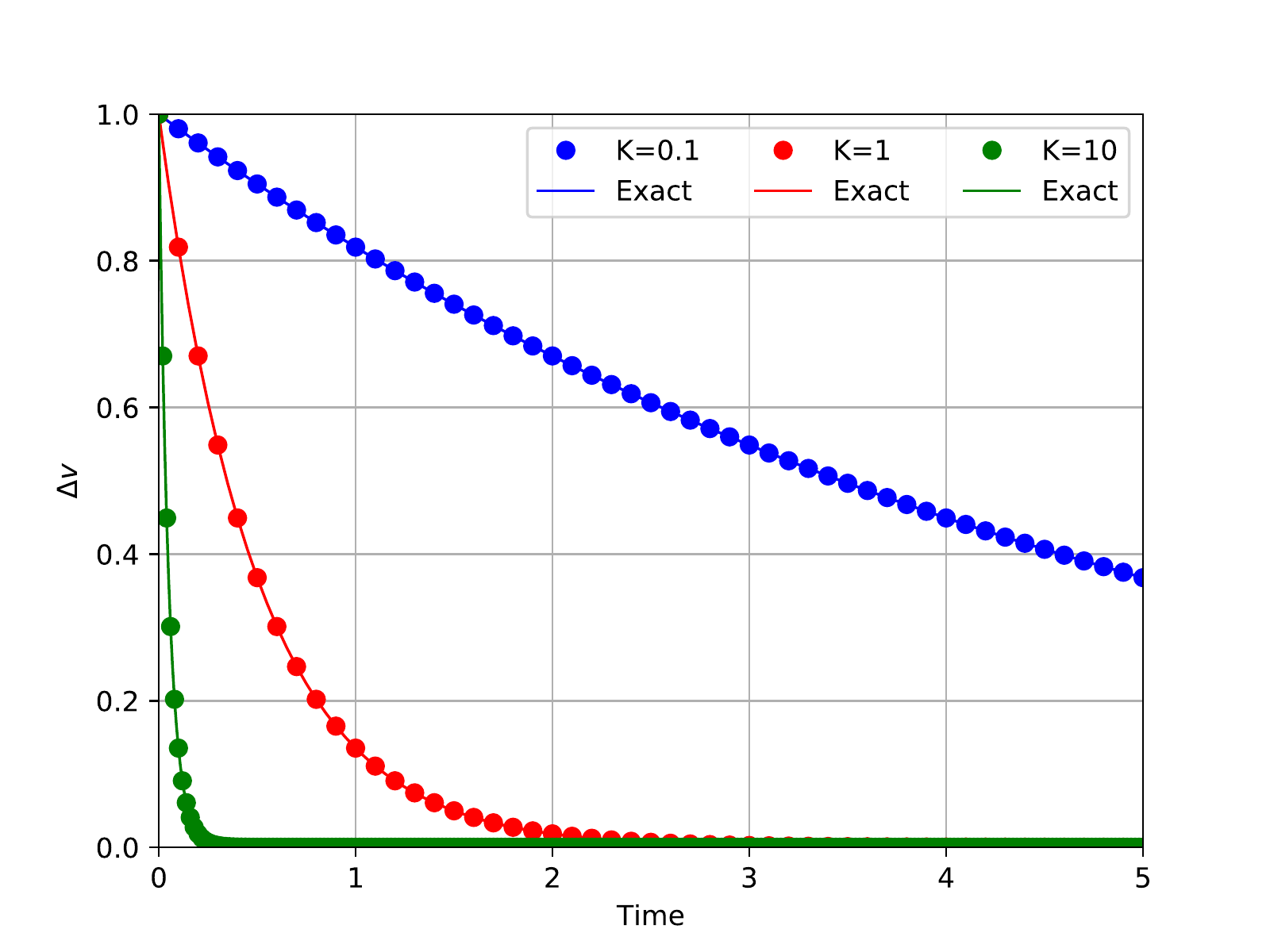}
}
\caption{Differential velocity as a function of time for 3 different values of the drag coefficient $K$ in \textsc{Dustybox}. The solid lines represent the analytical solutions while the dots represent interpolated values during the test. All the values are in code units throughout the test.}
\label{dusty-dv}
\end{figure}

\textsc{Dustybox} is a simple two-fluid test in which gas and dust have an initial differential velocity in a uniform box. This test ensures that the gas drag suppresses the differential velocity between the two phases according to an analytical solution that is relatively simple (the drag coefficient is constant). This test has been extensively described in \citet{laibeprice2012a} and we refer the reader to this paper for more details about the setup.
We extended this test by adding the verification of the interpolated differential velocity (Equation~\ref{eq:dvinterp}).
With a constant drag coefficient $K$, the differential velocity between gas and dust decreases exponentially in time, with a characteristic timescale that is the stopping time:
\begin{equation}
\Delta v = \Delta v_0 e^{-t/t_\mathrm{s}} = \Delta v_0 e^{-2Kt},
\label{eq:dv-analytical}
\end{equation}
where $\Delta v_0$ is the initial differential velocity and the factor 2 comes from the fact that $\rho_\mathrm{g} = \rho_\mathrm{d}$ in the test (see Equation~\ref{eq:tsglobal}).\\
We present a few examples of this test with different values of the drag coefficient $K$ in Fig.~\ref{dusty-dv}.
As the drag coefficient increases (from blue to green), the grains are more coupled to the gas. As a result the initial differential velocity is more efficiently damped and can reach a steady state (i.e. $\Delta v = 0$) faster.
The differential velocities of all the particles are tested at every timestep and the results show maximum relative errors of about $10^{-4}$ with the analytical solution.

\subsection{\textsc{Farmingbox}: motionless dust growth in the Epstein regime}
\label{subsec:farmingbox}

The main goal of \textsc{Farmingbox} is to simulate the motionless growth or fragmentation of dust particles and compare the integrated size with its analytical solution. To attain an analytically solvable growth rate, we fix the particles at their initial position (which cancels $\Delta v$).
Similarly to \textsc{Dustybox}, we set $\sim 7000$ gas and dust particles on a 3D lattice inside of a 1 $\times$ 0.5 $\times$ 0.3~au$^3$ box.
Gas and dust are set so that the dust experiences the Epstein regime throughout its growth or fragmentation. We use this test in both pure growth and pure Hard fragmentation models. The physical parameters used for each mode are shown in Table~\ref{fbox-setup}. We only present the results of the two-fluid algorithm in this Section, although the one-fluid version {was} also tested and {gave} sensible results.

\begin{table}
\caption{List of parameters used in \textsc{Farmingbox} for both pure growth and pure fragmentation modes. All parameters are uniform throughout the box.}
\label{fbox-setup}
\centering
\begin{tabular}{lcll}
\hline
Parameter & Unit & Growth & Fragmentation \\
\hline
$c_\mathrm{s}$ & m.s$^{-1}$ & $942$ 	& $942$ \\
$\rho_\mathrm{g}$ & kg.m$^{-3}$ & $10^{-8}$ 	& $10^{-8}$ \\
$\rho_\mathrm{d}$ & kg.m$^{-3}$ & $10^{-8}$ 	& $5 \times 10^{-9}$ \\
$\rho_\mathrm{s}$ & kg.m$^{-3}$ & $1000$ 	& $1000$ \\
$s_0$ & m & $10^{-4}$ & $10^{-2}$ \\
$\alpha$ & $\varnothing$ & $10^{-2}$ & $2.5\times 10^{-2}$ \\
\hline
\end{tabular}
\end{table}

\subsubsection{Growth}
Firstly, we test the pure growth case.
Gas and dust particles being motionless, the Schmidt number is simply $\mathrm{Sc} = 1 + \mathrm{St}$. Following \citetalias{laibe08}, we can rewrite Equation~(\ref{eq:dsdt}) with a change of variables such that
\begin{equation}
T = \dfrac{t}{\tau} + 2\sqrt{\mathrm{St}_0}\left(1+\dfrac{\mathrm{St}_0}{3}\right),
\end{equation}
where $\mathrm{St}_0$ is the initial Stokes number of the considered particle, of initial size $s_0$, and $\tau$ is a timescale defined as
\begin{equation}
\tau =  \sqrt{\dfrac{8}{\pi \gamma}}\dfrac{1}{\sqrt{2^{3/2}\mathrm{Ro}\alpha}\Omega_\mathrm{k}}\dfrac{\rho}{\rho_\mathrm{d}}.
\label{eq:tau}
\end{equation}
The Keplerian frequency $\Omega_k$ is set as $r^{-3/2}$ in code units, which emulates a $1$~$M_\odot$ central star. Even though there is no such star in the test, we keep this functional form to maintain the property that growth is different for particles at different distances to the center of the box. This is also appropriate in order to have values of sizes and Stokes numbers that would be found in real simulations.

With these new variables we can rewrite Equation~(\ref{eq:dsdt}) in terms of Stokes number and dimensionless time
\begin{equation}
\dfrac{\mathrm{dSt}}{\mathrm{d}T} = \dfrac{\sqrt{\mathrm{St}}}{1+\mathrm{St}}.
\label{eq:dsdt-St}
\end{equation}
The solution to this equation is \citepalias{laibe08}
\begin{equation}
\mathrm{St} = \dfrac{\sigma}{2} + \dfrac{2}{\sigma} - 2,
\label{eq:St-sol}
\end{equation}
where $\sigma = \left(8 + 9T^2+3T\sqrt{16+9T^2}\right)^{1/3}$.
Note that this solution has been adapted from \citetalias{laibe08} with our definition of the Stokes number (Equation~\ref{eq:tau}).

Finally, this gives the time evolution of the grain size
\begin{equation}
s = \sqrt{\dfrac{8}{\pi \gamma}} \dfrac{\rho c_\mathrm{s}}{\rho_\mathrm{s}\Omega_k} \left(\dfrac{\sigma^2 - 4\sigma + 4}{2\sigma}\right) = s_0 \left(\dfrac{\sigma^2 - 4\sigma + 4}{2\mathrm{St}_0\sigma}\right).
\label{eq:s-new}
\end{equation} 
\begin{figure}
\centering
\resizebox{\hsize}{!}{
\includegraphics[width=\columnwidth]{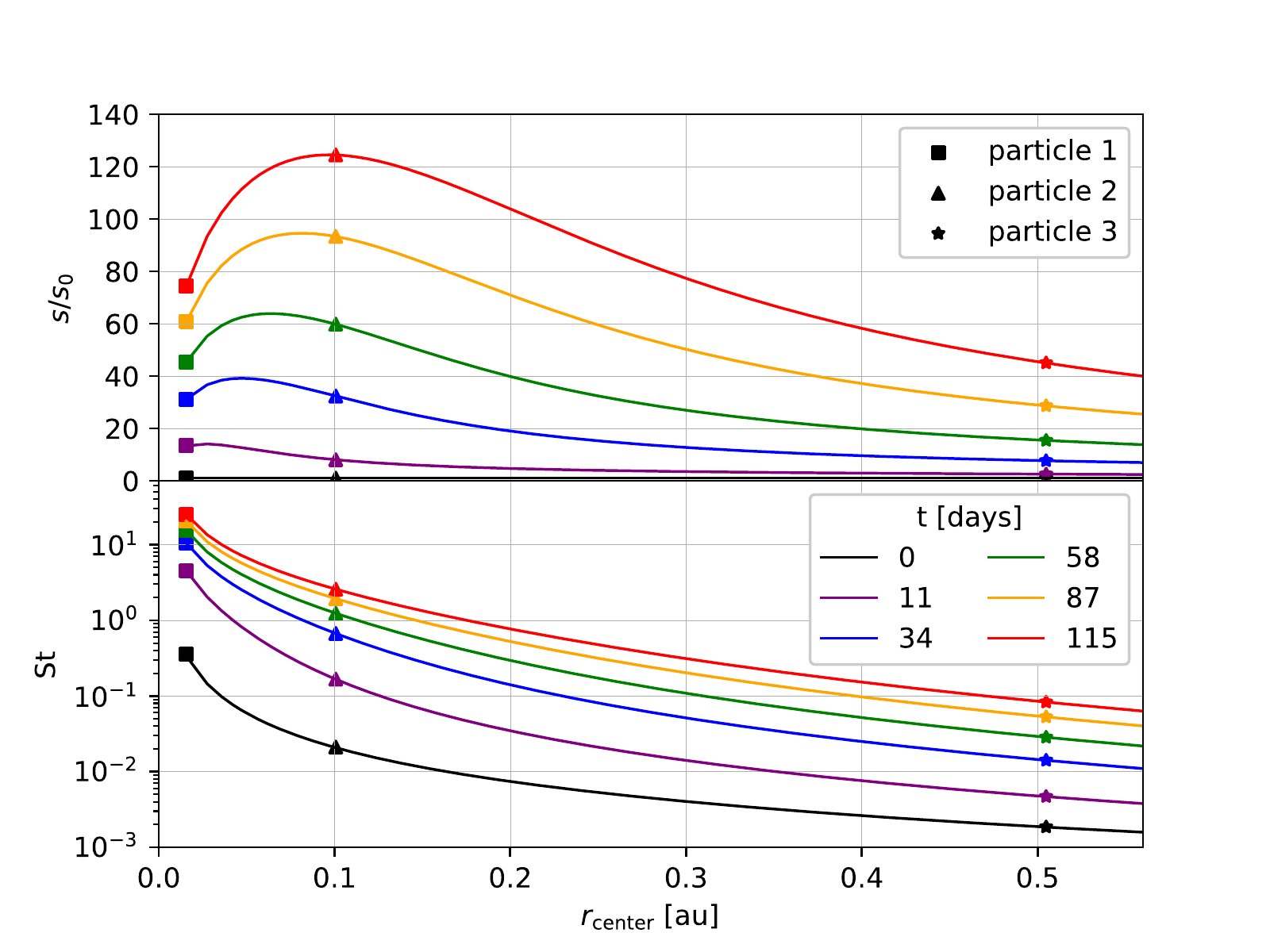}
}
\caption{Dust size (top) and Stokes number (bottom) analytical distributions as a function of the distance to the center of the \textsc{Farmingbox} for 6 different times in the pure growth case. Three specific particles are plotted on top (squares, triangles and stars). The error in L1 norm is approximately $6.13 \times 10^{-3}$.}
\label{fbox-g-allp}
\end{figure}
\begin{figure}
\centering
\resizebox{\hsize}{!}{
\includegraphics[width=\columnwidth]{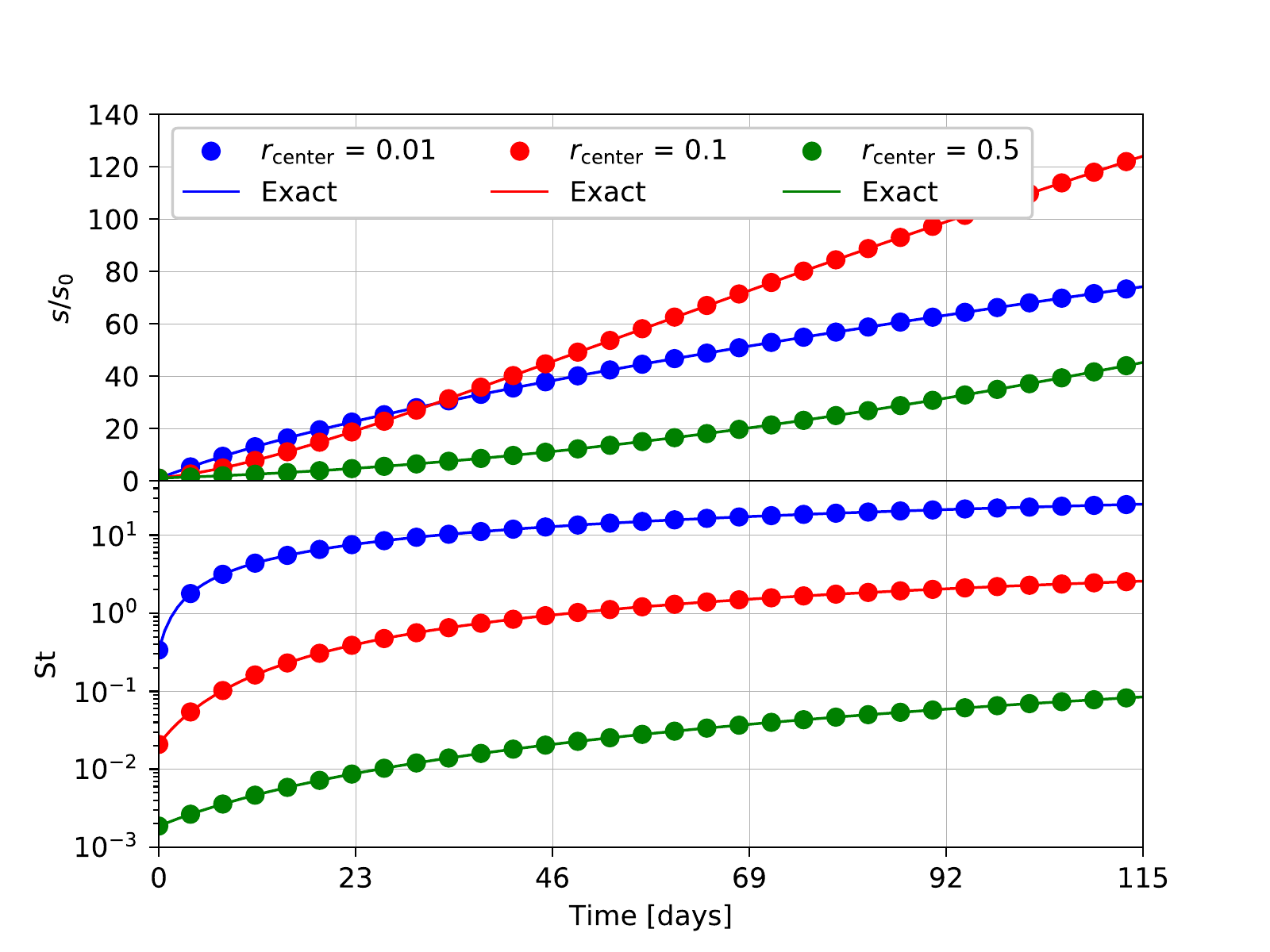}
}
\caption{Size (top) and Stokes number (bottom) evolution of the 3 particles plotted in Fig.~\ref{fbox-g-allp} as a function of time. The lines represent the exact solutions while the dots are the values outputted by the code.}
\label{fbox-g-3p}
\end{figure}
Figs.~\ref{fbox-g-allp} and~\ref{fbox-g-3p} show the results of \textsc{Farmingbox} in pure growth mode. Initially, all the particles have the same size (top panel of Fig.~\ref{fbox-g-allp}, black line), which results in the Stokes number profile to be the same as the Keplerian frequency (bottom panel, black line) and spanning between 2 and 3 orders of magnitude from $10^{-3}$ and up. As growth starts, particles at different radii grow differently due to their different Stokes number. More specifically, the growth rate is maximum for Stokes unity and decreases when grains are both very coupled or decoupled from the gas. As a result, the final size distribution (top panel, red curve) shows a maximum at $r \sim 0.1$~au, corresponding to the location where the grains have spent an optimal amount of time around $\mathrm{St} = 1$. Fig.~\ref{fbox-g-3p} demonstrates that in more details, as particle 1 (bottom panel, blue dots) crosses $\mathrm{St} = 1$ rapidly and decreases its growth rate (top panel). Conversely, particle 2 (red dots) starts with a lower Stokes number but spends most of its time close to Stokes unity, which overall is more efficient at making the particle grow. Finally, particle 3 (green dots) spends the entire test at low Stokes numbers due to its larger distance from the center of the box, which is growth inefficient.

\subsubsection{Fragmentation}

{We} now consider pure fragmentation.
Similarly to the pure growth case, we introduce the same change of variable but with the time going in the opposite direction
\begin{equation}
T = -\dfrac{t}{\tau} + 2\sqrt{\mathrm{St}_0}\left(1+\dfrac{\mathrm{St}_0}{3}\right),
\end{equation}
which also gives Equation~(\ref{eq:dsdt-St}) and the subsequent solution (equations~\ref{eq:St-sol} and~\ref{eq:s-new}).
In this particular case of pure fragmentation, $T$ can reach negative values, which would correspond to a negative size, which is not possible. As this is a physical limit, we ensure that $T$ stays positive throughout the test, i.e. we set the physical parameters in the box so that (see Table~\ref{fbox-setup})
\begin{equation}
t_\mathrm{max} < 2\sqrt{\mathrm{St}_0}\left(1+\dfrac{\mathrm{St}_0}{3}\right)\tau,
\end{equation}
where $t_\mathrm{max}$ is the maximum duration of the test, i.e. the time after which $T < 0$.
\begin{figure}
\centering
\resizebox{\hsize}{!}{
\includegraphics[width=\columnwidth]{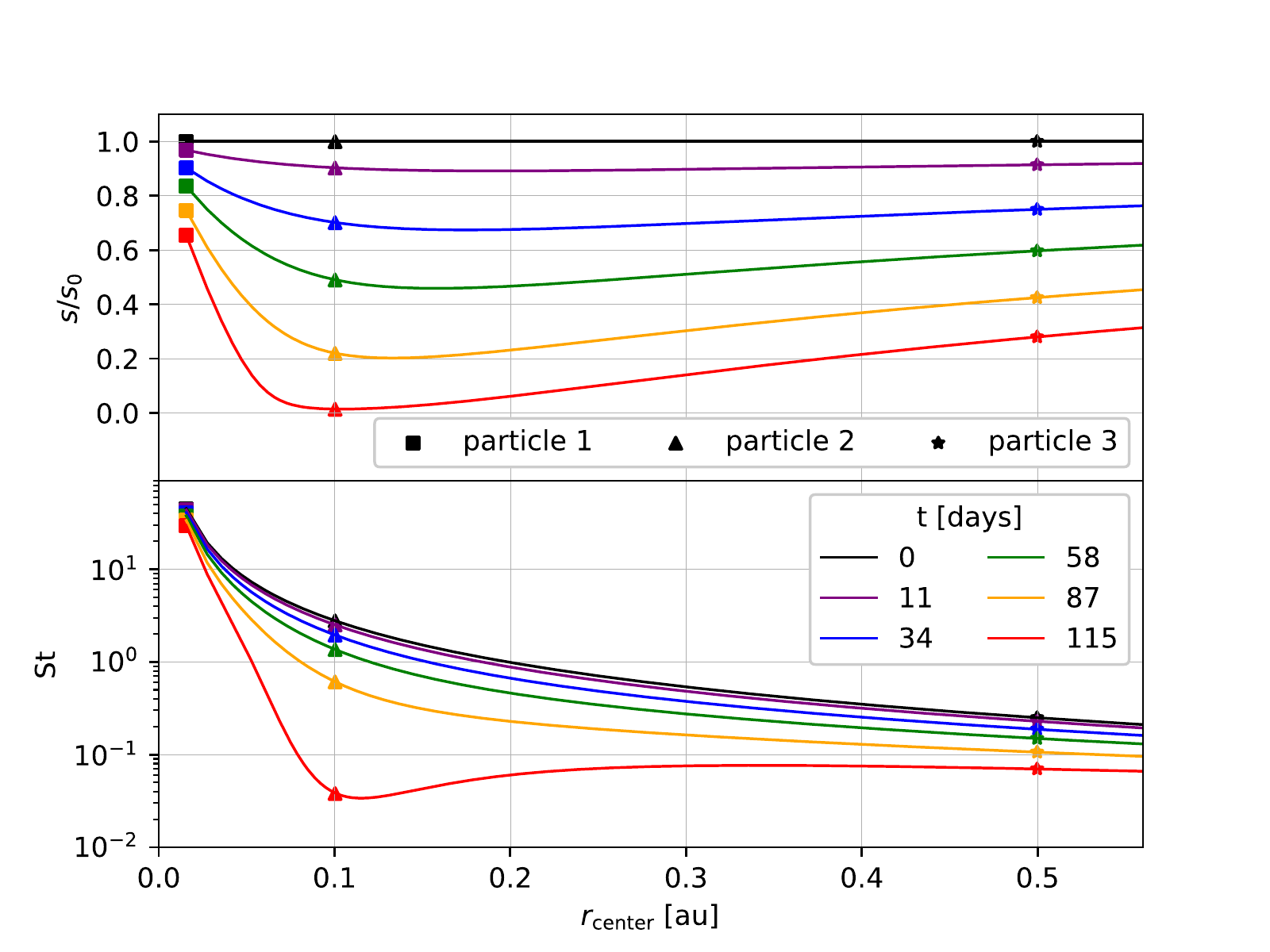}
}
\caption{Similar to Fig.~\ref{fbox-g-allp} but for pure fragmentation.}
\label{fbox-f-allp}
\end{figure}
\begin{figure}
\centering
\resizebox{\hsize}{!}{
\includegraphics[width=\columnwidth]{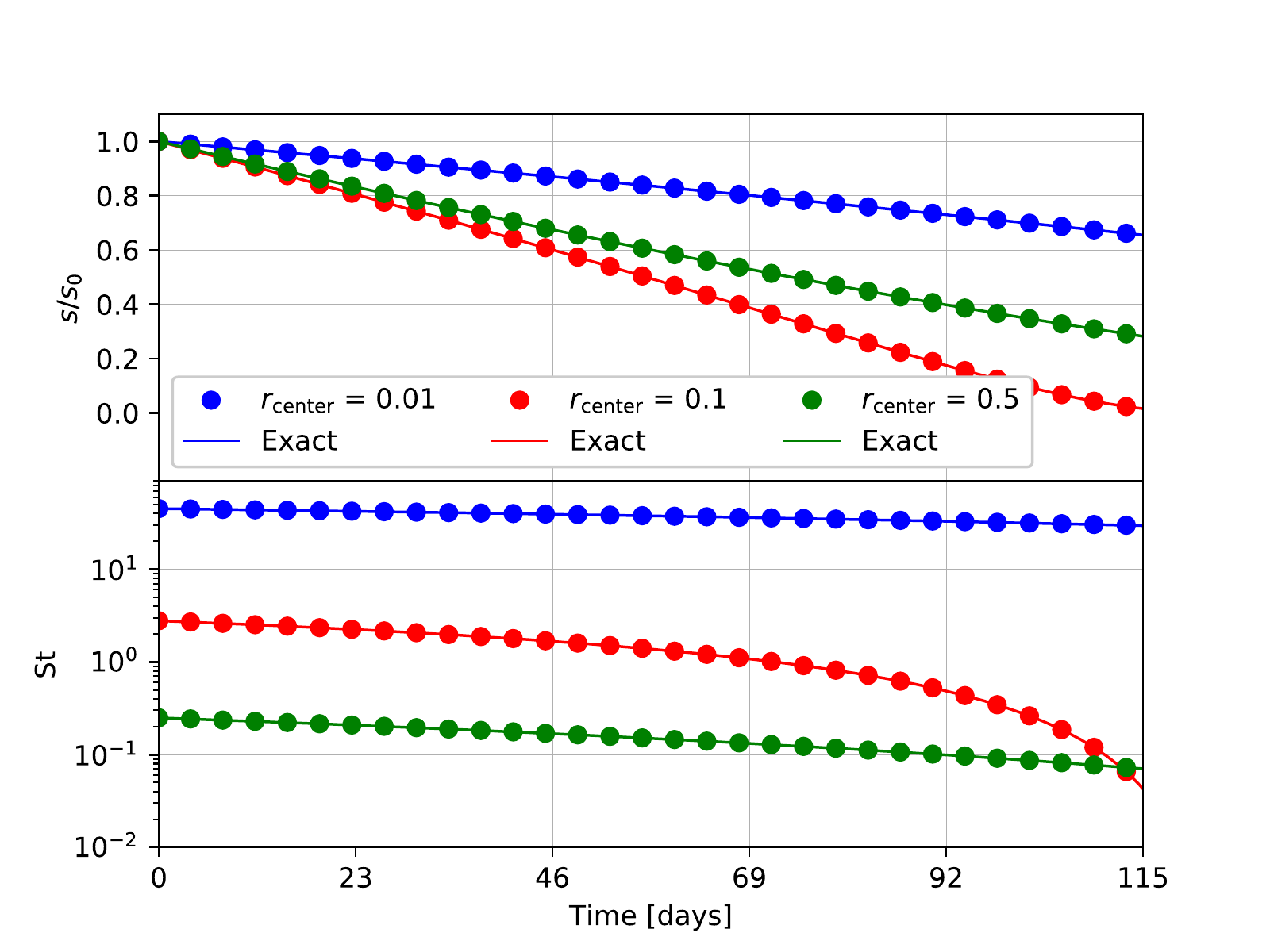}
}
\caption{Similar to Fig.~\ref{fbox-g-3p} but for pure fragmentation. The error in L1 norm is approximately $1.16 \times 10^{-3}$.}
\label{fbox-f-3p}
\end{figure}
Figs.~\ref{fbox-f-allp} and~\ref{fbox-f-3p} show the results of \textsc{Farmingbox} in pure fragmentation mode.
The behaviour is very similar to the growth case, in the sense that fragmentation is also the most efficient near Stokes unity, which is again observed for grains at $r\sim0.1$~au.\\

During \textsc{Farmingbox}, the Stokes number, size, and sound speed of all dust particles are tested against the analytical values at each timesteps. Fig.~\ref{fbox-g-allp} to~\ref{fbox-f-3p} only display a subset of these data points to keep the plots readable. The relative errors between the quantities and their analytical values throughout \textsc{Farmingbox} stay {in the range of $10^{-3} - 10^{-4}$}.

\section{Circumstellar disc simulations}
\label{sec:simus}
In this Section, we perform several 3D circumstellar disc simulations to compare the results given by \textsc{Phantom} with previous studies. In that regard, we will simulate a single disc model, that we will refer to as `Standard' (Std) and that is similar to the one used in previous papers \citep[e.g.][]{barriere05,laibe08}. We only consider the two-fluid algorithm from now on.

\subsection{Numerical setup}

We model a $1$~$M_\odot$ star around which orbits a $0.01$~$M_\odot$ disc represented by $1.2 \times 10^6$ particles: 1M gas and 0.2M dust. This ratio has been chosen to resolve the gas scale height and to limit dust over-concentrations (see Section~\ref{sec:discu_num}). The gas particles are positioned between $r_{\rm in}$ and $r_{\rm out}$ (see Table~\ref{tab:discmodel}), whereas the dust particles only extend up to $5/6\,r_{\rm out}$. We found having a dust disc smaller than the gas disc to be quite important in order to avoid numerical artefacts linked to the gas relaxation in the outer disc. This numerical choice is also justified by the radial drift, which inevitably tends to reduce the size of the dust disc.
To assist the disc relaxation in the inner parts of the disc, the gas initially follows a smoothed surface density profile such that
\begin{equation}
\Sigma_\mathrm{g}(r) = \Sigma_0 \left(1 - \sqrt{\dfrac{r_\mathrm{in}}{r}}\right)\left(\dfrac{r}{r_0}\right)^{-p},
\end{equation}
where $r$ refers to the cylindrical radius and $\Sigma_0$ is set by the total disc mass and is then allowed to evolve. The dust disc initially follows the same law, with an initial and uniform dust-to-gas ratio of 1\%. Both gas and dust particles are removed from the simulation if they cross an accretion radius $r_\mathrm{acc}$ that is set to be equal to $r_\mathrm{in}$ given our interest for the regions where $r > r_{\rm in}$.
The gas temperature is supposed locally isothermal and follows the usual power-law prescription
\begin{equation}
T(r) = T_0 \left(\dfrac{R}{r_0}\right)^{-q},
\end{equation}
where $R$ refers to the spherical radius and $T_0$ is set by the disc aspect ratio at $r_0$ (see Table.~\ref{tab:discmodel}).
In order to represent a \citet{shaksuny73} viscosity parameter $\alpha_{\rm SS} = 5 \times 10^{-3}$, we follow the formalism of \citet{lodatoprice10}, in which the SPH artificial viscosity parameter can be related to $\alpha_{\rm SS}$ as
\begin{equation}
\alpha^{\rm AV} \sim \dfrac{\alpha_{\rm SS}}{10} \dfrac{\langle h \rangle}{H},
\end{equation}
where $<h>$ is the average smoothing length on a given portion of the disc, with $\alpha^{\rm AV} \sim 0.2$. We take the coefficient preventing particle interpenetration $\beta^{\rm AV} = 2$.

The dust initially has a uniform size $s_0$ of $30$~$\mu$m throughout the disc and is allowed to grow or fragment in agreement with the model presented in Section~\ref{sec:model}. The influence of the initial grain size is explained in Appendix~\ref{app:s0}.
Considering the low disc mass, we neglect the disc self-gravity but do take into account the dust back-reaction onto the gas.
\begin{table}
\caption{The disc model used in our simulations, with $r_0 = 1$ $\mathrm{au}$.} 
\label{tab:discmodel}
\centering
\begin{tabular}{lcccccc}
\hline
Name & $p$ & $q$ & $M_\mathrm{disc}$ & $H/r_0$ & $r_{\mathrm{in}}$ & $r_{\mathrm{out}}$  \\
\hline
Std & $3/2$ & $3/4$ & $0.01$~$M_\odot$ & $0.0281$ & $20$~au & $300$~au \\
\hline
\end{tabular}
\end{table}

The short list of circumstellar disc simulations performed for this paper is showed in Table~\ref{tab:sims}.
\begin{table}
\caption{Simulation suite.} 
\label{tab:sims}
\centering
\begin{tabular}{lccc}
\hline
Label & Fragmentation & $V_{\rm frag}$ [m.s$^{-1}$] & Back-reaction \\
\hline
noF & No & $\varnothing$ & Yes \\
F-Hard & Hard & 15 & Yes \\
F-Hard-noBR & Hard & 15 & No \\
F-Smooth & Smooth & 15 & Yes \\
\hline
\end{tabular}
\end{table}

\subsection{Pure growth}
\label{subsec:PG}
{We first consider and analyse a} simulation without fragmentation (noF). Pure dust growth in a similar disc and with a similar growth model has been studied by \citetalias{laibe08}, therefore our results will be compared to theirs.

\subsubsection{Radial drift and dust concentration}

\begin{figure*}
\centering
\resizebox{\hsize}{!}{
\includegraphics[]{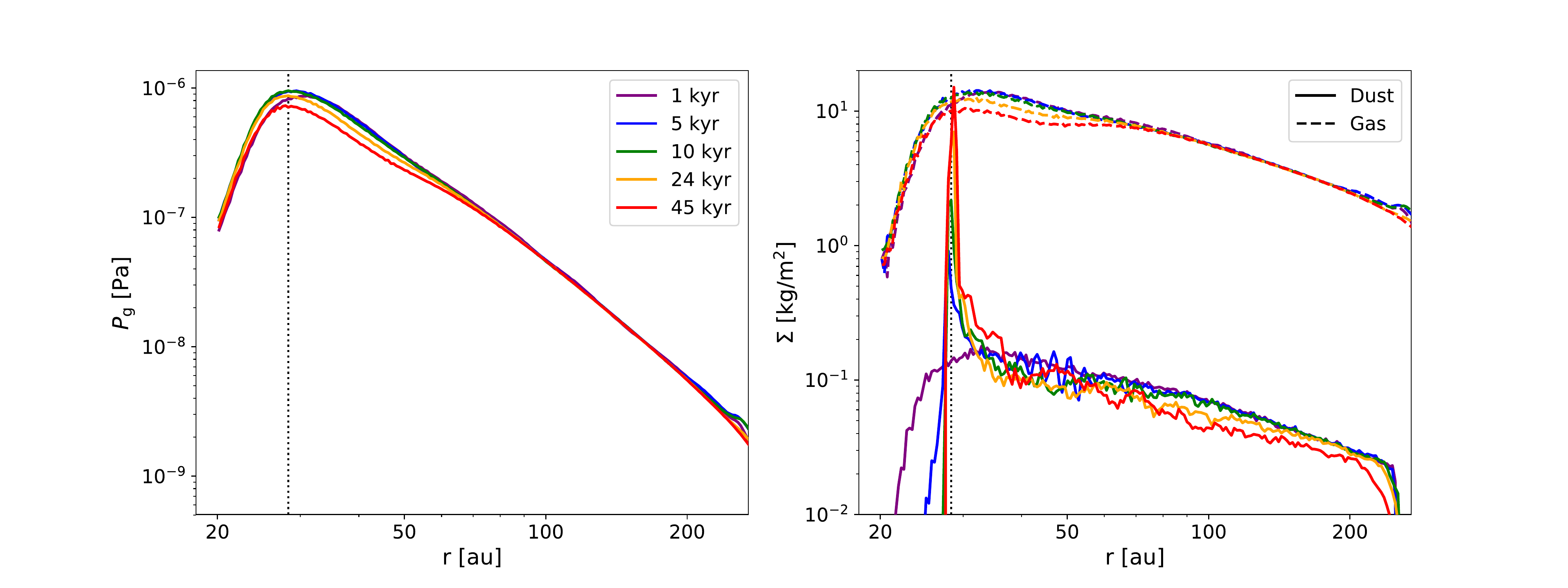}
}
\caption{\textbf{Left}: Pressure radial profile at 5 different times between 1 and 45~kyr in simulation noF. The position of the pressure maximum is showed with the vertical black dotted line. \textbf{Right}: Dust (solid lines) and gas (dashed lines) surface density radial profiles for the same times as the left plot. The position of the pressure maximum is also shown with the vertical black dotted line at $r\sim 28$~au.}
\label{PG:Psig}
\end{figure*}

Fig.~\ref{PG:Psig} shows the pressure and surface density radial profiles at different times.
The gas reaches a steady state in a few thousand years, resulting in a pressure maximum located near 28~au (left panel), which corresponds to the inner edge of the disc.
The negative pressure gradient between $\sim 28$~au and the outermost parts of the disc induces a headwind on the dust that leads to its radial drift towards the star. As a result, the dust mass is transferred towards the pressure maximum (right panel), where the headwind cancels itself out.
At this location the dust concentrates, while the outer parts of the disc are slowly being emptied. The dust radial drift also reduces the extent of the dust disc as one can see from the dust surface density profile (right panel). This whole process is accelerated by dust growth, which increases radial drift velocities up to the point where $\mathrm{St = 1}$.

\begin{figure}
\centering
\resizebox{\hsize}{!}{
\includegraphics[width=\columnwidth]{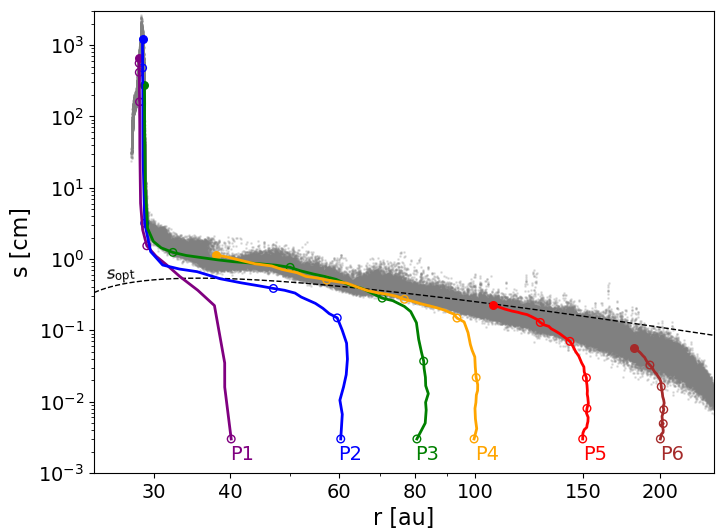}
}
\caption{Trajectories in the ($r, s$) plane of 6 particles initially in the mid-plane of the disc (solid lines) between 0 and 50~kyr in simulation noF. Empty circles of the same colour mark each particle's position every 10~kyr, with a filled circle for the final state. They grey points in the background represent the dust distribution at 50~kyr. The black dashed line represents $s_{\rm opt}$ (Equation~\ref{eq:sopt}) at $t=0$ ($\varepsilon=0.01$), i.e. where $\mathrm{St}=1$.}
\label{PG:P1-6}
\end{figure}

To study the coupled effects of dust growth and radial drift, we follow in Fig.~\ref{PG:P1-6} the evolution of 6 particles initially in the mid-plane of the disc with a size $s_0=30$~$\mu$m. Their initial distance to the star are 40, 60, 80, 100, 150 and 200~au respectively. We also over-plotted an estimation of $s_{\rm opt}$ at $t=0$, for which we considered the dust-to-gas ratio to be uniform and the differential velocity to be subsonic (such that $f\sim1$, see Equation~\ref{eq:kwok}).
While this estimation is limited near local dust concentrations, it still provides a useful comparative basis for the rest of the simulation.
Dust particles shown in Fig.~\ref{PG:P1-6} have similar behaviours that can be decomposed into 3 steps:
\begin{enumerate}
    \item{A growth phase with very little radial drift: $\mathrm{St} \ll 1$.}
    \item{A phase of radial drift and growth where radial drift is more efficient: $\mathrm{St \sim 1}$.}
    \item{A slower growth without radial drift: $\mathrm{St} \gg 1$.}
\end{enumerate}
The further away the particles are, the longer it takes for them to experience all three phases. For example, after 50~kyr, particles P1 to P3 experienced these different steps, whereas P4 is on the verge of entering phase (iii). Particles P6 is only entering phase (ii).

These different steps shape the dust distribution as we can see on the background (grey points, shown at 50~kyr). Grains in phase (i) form a tail at $r \gtrsim 200$~au, while grains in phase (ii) form a diagonal between $\sim 40$ and $\sim 200$~au. Particles that are decoupled (phase (iii)) form a reservoir of material at the pressure maximum, that is at $\sim 28$~au. The three specified evolutionary stages are the same as those identified in \citetalias{laibe08}.

The diagonal formed by drifting and growing grains (phase (ii)) is steeper than the curve corresponding to $s_{\rm opt}$. This result, also found by \citetalias{laibe08}, shows the effect of dust growth during the radial drift process, which increases their Stokes number faster than radial drift decreases it. {This behaviour was also found analytically by \citet{laibe14drift}.}

\subsubsection{Vertical settling}

\begin{figure}
\centering
\resizebox{\hsize}{!}{
\includegraphics[width=\columnwidth]{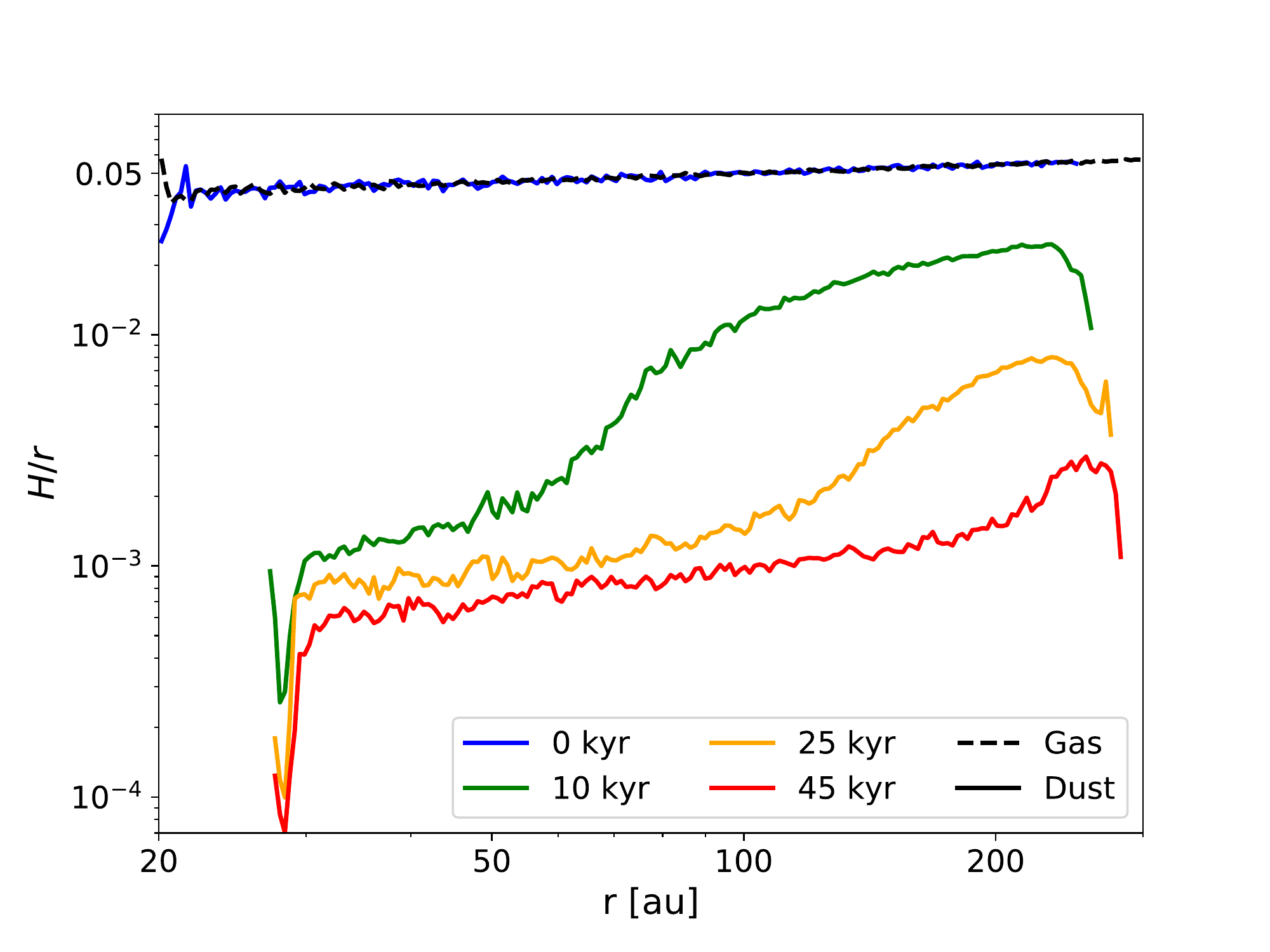}
}
\caption{Scale height radial profiles of the gas (black dashed line) and dust (coloured solid lines) discs at 6 different times between 0 and 45~kyr in simulation noF.}
\label{PG:hoverr}
\end{figure}

Another important aspect of dust growth that we want to explore is its effect on vertical settling. To this end, Fig.~\ref{PG:hoverr} shows the evolution of the disc scale height.
Pure growth is very efficient at transforming the initially flared dust disc into a thin stratified layer in the mid-plane. While the gas disc has a roughly constant aspect ratio of about 0.05 throughout the simulation, the dust disc aspect ratio tends to be less than $10^{-3}$ in a few tens of thousands of years.
This process is efficient because dust decouples from the gas as it grows. More precisely, the aerodynamic drag efficiently {damps} the vertical oscillations of the dust particles around the mid-plane, which tends to form layers with a thickness limited by the disc turbulence \citep{dubrulle95,fromang06}. These studies predict that the dust layer thickness is roughly proportional to $\mathrm{St
^{-1/2}}$. This is consistent with what we find, since the average Stokes number decreases when the distance to the star increases (see Fig.~\ref{PG:P1-6}, particles above $s_{\rm opt}$ are in the inner disc {in contrast to} the particles in the outer disc).

\begin{figure*}
\centering
\resizebox{\hsize}{!}{
\includegraphics[]{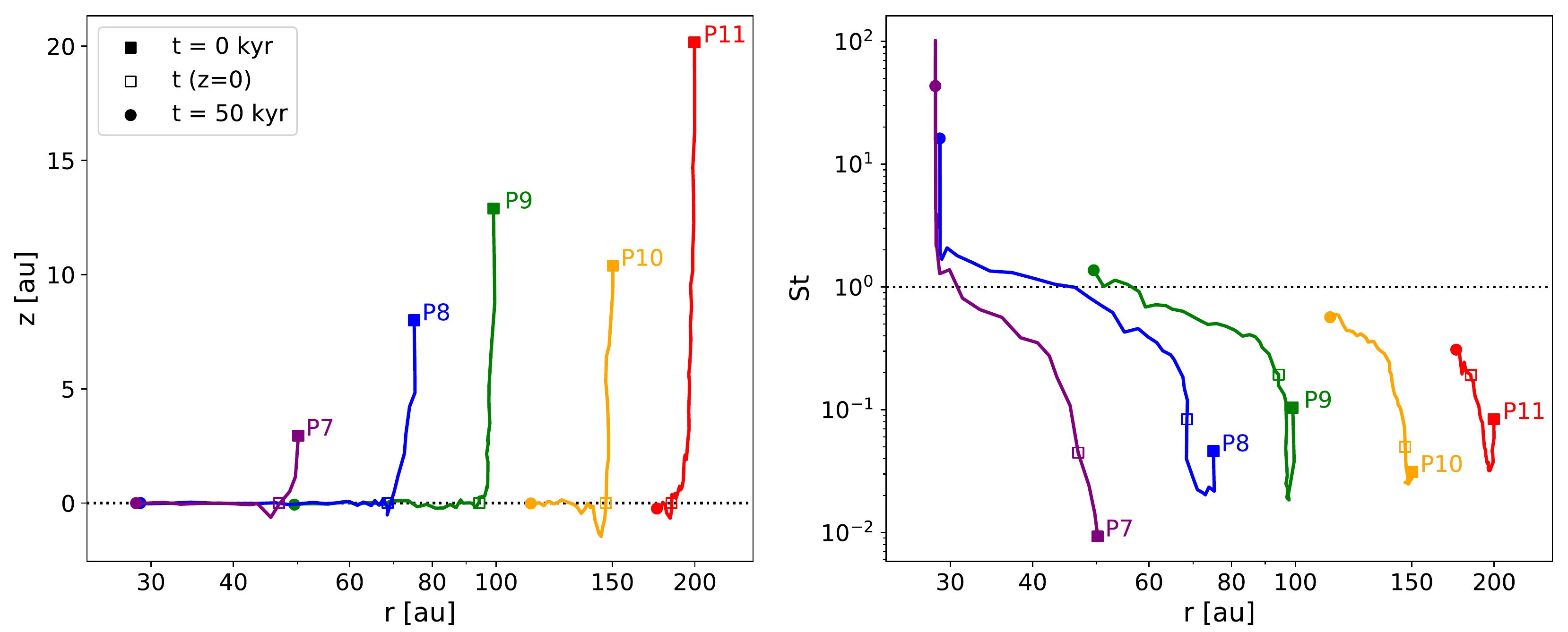}
}
\caption{\textbf{Left}: Trajectories in the ($r, z$) plane of 5 particles between 0 and 50~kyr in simulation noF. The filled squares, empty squares and filled circles represent the initial time, the time of fist contact with the mid-plane and the final time respectively. \textbf{Right}: Same as the left panel but for the ($r, \mathrm{St}$) plane.}
\label{PG:P7-11}
\end{figure*}

{We} take advantage of the Lagrangian formalism and follow the vertical settling process of a few particles in more detail. We select 5 particles, noted P7 to P11, with an initial distance to the star of 50, 70, 100, 150 and 200~au respectively. Fig.~\ref{PG:P7-11} plots their evolution in altitude and Stokes number.
Particle trajectories in the ($r, z$) plane are first vertical, and then radial, allowing us to consider both movements as decoupled from each other. The settling timescale is smaller than the drift timescale by a factor of the order of $(H/r)^2 \partial \ln P/\partial \ln r \ll 1$ \citep{garaud04,barriere05}.
The evolution of their Stokes number is another interesting point, as it {reveals two opposing processes} during the settling phase:
\begin{enumerate}
    \item{The settling of grains to denser layers of gas and dust, which strengthens the coupling between the two ($\mathrm{St} \propto (\rho_{\rm d} + \rho_{\rm g})^{-1} \searrow$).}
    \item{Dust growth that tends to decouple the dust from the gas ($\mathrm{St} \propto s \nearrow$).}
\end{enumerate}
For P8, P9 and P11, this competition is clear. The settling and growth dominate the Stokes number evolution one after the other. For P7, its relatively close proximity to the star and to the mid-plane makes the growth process very efficient and therefore it always dominates.
%
The pure growth simulation has shown the coupled effect of dust growth, vertical settling and radial drift. Our results {highlight the exact same dust regimes of evolution as} those {described in} \citetalias{laibe08}.

Although pure growth is an interesting academic case in order to understand the interplay between these different processes, dust growth in discs is more realistically limited by fragmentation when the collision velocities become larger than a given threshold. In the next section, we perform more realistic simulations by taking fragmentation into account.

\subsection{Adding fragmentation}
\label{subsec:G+FRAG}

Considering the same disc, we add fragmentation as presented in Section~\ref{sec:model}. We firstly analyse the simulation with the Hard fragmentation model (F-Hard). We consider a fragmentation threshold of 15~m.s$^{-1}$, value often used for wet icy aggregates \citep[e.g.][]{wada09,gonzalez15,vericelgonzalez20}.
Simulations with fragmentation {were evolved} twice as long as those with pure growth --- that is $\sim 100$~kyr --- because they reach a quasi steady state more slowly than their counterpart.

\subsubsection{General evolution}

\begin{figure*}
\centering
\resizebox{\hsize}{!}{
\includegraphics[]{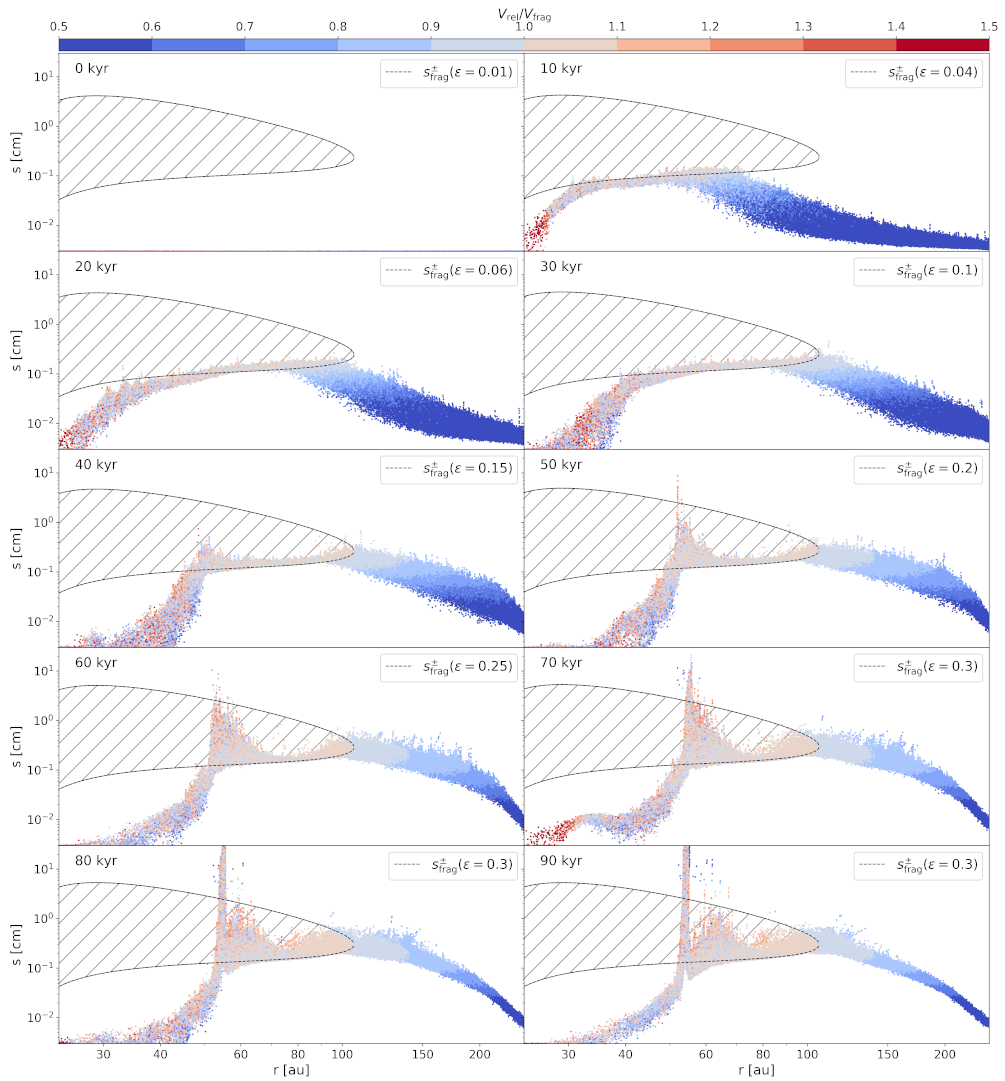}
}
\caption{Dust size radial distribution between 0 and 90~kyr in simulation F-Hard. Particles are coloured with their ratio $V_{\rm rel}/V_{\rm frag}$. The hatched zone corresponds to the fragmentation zone, which is delimited by the fragmentation sizes $s_{\rm frag}^\pm$. These sizes are evaluated with increasing dust-to-gas ratios over time (see legend on each panel).}
\label{F:srevol}
\end{figure*}

The evolution of the dust size radial distribution is firstly displayed in Fig.~\ref{F:srevol}, colour-coded with $V_\mathrm{rel}/V_\mathrm{frag}$. The fragmentation sizes $s_{\rm frag}^{\pm}$ (Equation~\ref{eq:sfrag}) are also over-plotted: they are estimated using the form of $s_{\rm opt}$ computed in Section~\ref{subsec:PG} with the power laws describing the initial disc structure and thus do not reflect the evolution of the gas profile.
The curves corresponding to $s_{\rm frag}^\pm$ define a `fragmentation zone' that is hatched in Fig.~\ref{F:srevol}. 
The fragmentation zone is shown using increasing dust-to-gas ratios over time, which are estimated using the surface density profiles. Notice that the fragmentation zone extends out to $\sim 100$~au, which fits with the calculation of $r_{\rm frag}$ (Equation~\ref{eq:rfrag}) that gives $\sim 105$~au.
%

Dust at $50 \lesssim r \lesssim 100$~au reaches fragmentation sizes $s_{\rm frag}^-$ relatively rapidly ($\sim 20$~kyr, third panel). Meanwhile, dust in the innermost parts of the disc is incapable of growing, which creates a steep size gradient at $r \sim 50$~au ($40-50$~kyr). This size contrast induces a gradient of radial drift velocities, i.e. a traffic jam. In other words, dust concentrates radially near the gradient of sizes, which in the figure can be seen from 40~kyr onward. A second -- but slight -- size gradient can also be observed at the entry of the fragmentation zone ($\sim 100$~au). Nevertheless, this gradient is less important and does not create a significant traffic jam in the disc. Note that once dust grains are trapped at $r \sim 50$~au, they continue to grow even when they appear to be in the hatched area. This is because the latter represents the initial shape of the fragmentation zone, not reflecting the changes in the gas profile or the reduction of the relative velocities of grains due to their confinement in the trap. Finally, the runaway growth seen after $\sim 80$~kyr results from artificial clumping, which locally overestimates the growth rate (see Section~\ref{sec:discu_num}.)

\begin{figure*}
\centering
\resizebox{\hsize}{!}{
\includegraphics[]{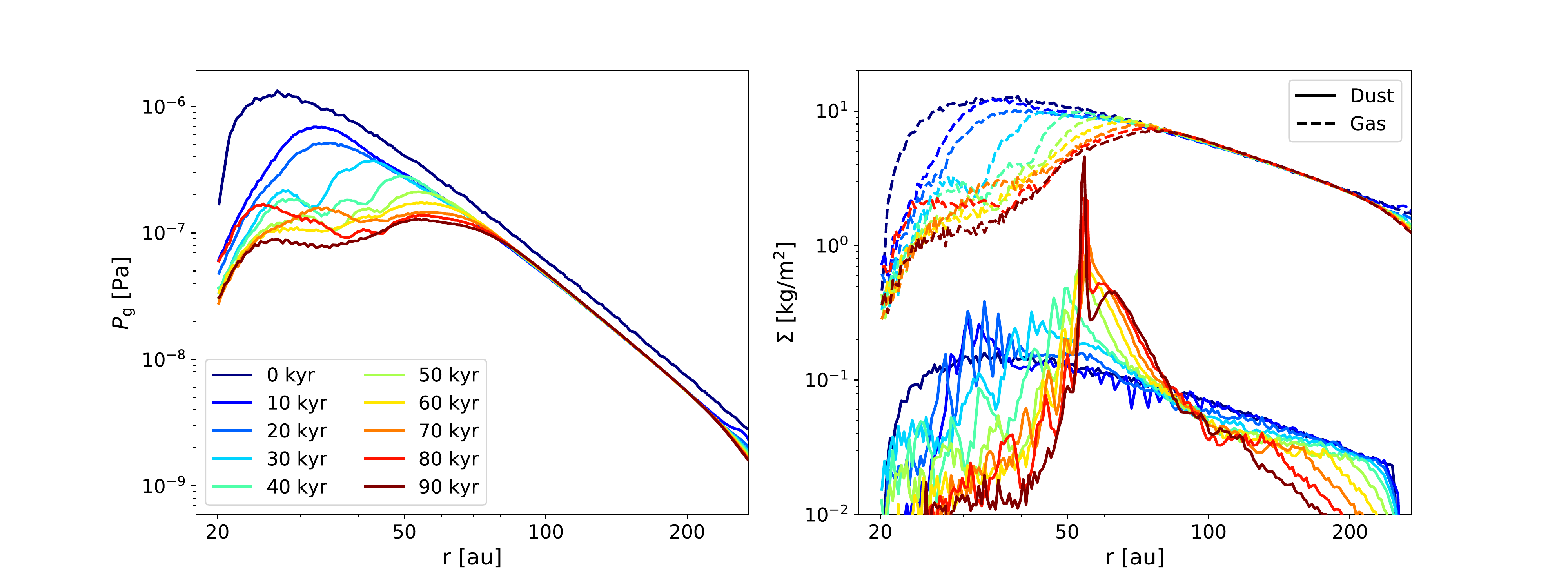}
}
\caption{\textbf{Left}: Pressure radial profiles between 0 and 90~kyr in simulation F-Hard. \textbf{Right}: Same, but for the dust (solid lines) and gas (dashed lines) surface densities.}
\label{F:Psig}
\end{figure*}

To understand in more detail the process taking place in this simulation, we show the evolution of the pressure and surface density profiles in Fig.~\ref{F:Psig}. The dust concentration is clearly observable in the surface density profiles at $\sim 50$~au.
Additionally, a pressure maximum develops near the dust concentration after a few thousand years, which hints at the back-reaction onto the gas as the source of this feature.

The combination of dust growth/fragmentation, its pile-up and the formation of a local pressure maximum in the disc might be the manifestation of the `self-induced dust trap' mechanism proposed by \citet{gonzalez17}. The next section focuses on this hypothesis.
%

\subsubsection{Self-induced dust trap: the importance of the dust back-reaction}

The `self-induced dust trap' mechanism has been proposed as a means for dust to overcome both the radial drift and fragmentation barriers in discs: by taking into account dust growth and fragmentation as well as large-scale gradients, dust particles pile-up and concentrate in the disc, such that the back-reaction reverses the gas flow locally to create a pressure maximum that further traps the dust.

To estimate the importance of the back-reaction, \citet{gonzalez17} considered the stationary solution for the gas radial velocity, which gives \citep{dipierrolaibe17,kanagawa17,gonzalez17}
\begin{equation}
v_{\mathrm{g},r} = - \dfrac{1}{1+\varepsilon} \left[ \dfrac{\varepsilon \mathrm{St}}{1+\mathrm{St}^2}v_{\rm drift} - \left(1 + \dfrac{\varepsilon \mathrm{St}^2}{1+\mathrm{St}^2}\right) v_{\rm visc} \right],
\label{eq:v_gr}
\end{equation}
where
\begin{equation}
v_{\rm drift} = \left(\dfrac{H}{r}\right)^2 \dfrac{\partial \ln P}{\partial \ln r} v_{\rm k}
\end{equation}
corresponds to the drift velocity caused by the pressure gradient \citep{naka86} and
\begin{equation}
v_{\mathrm{visc}} = \dfrac{\dfrac{\partial}{\partial r}\left(\rho_{\mathrm{g}}\nu r^3\dfrac{\partial \Omega_{\mathrm{k}}}{\partial r}\right)}{r\rho_{\mathrm{g}}\dfrac{\partial}{\partial r}(r^2\Omega_{\mathrm{k}})}
\end{equation}
is the usual viscosity-induced velocity \citep{lynden74}.
Notice that the form used by \citet{gonzalez17} is slightly different, because their Stokes number definition concerns the dust only and not the mixture, which differs by a factor $1+\varepsilon$.
The viscosity term tends to push the gas inward, whereas the drag term does the opposite. If the drag term is greater than its counterpart, the net motion is towards the outer parts of the disc, which changes the gas structure locally.

\begin{figure*}
\centering
\resizebox{\hsize}{!}{
\includegraphics[]{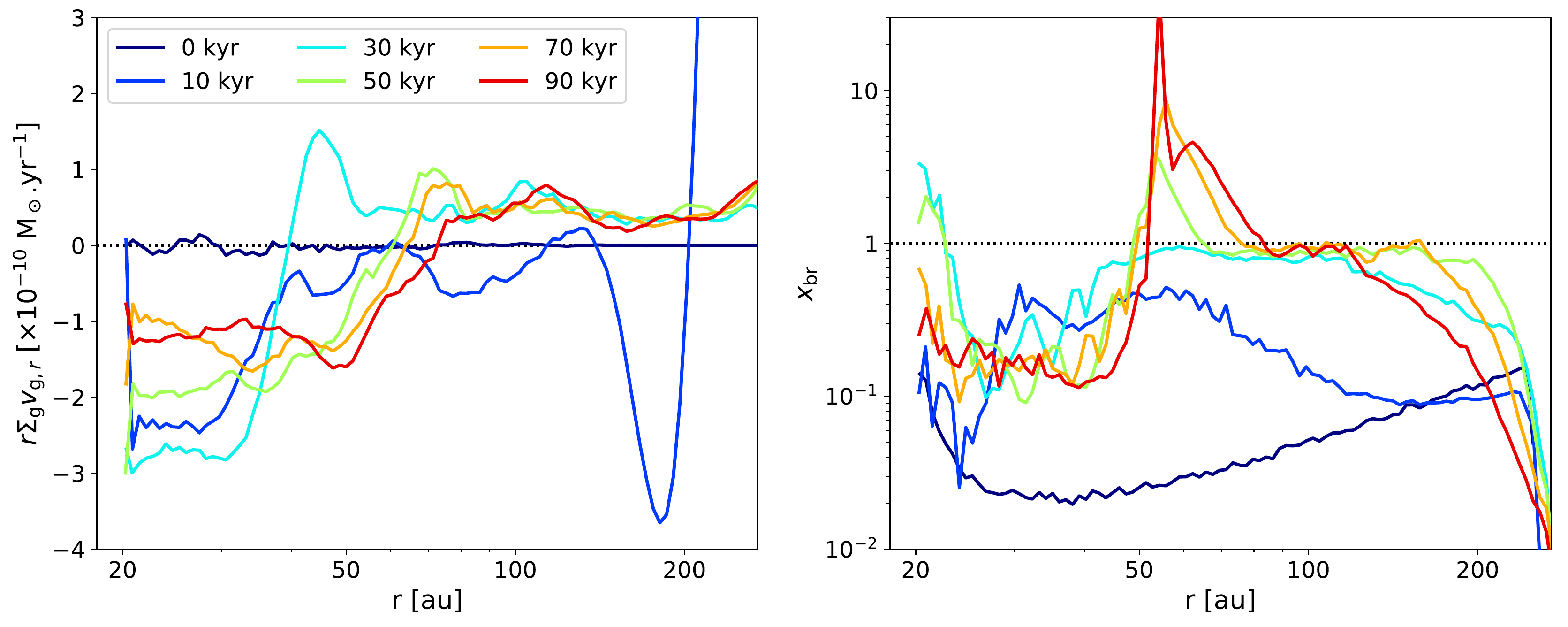}
}
\caption{\textbf{Left}: Gas mass flux radial profiles at different times between 10 and 90~kyr. \textbf{Right}: Same, but for the estimation of $x_{\rm br}$, which is evaluated with the vertically integrated dust-to-gas ratio, that is $\varepsilon = \Sigma_{\rm d}/\Sigma_{\rm g}$.}
\label{F:xbr}
\end{figure*}

To measure the impact of back-reaction on the gas motion, we use a parameter, called $x_{\rm br}$, which measures the balance between the viscous and drag terms in equation~(\ref{eq:v_gr}). Following \cite{gonzalez17,sidterrat}, i.e. assuming $v_{\rm visc}/v_{\rm drift} \sim 1/\alpha$, we subsequently find
\begin{equation}
x_{\rm br} \simeq \dfrac{\alpha^{-1}\varepsilon \mathrm{St}}{1+\mathrm{St}^2(1+\varepsilon)}.
\end{equation}
Where $x_{\rm br} > 1$, $v_{\mathrm{g},r} > 0$ and the gas flows outwards, which corresponds to the formation of a local pressure maximum.
The gas mass flux as well as the estimation of $x_{\rm br}$ are plotted in Fig.~\ref{F:xbr}. When the dust concentration develops ($\sim 30$~kyr), the gas starts to be pulled outwards ($r \Sigma_\mathrm{g} v_{\mathrm{g},r} > 0$, cyan). Moreover, we observe a tendency where the gas seems to be pulled further away with increasing time. This is consistent with the nature of the self-induced dust trap mechanism, because the dust concentration extends at increasing distances to the star over time. Indeed, the regions where $x_\mathrm{br} \ge 1$ becomes more extended as time goes on. The estimation of $x_{\rm br}$ also shows that the back-reaction should be able to reverse the gas flow. This is also consistent with the findings of \citet{garate20}.
%

\begin{figure*}
\centering
\resizebox{\hsize}{!}{
\includegraphics[]{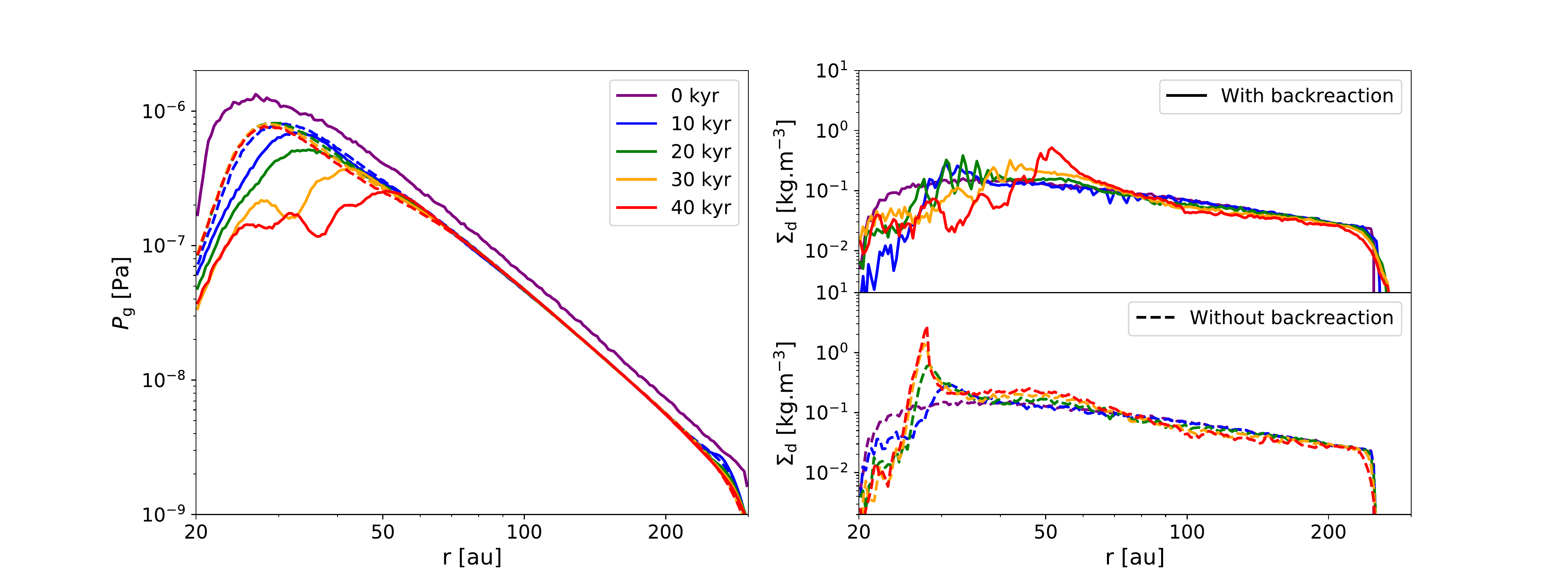}
}
\caption{Same as Fig.~\ref{F:Psig} but with a comparison between simulations with (F-Hard, solid lines) and without (F-Hard-noBR, dashed lines) back-reaction {and only showing $\Sigma_\mathrm{d}$}.
}
\label{F:noBR}
\end{figure*}

\begin{figure*}
\centering
\resizebox{\hsize}{!}{
\includegraphics[]{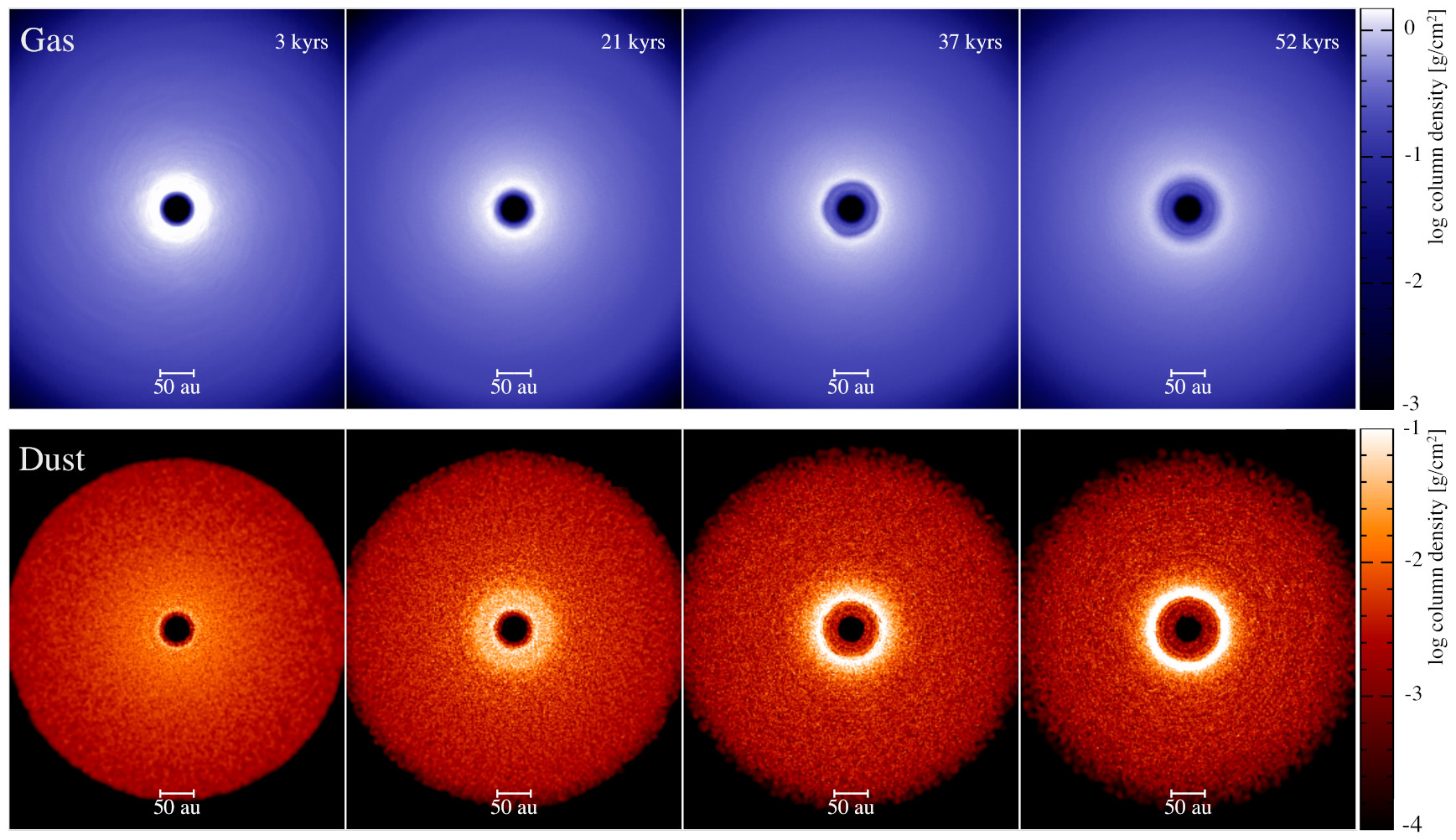}
}
\caption{{Face-on rendered view of the gas (top) and dust (bottom) column densities at 4 different times during the formation of the self-induced dust trap in simulation F-Hard.}}
\label{F:render}
\end{figure*}

{To confirm this hypothesis}, we finally performed a simulation without back-reaction as verification. Fig.~\ref{F:noBR} compares the corresponding radial profiles of pressure and surface density. Without back-reaction, the pressure maximum is the same as the pure growth simulation, i.e. at the inner edge of the disc. In this simulation, no other pressure maximum develops in the disc, even though the dust experiences growth and fragmentation. As a result, grains drift towards the inner edge of the disc and are trapped there, as opposed to the simulation with back-reaction that shows self-trapping at $\sim 50$~au. With this comparison, we are confident that the mechanism observed is a self-induced dust trap. Although this mechanism has been observed several times \citep{gonzalez17,garciathesis18,pignatale19,vericelgonzalez20}, this is the first time it is observed in another, independent code, which validates its existence. {A face-on rendered view of the gas and dust density evolution during the self-induced dust trap formation is shown in Fig.~\ref{F:render}.}
%
%

\subsubsection{Comparison with the pure growth case}

\begin{figure*}
\centering
\resizebox{\hsize}{!}{
\includegraphics[]{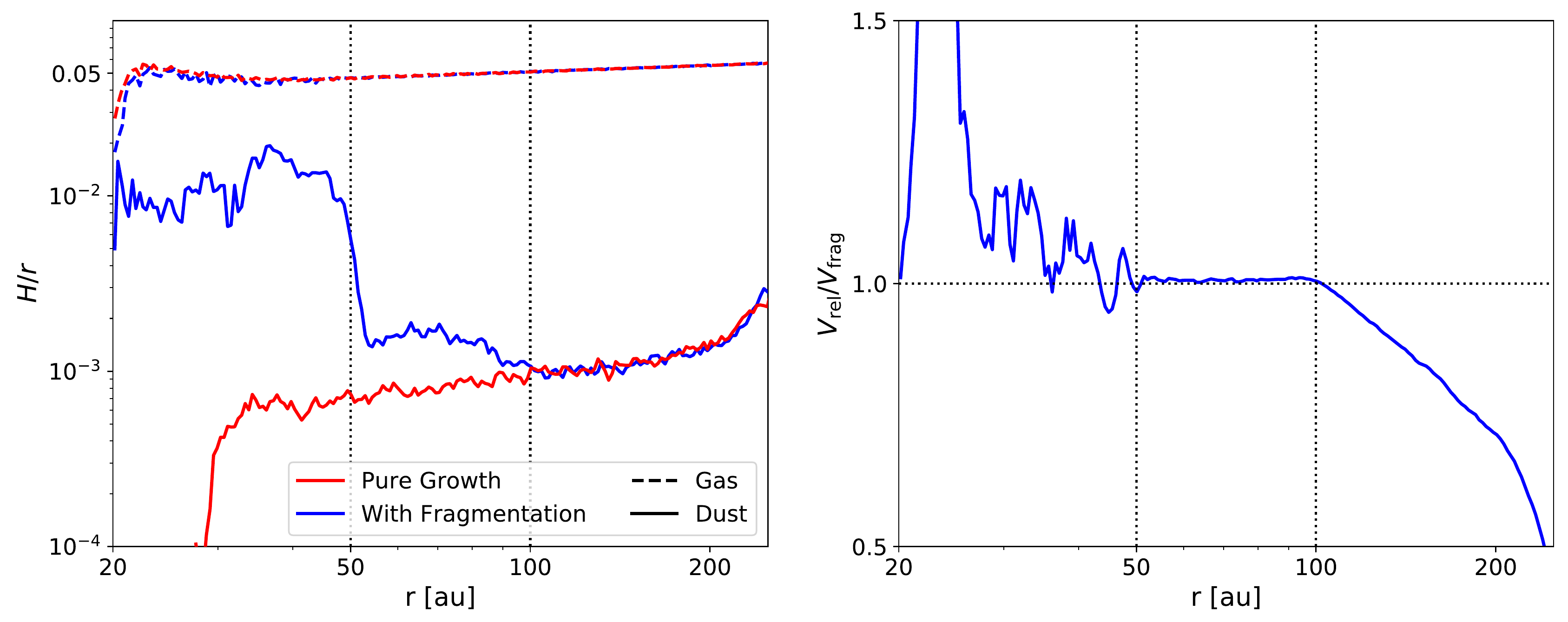}
}
\caption{\textbf{Left}: Disc aspect ratio radial profiles with (F-Hard, blue) and without (noF, red) fragmentation at 50~kyr. \textbf{Right}: Radial profile of the relative-to-fragmentation velocity ratio for simulation with fragmentation at the same time as the left panel. The vertical lines delimit the zone where the dust relative velocity is in equilibrium with the fragmentation velocity.}
\label{F:vsPG}
\end{figure*}

\begin{figure}
\centering
\resizebox{\hsize}{!}{
\includegraphics[]{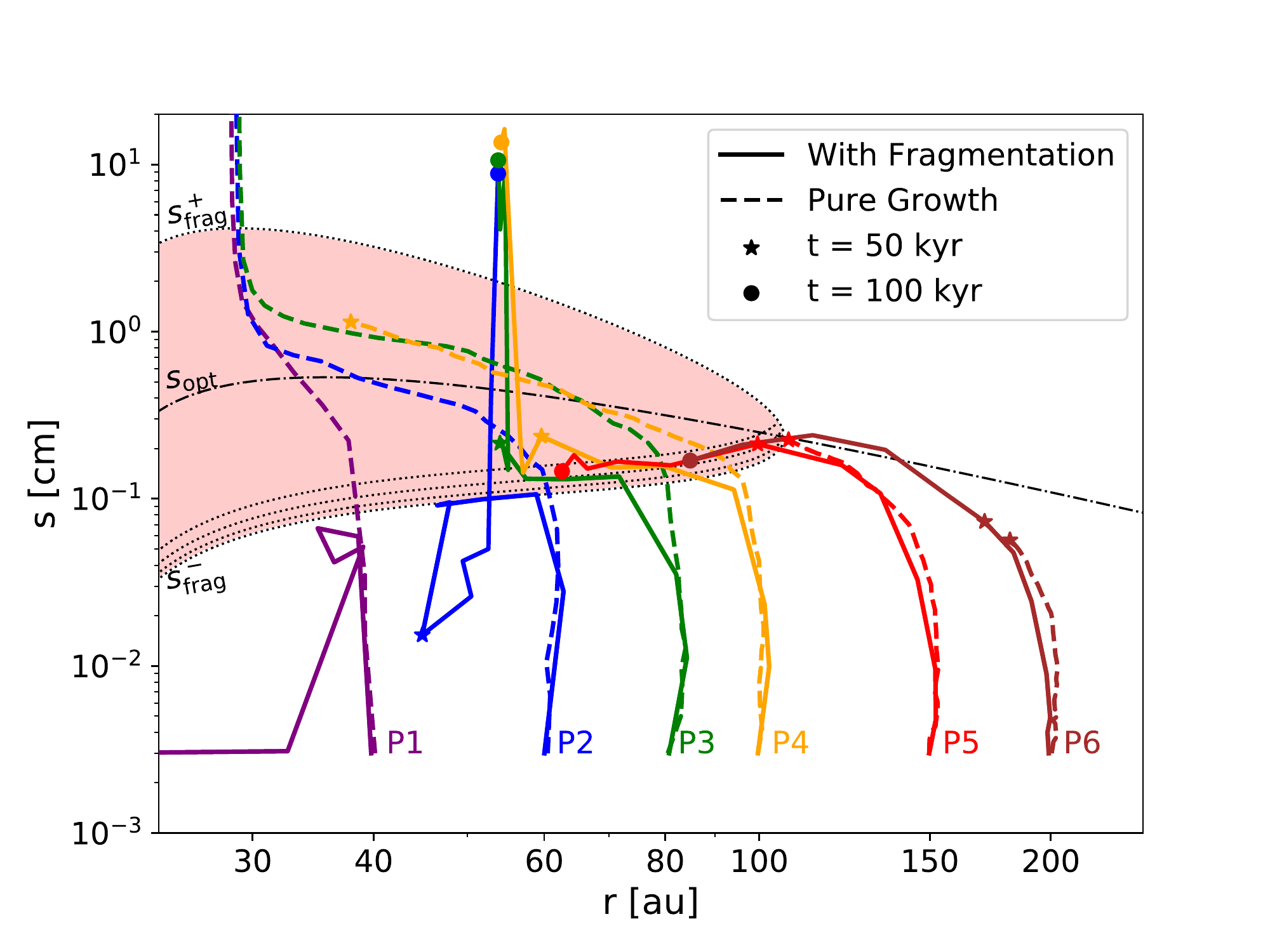}
}
\caption{Trajectories in the ($r, s$) plane of 6 particles initially at 40, 60, 80, 100, 150 and 200~au from the star for simulations with (F-Hard, solid lines) and without (noF, dashed lines) fragmentation. The fragmentation zone (salmon colour, computed using the power laws initially describing the disc structure) is delimited by the fragmentation sizes $s_{\rm frag}^\pm$ (dotted lines) and estimated with $\varepsilon=0.01$. $s_{\rm frag}^-$ is also plotted for larger dust-to-gas ratios ($\varepsilon=0.1, 0.25$ and 0.5, from smaller to larger sizes). The optimal drift size $s_{\rm opt}$ is also added (dash dotted line). The simulation with fragmentation extends up to 100~kyr, that is twice as long as the pure growth case, which reaches a quasi steady state at this time. To compare them efficiently, the stars represent the particles state at 50~kyr, while the filled circles mark their state at 100~kyr.}
\label{F:P1-6}
\end{figure}

\begin{figure}
\centering
\resizebox{\hsize}{!}{
\includegraphics[]{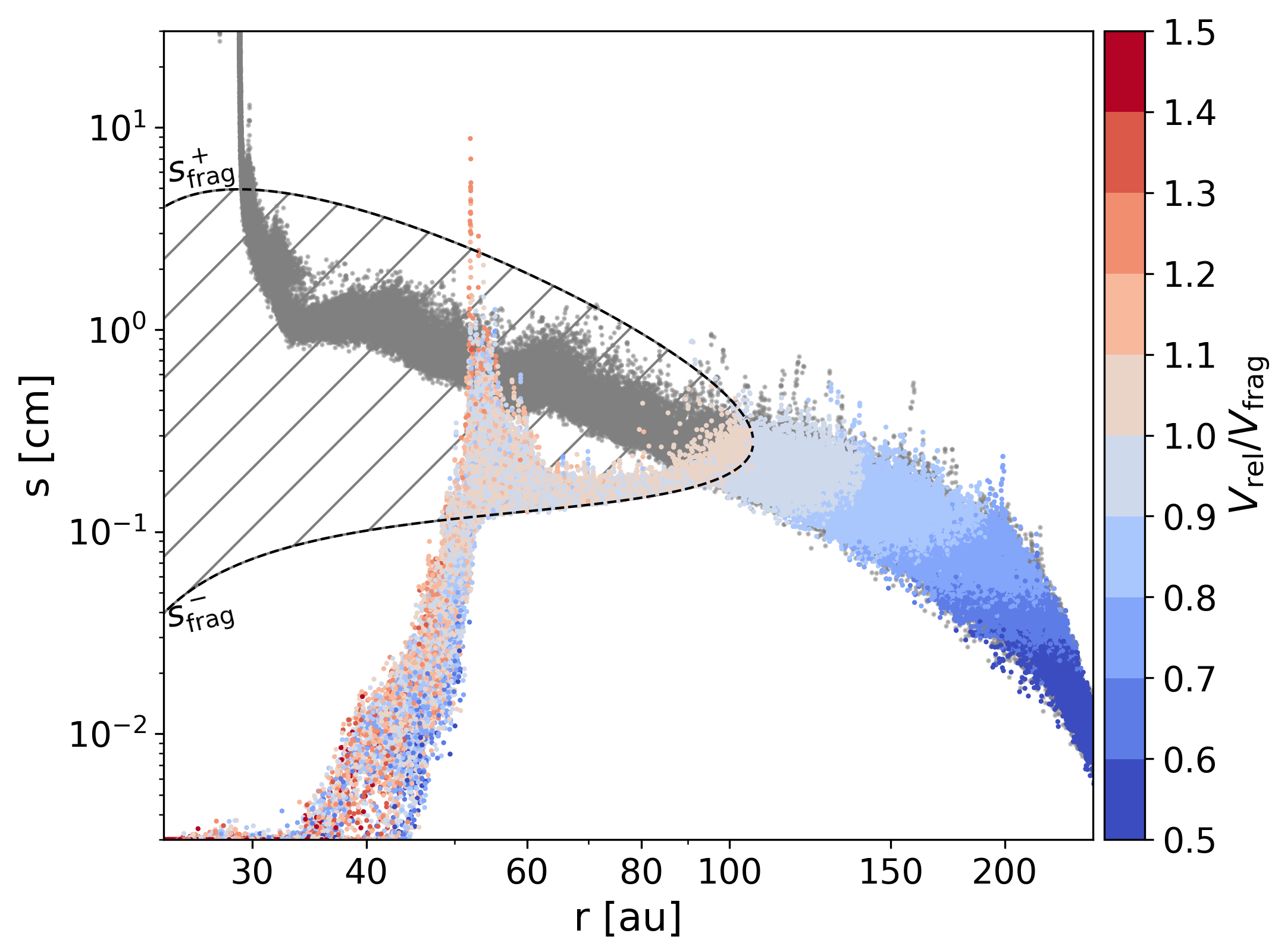}
}
\caption{Dust size radial distribution in simulations with (F-Hard, coloured particles) and without (noF, grey particles) fragmentation at 50~kyr. The colour represents the ratio $V_{\rm rel} / V_{\rm frag}$. The fragmentation zone is hatched and estimated using $s_{\rm frag}^\pm$ and $\varepsilon=0.2$ (see also Fig.~\ref{F:srevol}).}
\label{F:distribcomp}
\end{figure}

To further point out the major differences that fragmentation produces, we compare simulations with (F-Hard) and without (noF) fragmentation in this section. A first example can be seen in Fig.~\ref{F:vsPG}, where we show the difference in dust settling.
With fragmentation, the internal parts of the disc are, as one would expect, less settled than the pure growth case, mainly because the dust cannot reach large sizes.
More generally, the disc can be separated into 3 zones (see black vertical dashed lines):
\begin{enumerate}
\item{An external one ($r \gtrsim r_{\rm frag}$) where fragmentation is impossible. In this region, fragmentation has no impact and the dust discs reach similar aspect ratios.}
\item{An intermediate one ($50 \lesssim r \lesssim r_{\rm frag}$~au) where the dust relative velocity is in equilibrium with the fragmentation velocity (see right panel). In this region, fragmentation limits dust growth, thus the grains are more coupled to the gas and the disc is less settled than in the pure growth case.}
\item{An inner one ($r \lesssim 50$~au) where the grains are growing and fragmenting on short timescales and do not reach an equilibrium with the fragmentation threshold. In this region, grains have even smaller sizes and the dust disc is thinner than the gas disc{, but much thicker than in the pure growth case}.}
\end{enumerate}
%
A comparison of the trajectories of similar particles is also shown in Fig.~\ref{F:P1-6}.
Trajectories outside of the fragmentation zone are extremely similar since the particles experience the same growth path. On the other hand, when particles enter this zone (salmon colour), their size first evolve along the $s_{\rm frag}^-$ line, that is they tend to fragment slowly (see P2 for example).
Note that P5 or P6 enter the fragmentation zone after typically 50~kyr or more and follow a fragmentation size corresponding to a larger dust-to-gas ratio. This makes perfect sense, since dust settling and pile-up tend to concentrate dust in the mid-plane of the disc and increase the dust-to-gas ratio over time.
The formation of the self-induced dust trap can be seen from the point of views of P2, P3 {and P4},
 as after some time ($\sim 50$~kyr) their radial drift stops and their size increases near $\sim 50$~au. They are then able to grow inside the estimated fragmentation zone because the gas structure is modified locally and the grains relative velocity decreases as they concentrate.
Finally, the shape of the dust size distribution carries these differences, as one can see in Fig.~\ref{F:distribcomp}.
Following the previous explanations, it is very clear that the distribution only differs in the fragmentation zone, which we estimated rather well.
%

\subsubsection{Effects of the fragmentation model}

\begin{figure*}
\centering
\resizebox{\hsize}{!}{
\includegraphics[]{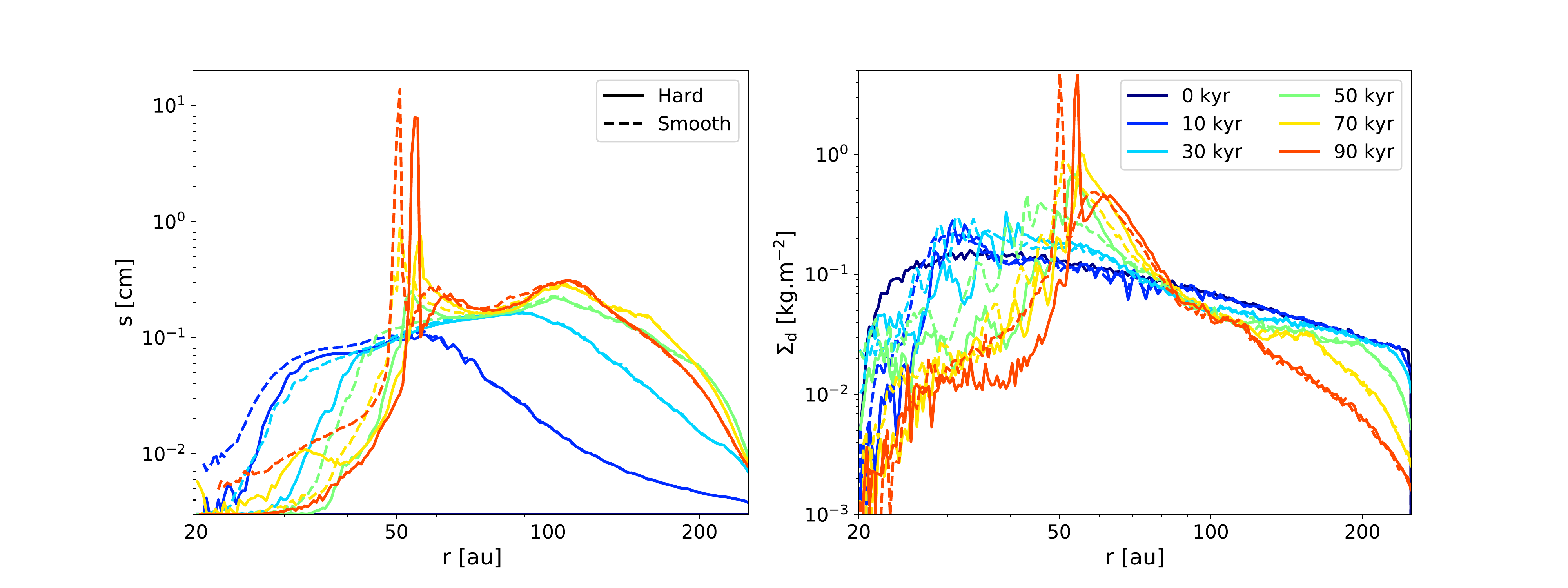}
}
\caption{\textbf{Left}: Dust size radial profiles between 0 and 90~kyr for fragmentation models Hard (solid lines) and Smooth (dashed lines). \textbf{Right}: Same but for the dust surface density.}
\label{F:fragmodels}
\end{figure*}

We implemented two fragmentation models to measure their impact on dust evolution. Here, we compare the Hard model already presented to the Smooth one (simulation F-Smooth).
The comparison can be seen in Fig.~\ref{F:fragmodels}.
What we gather from the comparison is that the Smooth model also forms a self-induced dust trap in the disc. It also tends to form this dust trap slightly closer to the star. Indeed, this model is less harsh on the dust fragmentation, which shifts the size gradient towards the star (it allows slightly larger grains at similar radii). As a consequence, the pressure maximum formed, which is intimately linked to the position of the dust size gradient, is also located closer to the star, finally leading to the observed position of the trap. Even though this tendency is observed, we stress that the differences are extremely minor, with a radial shift of a few au only. We thus argue that the two fragmentation models produce largely similar results.
%

The effects of different disc parameters are considered in Appendix~\ref{app:disc_model}.

\subsubsection{Synthetic images}
\label{sec:synth}

We transformed our simulations with fragmentation at 50~kyr into synthetic observations using the radiative transfer code \textsc{mcfost} \citep{mcfost06,pinte09}. At this time, the traps have developed, but the artificial clumping has not started yet (see Fig.~\ref{F:P1-6} and Section~\ref{sec:discu_num}.)
To achieve this, we recreated a local dust size distribution on each gas particle from our single set of dust particles. More precisely, we separated a given simulation into 20 grain size bins and recomputed using the SPH kernel the density structure at the location of the gas particles for each bin independently. For each of these locations and from the 20 dust density fields, \textsc{mcfost} then interpolated the grain size distributions over 100 bins uniformly distributed in log and sampling grain sizes from $0.03$ to $1000$~$\mu$m. We assume dust grains smaller than 1~$\mu$m follow the gas. The total grain size distribution integrated over the whole disc is finally normalised, assuming a power-law distribution $\mathrm{d}n(s) \propto s^{-3.5}\mathrm{d}s$ and the total disc mass. Grains optical properties were calculated using the Mie theory, assuming astrosilicates composition \citep{Weingartner01}.

\begin{figure*}
\centering
\resizebox{\hsize}{!}{
\includegraphics[]{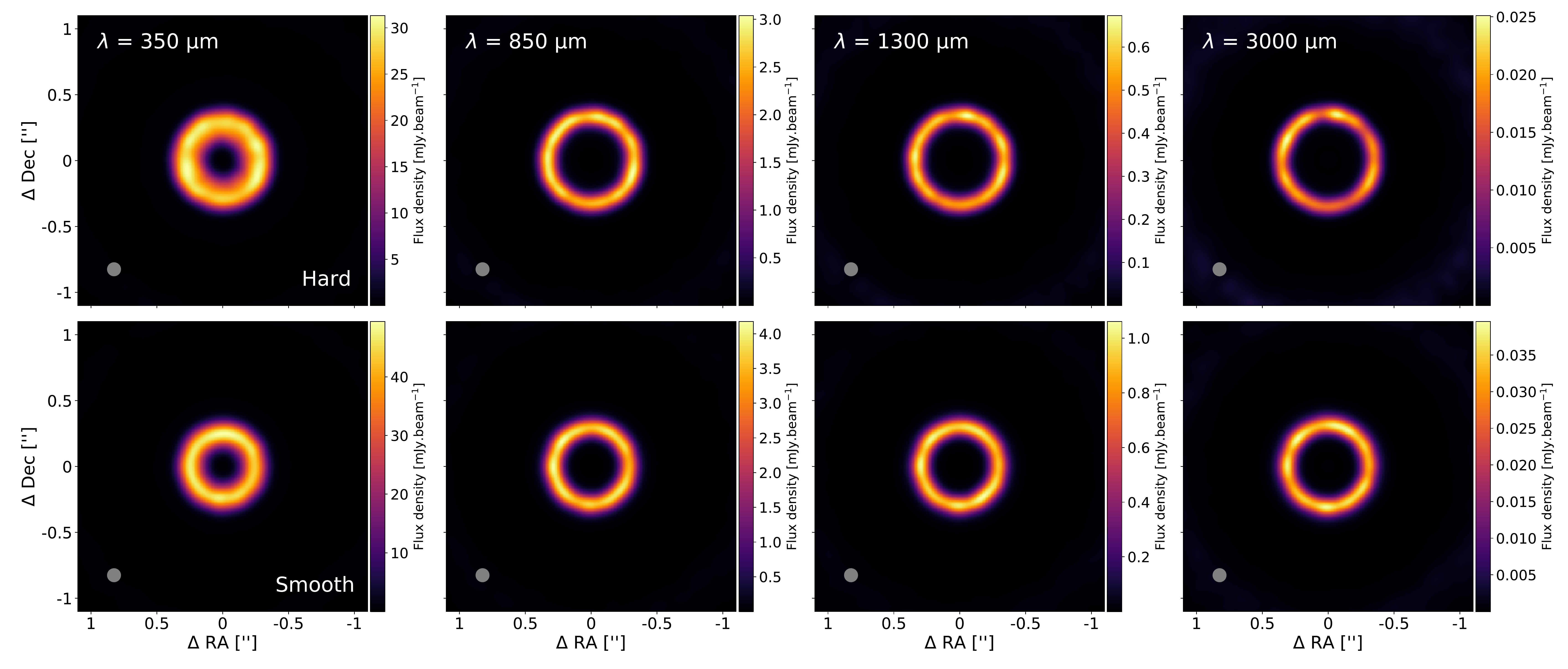}
}
\caption{Synthetic images at $\lambda = 350, 850, 1300$ and 3000~$\mu$m (left to right) of simulations with Hard (top) and Smooth (bottom) fragmentation models at 50~kyr. Images are {convolved} using a $0.1 \times 0.1$ arcsec$^2$ {Gaussian} beam (displayed on the bottom left of each panel). The system is assumed to be at a distance of $140$~pc and seen pole-on.}
\label{F:synth}
\end{figure*}

The central star was represented by a sphere of radius $R_\star = 2$~$R_\odot$ and effective temperature $T_{\rm eff} = 4000$~K radiating as a black body. The disc temperature structure was first computed using $\sim 10^6$ photon packets. The images were then computed at $\lambda = 350, 850, 1300$ and 3000~$\mu$m using $10^7$ photon packets at each wavelength to sample the specific intensity, followed by a ray-tracing integration to generate the synthetic maps. The system {was} assumed to be at a distance of $140$~pc and seen pole-on. The synthetic images are shown in Fig.~\ref{F:synth}.
The dust concentration at $\sim 50$~au translates into a bright and thin ring at millimetre wavelengths. The observed ring also shows a thinner structure as the wavelength increases, which is a manifestation of the dust size distribution being more peaked for higher sizes (the thermal emission of grains of size $s$ peaks at $\lambda \sim 2\pi s$).
Both fragmentation models give similar images. The major difference is seen at lower wavelengths, where the Hard model shows a thicker ring than its counterpart.

Considering these synthetic images, the self-induced dust trap mechanism observed in our simulations should be detectable by instruments such as ALMA, which is sensitive to the dust thermal emission in the mid-plane of the disc. This is particularly interesting given the commonness of axisymmetric rings in recent disc observations \citep{andrews18,huang18}. 
However, we stress that it might be hard to disentangle this mechanism from others that produce bright axisymetric rings \citep[e.g. at the edge of a massive planet's gap {or because of dust sintering,}][]{rice06,oku16,toci20}.

\section{Discussion}
\label{sec:discu}

\subsection{The monodisperse approximation}

\begin{figure}
\centering
\resizebox{\hsize}{!}{
\includegraphics[]{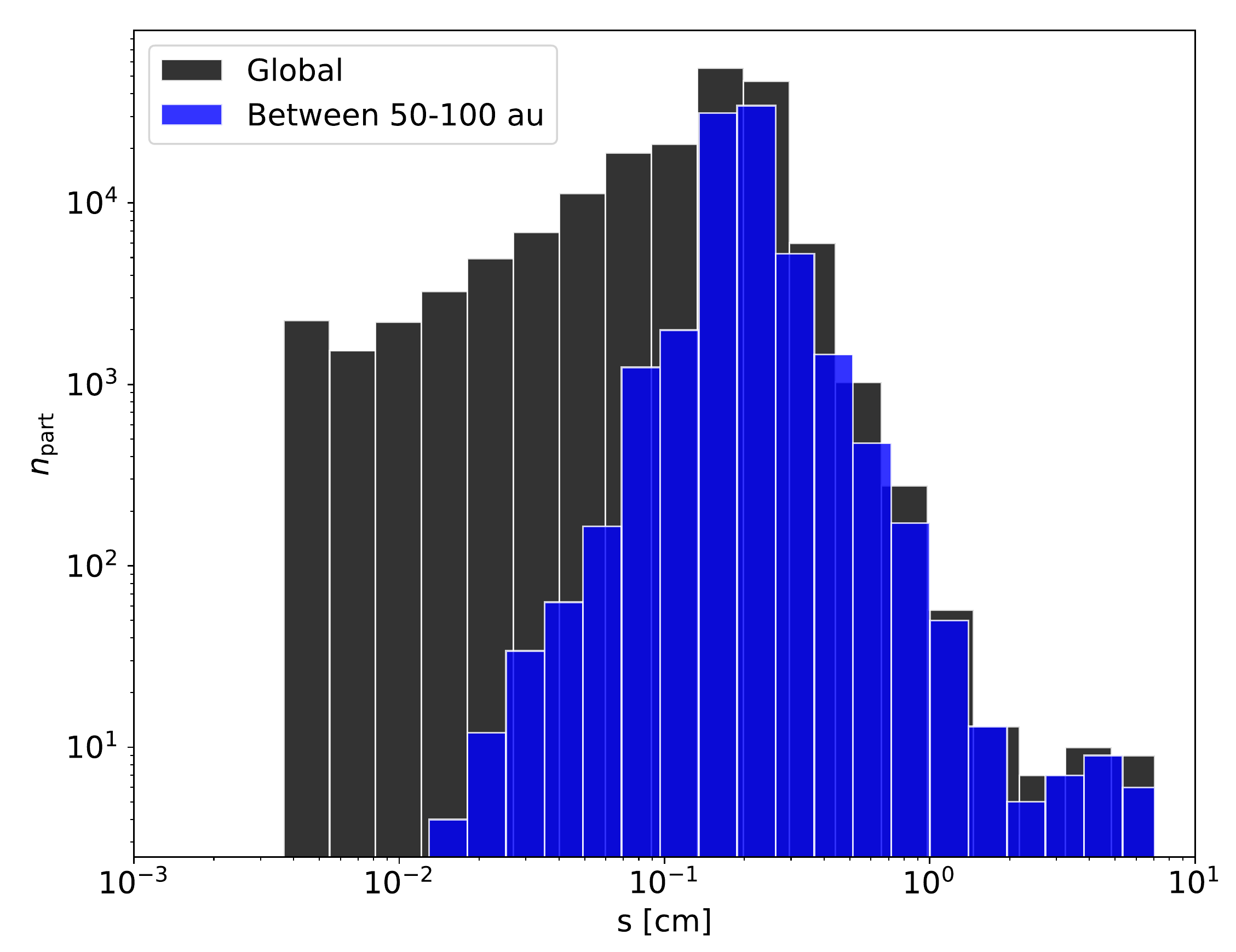}
}
\caption{Grain size distribution in simulation F-Hard at 50~kyr in the whole disc (black) and in the 50--100~au region only (blue).}
\label{F:distrib}
\end{figure}

The dust growth module presented in this paper does not solve the Smoluchowski coagulation equation \citep{smolu16} but rather considers locally monodisperse size distributions. This facilitates the implementation as it becomes a single differential equation to solve numerically for every particle of the simulation.
The monodisperse approximation considers the local distribution to be highly peaked around a single value that is carried by the dust SPH particle. By construction, this makes it impossible to track all sizes at the same time and therefore we only keep the largest ones in memory, which creates a top-heavy dust size distribution. This is particularly the case in regions of efficient growth, as illustrated in Fig.~\ref{F:distrib} between 50 and 100~au in simulation F-Hard.
In that regard, small sizes are less represented, which physically is an acceptable approximation since most of the dust mass is carried by the larger populations. However, this becomes more concerning when simulating synthetic images, as most of the stellar light is absorbed and scattered  by tightly coupled grains (i.e. small grains $\lesssim$ 1\,$\mu$m). This directly impacts the scattered light images, but also the disk thermal structure (and sub-mm images) as the disk midplane is heated by photons scattered from the disk upper layers. We alleviate part of this issue at the stage where \textsc{mcfost} reconstructs the size distribution (see Section~\ref{sec:synth}) by adding grains smaller than 1~$\mu$m, considering that they follow the gas, which limits the impact of this caveat.
One might also argue that the SPH method is already poorly adapted to images in scattered light, since it comes from the upper layers of the disc, which, by effect of settling and overall vertical density distributions, have a lower resolution.
On the contrary, the mid-plane of the disc is better resolved and contains mm to cm grain sizes, which makes the synthetic images at millimeter wavelengths more adapted to this kind of dust growth model.
Adding the sub-micron sized grains ensures that our disk temperature structure is correct and ALMA synthetic images of self-induced dust traps, such as shown in Fig.~\ref{F:synth}, should be mostly unaffected by our dust treatment, with the possible exception of its shortest wavelengths.
In the mid- to far-infrared, under-representation of the 1--100~$\mu$m grain population would mostly affect synthetic images of regions of efficient growth, and in particular the contrast between bright and dark rings. We do not compute images in this wavelength range.

The resolution of the Smoluchoswki equation in a code that directly integrates the equations of motion self-consistently for both gas and dust is extremely challenging, although methods suitable for use in {\sc phantom} are under development \citep{lombart21}.
The results of such an implementation\, providing a continuous size distribution, will precise the extent of the validity of the monodisperse approximation.

Lastly, the growth model presented in this paper and based on \citet{stepvala97} requires an explicit formulation of the Stokes number --- presently expressed with the Keplerian frequency. 
This can be a source of limitation, as it assumes the disc is Keplerian by nature. Significant non-Keplerian rotations are a possibility, e.g. when the disc is perturbed by massive companions, which would limit the use of such growth model.
Although alternatives to this issue are beyond the scope of this paper, the Stokes number would probably be best computed in that scenario with a dynamical timescale estimated within the simulation itself.

\subsection{Numerical limitations}
\label{sec:discu_num}

As mentioned in section~\ref{subsec:twofluid}, the dust growth model is also limited by the numerical formalism that we employ. Here, we will discuss the {one we} used in our simulations: the two-fluid algorithm.
More precisely, the handling of two sets of resolution can become delicate to manage when the two resolutions stop being comparable. In practice, the gas structure is relatively steady because of the pressure support. On the other hand, dust particles are pressureless, which in case of local dust concentrations can become a problem and lead to numerical clumps. Dust concentrations arise naturally at pressure bumps since the radial drift vanishes, which increases locally the dust-to-gas ratio. At a certain point, the dust smoothing length becomes so small that the gas is effectively invisible to the dust particles, which underestimates the extent of the dust structures (see for example Fig.~\ref{PG:P1-6} and~\ref{F:srevol}).
The code used by \citetalias{laibe08}, {\textsc{LyonSPH},} seems to be less {prone} to this limitation. We suspect this difference to come from two main reasons:

\begin{enumerate}
\item{The way the smoothing length is handled. In \textsc{Phantom}, the smoothing length is carefully computed and adjusted using iterations on $h-\rho$, while in {\textsc{LyonSPH}} 
$h$ is independently estimated and thus {changes} less over time, which might smooth out the density in the disc. While the method used by \textsc{Phantom} produces better resolved density structures, it could also be more affected by numerical clumping.}
\item{The viscosity treatment. While we used the formalism proposed by \citet{lodatoprice10}, {\textsc{LyonSPH}} uses the older \citet{mona92} formulation. We have noticed that in {\textsc{LyonSPH}}, particles tend to be more accreted by the star, which increases the effective viscosity of the disc over time. As a result, an effectively more dissipating disc could smooth out the disc structures and thus limit local dust over-resolutions.}
\end{enumerate}

Numerical clumping is hard to detach from in the two-fluid formalism, especially when considering dust growth that further decouples dust from gas. However, to limit its effects as much as possible, the best solution is to adjust the ratio between dust and gas particles, as we already did in our setup. For example, we want a high enough number of gas particles to resolve the disc scale height ($\gtrsim 1$M) while having a large gas-to-dust particles ratio such that the gas smoothing length is as small as possible compared to the dust's ($\sim 5-10$).
By simple analytical arguments, if we consider the condition $h_{\rm g} \lesssim h_{\rm d}$ to be fulfilled to avoid any numerical clumping, this gives a maximum safely attainable dust-to-gas ratio of
\begin{equation}
\varepsilon_{\rm max} \sim \varepsilon_0 \dfrac{n_{\rm g}^{\rm SPH}}{n_{\rm d}^{\rm SPH}},
\end{equation}
where $\varepsilon_0$ is the initial dust-to-gas ratio. With great safely attainable dust-to-gas ratio comes great computational time.

\subsection{Perspectives}

Beyond the presentation of the model implementation in \textsc{Phantom} and the associated tests, this paper mainly revisited and found similar results as previous studies that considered dust growth in protoplanetary discs.
The main objective of this paper is to present the algorithm and stress that it is public and usable by the community. With \textsc{Phantom}'s modularity, this means that dust growth can be studied {in 3D} in any given system {that was previously out of reach, all with better overall performances and a lower memory footprint than {\textsc{LyonSPH}}.}

For example, multiple systems are of much interest, since they can constitute between a few and about 60\% of systems \citep{ruben97,bella02,gue14}. Grasping planet formation in general surely means that dust growth in these poorly studied systems is needed. This point is especially of interest considering the recent substructures than have been theoretically and observationally discovered in circumbinary discs \citep[neglecting dust growth, e.g.][]{ragusa17,calcino19,poblete19}.

Dust properties are also another important element that can be improved in this model. While we consider perfectly spherical grains composed of a given species, we are aware that reality is vastly more complicated. For example, snow lines can change the dust chemical composition, affect their surface properties and thus their growth and dynamic \citep[e.g.][]{wada09,drazk17,vericelgonzalez20}.
Another important element could be the dust porosity, which can be significant considering for instance the very small filling factors of about $10^{-3}$ that have been measured on comet 67P/Churyumov–Gerasimenko \citep{fulle15}. Dust porosity can accelerate dust growth and help overcome planet formation barriers \citep{garciathesis18,anto20}, which is promising for planetesimal formation. The implementation of a porosity model is underway in \textsc{Phantom} and will be the subject of further investigations in the future.

\section{Conclusion}
\label{sec:conclu}

We presented a dust growth model that considers local monodisperse size distributions. After specifying the numerical implementation in \textsc{Phantom}, we presented two tests that ensure the algorithm works as intended. We then proceeded to circumstellar disc simulations, while first focusing on pure growth and adding fragmentation effects later on. We finally performed radiative transfer calculations to produce synthetic images.
Our main findings can be summarised as follows:

\begin{enumerate}
\item{The tests show that the parameters of interest are evaluated with a {maximum relative error of about $10^{-4}$, a }satisfying degree of precision.}
\item{Simulations without fragmentation develop a dust concentration at the inner edge of the disc, which corresponds to the usual gas pressure maximum. The dust concentrated at this location drifts from the outer parts of the disc and ends its course well decoupled from the gas. Results are consistent with those found by  \citetalias{laibe08}.}
\item{Simulations with growth and fragmentation show the formation of a self-induced dust trap as first proposed by \citet{gonzalez17}. This trap forms at several tens of au from the star and is the result of the dust piling-up in the disc and modifying the gas structure through back-reaction.}
\item{Simulations in which a self-induced dust trap forms produce synthetic images showing an axisymmetric bright ring at millimetre wavelengths. These should be detectable by instruments such as ALMA.}
\end{enumerate}

\section*{Acknowledgements}
The authors would like to thank the referee, Matías Gárate, for constructive comments and suggestions that helped improved this work.
AV would also like to thank Hossam Aly and Kieran Hirsh for their creative suggestions. This research was supported by the \'Ecole Doctorale PHAST (ED 52) of the Universit\'e de Lyon. The authors acknowledge funding from ANR (Agence Nationale de la Recherche) of France under contract number ANR-16-CE31-0013 (Planet-Forming-Disks) and thank the LABEX Lyon Institute of Origins (ANR-10-LABX-0066) of the Universit\'e de Lyon for its financial support within the programme `Investissements d'Avenir' (ANR-11-IDEX-0007) of the French government operated by the ANR. This project has received funding from the European Union's Horizon 2020 research and innovation program under the Marie Sk\l{}odowska-Curie grant agreement No 823823. DJP and CP acknowledge funding from the Australian Research Council via grants DP180104235, FT130100034 and FT170100040. SPH simulations were run at the Common Computing Facility (CCF) of LABEX LIO and at the PSMN (Pôle Scientifique de Modélisation Numérique) of the ENS de Lyon, and analysed with \textsc{Fantomanalysis}, which has been developed by AV. We acknowledge use of the Ozstar supercomputer, funded by Swinburne University and the Australian Government. Figures were made with \textsc{Matplotlib} \citep{hunter07}{, except for Fig.~\ref{F:render} which was made with \textsc{Splash} \citep{splash07,splash11}}.

\section*{Data Availability}
The \textsc{Phantom} SPH code is available from \url{https://github.com/danieljprice/phantom}. \textsc{Mcfost} is available for use on a collaborative basis from \url{https://ipag.osug.fr/~pintec/mcfost/docs/html/overview.html}. \textsc{Fantomanalysis} is available from \url{https://github.com/arnaudvericel/fantomanalysis}. The parameter files for generating our SPH simulations and radiative transfer models are available upon request.




\bibliographystyle{mnras}
\bibliography{refs-growth} 

\begin{thebibliography}{}
\makeatletter
\relax
\def\mn@urlcharsother{\let\do\@makeother \do\$\do\&\do\#\do\^\do\_\do\%\do\~}
\def\mn@doi{\begingroup\mn@urlcharsother \@ifnextchar [ {\mn@doi@}
  {\mn@doi@[]}}
\def\mn@doi@[#1]#2{\def\@tempa{#1}\ifx\@tempa\@empty \href
  {http://dx.doi.org/#2} {doi:#2}\else \href {http://dx.doi.org/#2} {#1}\fi
  \endgroup}
\def\mn@eprint#1#2{\mn@eprint@#1:#2::\@nil}
\def\mn@eprint@arXiv#1{\href {http://arxiv.org/abs/#1} {{\tt arXiv:#1}}}
\def\mn@eprint@dblp#1{\href {http://dblp.uni-trier.de/rec/bibtex/#1.xml}
  {dblp:#1}}
\def\mn@eprint@#1:#2:#3:#4\@nil{\def\@tempa {#1}\def\@tempb {#2}\def\@tempc
  {#3}\ifx \@tempc \@empty \let \@tempc \@tempb \let \@tempb \@tempa \fi \ifx
  \@tempb \@empty \def\@tempb {arXiv}\fi \@ifundefined
  {mn@eprint@\@tempb}{\@tempb:\@tempc}{\expandafter \expandafter \csname
  mn@eprint@\@tempb\endcsname \expandafter{\@tempc}}}

\bibitem[\protect\citeauthoryear{{Abod}, {Simon}, {Li}, {Armitage}, {Youdin}
  \& {Kretke}}{{Abod} et~al.}{2019}]{abod19}
{Abod} C.~P.,  {Simon} J.~B.,  {Li} R.,  {Armitage} P.~J.,  {Youdin} A.~N.,
  {Kretke} K.~A.,  2019, \mn@doi [\apj] {10.3847/1538-4357/ab40a3}, \href
  {https://ui.adsabs.harvard.edu/abs/2019ApJ...883..192A} {883, 192}

\bibitem[\protect\citeauthoryear{{Adachi}, {Hayashi}  \& {Nakazawa}}{{Adachi}
  et~al.}{1976}]{adachi76}
{Adachi} I.,  {Hayashi} C.,   {Nakazawa} K.,  1976, \mn@doi [Progress of
  Theoretical Physics] {10.1143/PTP.56.1756}, \href
  {https://ui.adsabs.harvard.edu/abs/1976PThPh..56.1756A} {56, 1756}

\bibitem[\protect\citeauthoryear{{Andrews} et~al.,}{{Andrews}
  et~al.}{2018}]{andrews18}
{Andrews} S.~M.,  et~al., 2018, \mn@doi [\apjl] {10.3847/2041-8213/aaf741},
  \href {https://ui.adsabs.harvard.edu/abs/2018ApJ...869L..41A} {869, L41}

\bibitem[\protect\citeauthoryear{{Auffinger} \& {Laibe}}{{Auffinger} \&
  {Laibe}}{2018}]{auffinger18}
{Auffinger} J.,  {Laibe} G.,  2018, \mn@doi [\mnras] {10.1093/mnras/stx2395},
  \href {http://adsabs.harvard.edu/abs/2018MNRAS.473..796A} {473, 796}

\bibitem[\protect\citeauthoryear{{Ballabio}, {Dipierro}, {Veronesi}, {Lodato},
  {Hutchison}, {Laibe}  \& {Price}}{{Ballabio} et~al.}{2018}]{ballabio18}
{Ballabio} G.,  {Dipierro} G.,  {Veronesi} B.,  {Lodato} G.,  {Hutchison} M.,
  {Laibe} G.,   {Price} D.~J.,  2018, \mn@doi [\mnras] {10.1093/mnras/sty642},
  \href {https://ui.adsabs.harvard.edu/abs/2018MNRAS.477.2766B} {477, 2766}

\bibitem[\protect\citeauthoryear{{Barri{\`e}re-Fouchet}, {Gonzalez}, {Murray},
  {Humble}  \& {Maddison}}{{Barri{\`e}re-Fouchet} et~al.}{2005}]{barriere05}
{Barri{\`e}re-Fouchet} L.,  {Gonzalez} J.-F.,  {Murray} J.~R.,  {Humble} R.~J.,
    {Maddison} S.~T.,  2005, \mn@doi [\aap] {10.1051/0004-6361:20042249}, \href
  {http://cdsads.u-strasbg.fr/abs/2005A%26A...443..185B} {443, 185}

\bibitem[\protect\citeauthoryear{{Bellazzini}, {Fusi Pecci}, {Messineo},
  {Monaco}  \& {Rood}}{{Bellazzini} et~al.}{2002}]{bella02}
{Bellazzini} M.,  {Fusi Pecci} F.,  {Messineo} M.,  {Monaco} L.,   {Rood}
  R.~T.,  2002, \mn@doi [\aj] {10.1086/339222}, \href
  {https://ui.adsabs.harvard.edu/abs/2002AJ....123.1509B} {123, 1509}

\bibitem[\protect\citeauthoryear{{Birnstiel}, {Dullemond}  \&
  {Brauer}}{{Birnstiel} et~al.}{2009}]{birnstiel09}
{Birnstiel} T.,  {Dullemond} C.~P.,   {Brauer} F.,  2009, \mn@doi [\aap]
  {10.1051/0004-6361/200912452}, \href
  {https://ui.adsabs.harvard.edu/abs/2009A&A...503L...5B} {503, L5}

\bibitem[\protect\citeauthoryear{{Birnstiel}, {Dullemond}  \&
  {Brauer}}{{Birnstiel} et~al.}{2010}]{birn10}
{Birnstiel} T.,  {Dullemond} C.~P.,   {Brauer} F.,  2010, \mn@doi [\aap]
  {10.1051/0004-6361/200913731}, \href
  {https://ui.adsabs.harvard.edu/abs/2010A&A...513A..79B} {513, A79}

\bibitem[\protect\citeauthoryear{{Birnstiel}, {Klahr}  \&
  {Ercolano}}{{Birnstiel} et~al.}{2012}]{birnstiel12}
{Birnstiel} T.,  {Klahr} H.,   {Ercolano} B.,  2012, \mn@doi [\aap]
  {10.1051/0004-6361/201118136}, \href
  {https://ui.adsabs.harvard.edu/abs/2012A&A...539A.148B} {539, A148}

\bibitem[\protect\citeauthoryear{{Blum} \& {Wurm}}{{Blum} \&
  {Wurm}}{2008}]{blumwurm08}
{Blum} J.,  {Wurm} G.,  2008, \mn@doi [\araa]
  {10.1146/annurev.astro.46.060407.145152}, \href
  {http://adsabs.harvard.edu/abs/2008ARA%26A..46...21B} {46, 21}

\bibitem[\protect\citeauthoryear{{Brauer}, {Dullemond}  \& {Henning}}{{Brauer}
  et~al.}{2008}]{brauer08}
{Brauer} F.,  {Dullemond} C.~P.,   {Henning} T.,  2008, \mn@doi [\aap]
  {10.1051/0004-6361:20077759}, \href
  {https://ui.adsabs.harvard.edu/abs/2008A&A...480..859B} {480, 859}

\bibitem[\protect\citeauthoryear{{Calcino}, {Price}, {Pinte}, {van der Marel},
  {Ragusa}, {Dipierro}, {Cuello}  \& {Christiaens}}{{Calcino}
  et~al.}{2019}]{calcino19}
{Calcino} J.,  {Price} D.~J.,  {Pinte} C.,  {van der Marel} N.,  {Ragusa} E.,
  {Dipierro} G.,  {Cuello} N.,   {Christiaens} V.,  2019, \mn@doi [\mnras]
  {10.1093/mnras/stz2770}, \href
  {https://ui.adsabs.harvard.edu/abs/2019MNRAS.490.2579C} {490, 2579}

\bibitem[\protect\citeauthoryear{{Courant}, {Friedrichs}  \& {Lewy}}{{Courant}
  et~al.}{1928}]{courant28}
{Courant} R.,  {Friedrichs} K.,   {Lewy} H.,  1928, \mn@doi [Mathematische
  Annalen] {10.1007/BF01448839}, \href
  {https://ui.adsabs.harvard.edu/abs/1928MatAn.100...32C} {100, 32}

\bibitem[\protect\citeauthoryear{{Cuello} et~al.,}{{Cuello}
  et~al.}{2019}]{cuello19}
{Cuello} N.,  et~al., 2019, \mn@doi [\mnras] {10.1093/mnras/sty3325}, \href
  {https://ui.adsabs.harvard.edu/abs/2019MNRAS.483.4114C} {483, 4114}

\bibitem[\protect\citeauthoryear{{Cuello} et~al.,}{{Cuello}
  et~al.}{2020}]{cuello20}
{Cuello} N.,  et~al., 2020, \mn@doi [\mnras] {10.1093/mnras/stz2938}, \href
  {https://ui.adsabs.harvard.edu/abs/2020MNRAS.491..504C} {491, 504}

\bibitem[\protect\citeauthoryear{{Dipierro} \& {Laibe}}{{Dipierro} \&
  {Laibe}}{2017}]{dipierrolaibe17}
{Dipierro} G.,  {Laibe} G.,  2017, \mn@doi [\mnras] {10.1093/mnras/stx977},
  \href {http://adsabs.harvard.edu/abs/2017MNRAS.469.1932D} {469, 1932}

\bibitem[\protect\citeauthoryear{{Dipierro}, {Price}, {Laibe}, {Hirsh},
  {Cerioli}  \& {Lodato}}{{Dipierro} et~al.}{2015}]{dipierro15}
{Dipierro} G.,  {Price} D.,  {Laibe} G.,  {Hirsh} K.,  {Cerioli} A.,   {Lodato}
  G.,  2015, \mn@doi [\mnras] {10.1093/mnrasl/slv105}, \href
  {https://ui.adsabs.harvard.edu/abs/2015MNRAS.453L..73D} {453, L73}

\bibitem[\protect\citeauthoryear{{Dipierro}, {Laibe}, {Alexander}  \&
  {Hutchison}}{{Dipierro} et~al.}{2018}]{2018MNRAS.479.4187D}
{Dipierro} G.,  {Laibe} G.,  {Alexander} R.,   {Hutchison} M.,  2018, \mn@doi
  [\mnras] {10.1093/mnras/sty1701}, \href
  {https://ui.adsabs.harvard.edu/abs/2018MNRAS.479.4187D} {479, 4187}

\bibitem[\protect\citeauthoryear{{Dominik} \& {Tielens}}{{Dominik} \&
  {Tielens}}{1997}]{dominiktielens97}
{Dominik} C.,  {Tielens} A.~G.~G.~M.,  1997, \mn@doi [\apj] {10.1086/303996},
  \href {http://adsabs.harvard.edu/abs/1997ApJ...480..647D} {480, 647}

\bibitem[\protect\citeauthoryear{{Dong}, {Fung}  \& {Chiang}}{{Dong}
  et~al.}{2016}]{dong16}
{Dong} R.,  {Fung} J.,   {Chiang} E.,  2016, \mn@doi [\apj]
  {10.3847/0004-637X/826/1/75}, \href
  {https://ui.adsabs.harvard.edu/abs/2016ApJ...826...75D} {826, 75}

\bibitem[\protect\citeauthoryear{{Dong} et~al.,}{{Dong} et~al.}{2018}]{dong18}
{Dong} R.,  et~al., 2018, \mn@doi [\apj] {10.3847/1538-4357/aac6cb}, \href
  {https://ui.adsabs.harvard.edu/abs/2018ApJ...860..124D} {860, 124}

\bibitem[\protect\citeauthoryear{{Dr{\k{a}}{\.z}kowska} \&
  {Alibert}}{{Dr{\k{a}}{\.z}kowska} \& {Alibert}}{2017}]{drazk17}
{Dr{\k{a}}{\.z}kowska} J.,  {Alibert} Y.,  2017, \mn@doi [\aap]
  {10.1051/0004-6361/201731491}, \href
  {https://ui.adsabs.harvard.edu/abs/2017A&A...608A..92D} {608, A92}

\bibitem[\protect\citeauthoryear{{Dr{\k{a}}{\.z}kowska}, {Windmark}  \&
  {Dullemond}}{{Dr{\k{a}}{\.z}kowska} et~al.}{2014}]{draz14}
{Dr{\k{a}}{\.z}kowska} J.,  {Windmark} F.,   {Dullemond} C.~P.,  2014, \mn@doi
  [\aap] {10.1051/0004-6361/201423708}, \href
  {https://ui.adsabs.harvard.edu/abs/2014A&A...567A..38D} {567, A38}

\bibitem[\protect\citeauthoryear{{Dr{\k{a}}{\.z}kowska}, {Alibert}  \&
  {Moore}}{{Dr{\k{a}}{\.z}kowska} et~al.}{2016}]{drazk16}
{Dr{\k{a}}{\.z}kowska} J.,  {Alibert} Y.,   {Moore} B.,  2016, \mn@doi [\aap]
  {10.1051/0004-6361/201628983}, \href
  {https://ui.adsabs.harvard.edu/abs/2016A&A...594A.105D} {594, A105}

\bibitem[\protect\citeauthoryear{{Dr{\k{a}}{\.z}kowska}, {Li}, {Birnstiel},
  {Stammler}  \& {Li}}{{Dr{\k{a}}{\.z}kowska} et~al.}{2019}]{draz19}
{Dr{\k{a}}{\.z}kowska} J.,  {Li} S.,  {Birnstiel} T.,  {Stammler} S.~M.,   {Li}
  H.,  2019, \mn@doi [\apj] {10.3847/1538-4357/ab46b7}, \href
  {https://ui.adsabs.harvard.edu/abs/2019ApJ...885...91D} {885, 91}

\bibitem[\protect\citeauthoryear{{Dubrulle}, {Morfill}  \&
  {Sterzik}}{{Dubrulle} et~al.}{1995}]{dubrulle95}
{Dubrulle} B.,  {Morfill} G.,   {Sterzik} M.,  1995, \mn@doi [\icarus]
  {10.1006/icar.1995.1058}, \href
  {https://ui.adsabs.harvard.edu/abs/1995Icar..114..237D} {114, 237}

\bibitem[\protect\citeauthoryear{Epstein}{Epstein}{1924}]{epstein24}
Epstein P.~S.,  1924, \mn@doi [Phys. Rev.] {10.1103/PhysRev.23.710}, 23, 710

\bibitem[\protect\citeauthoryear{{Fromang} \& {Papaloizou}}{{Fromang} \&
  {Papaloizou}}{2006}]{fromang06}
{Fromang} S.,  {Papaloizou} J.,  2006, \mn@doi [\aap]
  {10.1051/0004-6361:20054612}, \href
  {https://ui.adsabs.harvard.edu/abs/2006A&A...452..751F} {452, 751}

\bibitem[\protect\citeauthoryear{{Fulle} et~al.,}{{Fulle}
  et~al.}{2015}]{fulle15}
{Fulle} M.,  et~al., 2015, \mn@doi [\apjl] {10.1088/2041-8205/802/1/L12}, \href
  {https://ui.adsabs.harvard.edu/abs/2015ApJ...802L..12F} {802, L12}

\bibitem[\protect\citeauthoryear{{G{\'a}rate}, {Birnstiel},
  {Dr{\k{a}}{\.z}kowska}  \& {Stammler}}{{G{\'a}rate} et~al.}{2020}]{garate20}
{G{\'a}rate} M.,  {Birnstiel} T.,  {Dr{\k{a}}{\.z}kowska} J.,   {Stammler}
  S.~M.,  2020, \mn@doi [\aap] {10.1051/0004-6361/201936067}, \href
  {https://ui.adsabs.harvard.edu/abs/2020A&A...635A.149G} {635, A149}

\bibitem[\protect\citeauthoryear{{Garaud}, {Barri{\`e}re-Fouchet}  \&
  {Lin}}{{Garaud} et~al.}{2004}]{garaud04}
{Garaud} P.,  {Barri{\`e}re-Fouchet} L.,   {Lin} D.~N.~C.,  2004, \mn@doi
  [\apj] {10.1086/381385}, \href
  {https://ui.adsabs.harvard.edu/abs/2004ApJ...603..292G} {603, 292}

\bibitem[\protect\citeauthoryear{Garcia}{Garcia}{2018}]{garciathesis18}
Garcia A.,  2018, Phd thesis, {Universit{\'e} de Lyon}, \url
  {https://tel.archives-ouvertes.fr/tel-01977317}

\bibitem[\protect\citeauthoryear{{Garcia} \& {Gonzalez}}{{Garcia} \&
  {Gonzalez}}{2020}]{anto20}
{Garcia} A. J.~L.,  {Gonzalez} J.-F.,  2020, \mn@doi [\mnras]
  {10.1093/mnras/staa382}, \href
  {https://ui.adsabs.harvard.edu/abs/2020MNRAS.493.1788G} {493, 1788}

\bibitem[\protect\citeauthoryear{{Gonzalez}, {Laibe}, {Maddison}, {Pinte}  \&
  {M{\'e}nard}}{{Gonzalez} et~al.}{2015}]{gonzalez15}
{Gonzalez} J.-F.,  {Laibe} G.,  {Maddison} S.~T.,  {Pinte} C.,   {M{\'e}nard}
  F.,  2015, \mn@doi [\planss] {10.1016/j.pss.2015.05.018}, \href
  {http://adsabs.harvard.edu/abs/2015P%26SS..116...48G} {116, 48}

\bibitem[\protect\citeauthoryear{{Gonzalez}, {Laibe}  \& {Maddison}}{{Gonzalez}
  et~al.}{2017a}]{gonzalez17}
{Gonzalez} J.-F.,  {Laibe} G.,   {Maddison} S.~T.,  2017a, \mn@doi [\mnras]
  {10.1093/mnras/stx016}, \href
  {http://adsabs.harvard.edu/abs/2017MNRAS.467.1984G} {467, 1984}

\bibitem[\protect\citeauthoryear{{Gonzalez}, {Laibe}  \& {Maddison}}{{Gonzalez}
  et~al.}{2017b}]{sidterrat}
{Gonzalez} J.~F.,  {Laibe} G.,   {Maddison} S.~T.,  2017b, \mn@doi [\mnras]
  {10.1093/mnras/stx2024}, \href
  {https://ui.adsabs.harvard.edu/abs/2017MNRAS.472.1162G} {472, 1162}

\bibitem[\protect\citeauthoryear{{Gonzalez} et~al.,}{{Gonzalez}
  et~al.}{2020}]{gonzalez20}
{Gonzalez} J.-F.,  et~al., 2020, \mnras, submitted

\bibitem[\protect\citeauthoryear{{Guerrero}, {Orlov}, {Monroy-Rodr{\'\i}guez}
  \& {Voitsekhovich}}{{Guerrero} et~al.}{2014}]{gue14}
{Guerrero} C.~A.,  {Orlov} V.~G.,  {Monroy-Rodr{\'\i}guez} M.~A.,
  {Voitsekhovich} V.~V.,  2014, \mn@doi [\aj] {10.1088/0004-6256/147/2/28},
  \href {https://ui.adsabs.harvard.edu/abs/2014AJ....147...28G} {147, 28}

\bibitem[\protect\citeauthoryear{{Haghighipour}}{{Haghighipour}}{2005}]{haghi05}
{Haghighipour} N.,  2005, \mn@doi [\mnras] {10.1111/j.1365-2966.2005.09362.x},
  \href {http://adsabs.harvard.edu/abs/2005MNRAS.362.1015H} {362, 1015}

\bibitem[\protect\citeauthoryear{{Huang} et~al.,}{{Huang}
  et~al.}{2018}]{huang18}
{Huang} J.,  et~al., 2018, \mn@doi [\apj] {10.3847/2041-8213/aaf740}, \href
  {https://ui.adsabs.harvard.edu/abs/2018ApJ...869L..42H} {869, L42}

\bibitem[\protect\citeauthoryear{Hunter}{Hunter}{2007}]{hunter07}
Hunter J.~D.,  2007, \mn@doi [Computing in Science \& Engineering]
  {10.1109/MCSE.2007.55}, 9, 90

\bibitem[\protect\citeauthoryear{{Johansen} \& {Youdin}}{{Johansen} \&
  {Youdin}}{2007}]{joha07}
{Johansen} A.,  {Youdin} A.,  2007, \mn@doi [\apj] {10.1086/516730}, \href
  {https://ui.adsabs.harvard.edu/abs/2007ApJ...662..627J} {662, 627}

\bibitem[\protect\citeauthoryear{{Kanagawa}, {Ueda}, {Muto}  \&
  {Okuzumi}}{{Kanagawa} et~al.}{2017}]{kanagawa17}
{Kanagawa} K.~D.,  {Ueda} T.,  {Muto} T.,   {Okuzumi} S.,  2017, \mn@doi [\apj]
  {10.3847/1538-4357/aa7ca1}, \href
  {https://ui.adsabs.harvard.edu/abs/2017ApJ...844..142K} {844, 142}

\bibitem[\protect\citeauthoryear{{Kley} \& {Nelson}}{{Kley} \&
  {Nelson}}{2012}]{kley12}
{Kley} W.,  {Nelson} R.~P.,  2012, \mn@doi [\araa]
  {10.1146/annurev-astro-081811-125523}, \href
  {https://ui.adsabs.harvard.edu/abs/2012ARA&A..50..211K} {50, 211}

\bibitem[\protect\citeauthoryear{{Krijt} \& {Ciesla}}{{Krijt} \&
  {Ciesla}}{2016}]{kriciesla16}
{Krijt} S.,  {Ciesla} F.~J.,  2016, \mn@doi [\apj]
  {10.3847/0004-637X/822/2/111}, \href
  {https://ui.adsabs.harvard.edu/abs/2016ApJ...822..111K} {822, 111}

\bibitem[\protect\citeauthoryear{{Krijt}, {Ormel}, {Dominik}  \&
  {Tielens}}{{Krijt} et~al.}{2016}]{kri16}
{Krijt} S.,  {Ormel} C.~W.,  {Dominik} C.,   {Tielens} A.~G.~G.~M.,  2016,
  \mn@doi [\aap] {10.1051/0004-6361/201527533}, \href
  {https://ui.adsabs.harvard.edu/abs/2016A&A...586A..20K} {586, A20}

\bibitem[\protect\citeauthoryear{{Kwok}}{{Kwok}}{1975}]{kwok75}
{Kwok} S.,  1975, \mn@doi [\apj] {10.1086/153637}, \href
  {https://ui.adsabs.harvard.edu/abs/1975ApJ...198..583K} {198, 583}

\bibitem[\protect\citeauthoryear{{Laibe}}{{Laibe}}{2014}]{laibe14drift}
{Laibe} G.,  2014, \mn@doi [\mnras] {10.1093/mnras/stt1928}, \href
  {https://ui.adsabs.harvard.edu/abs/2014MNRAS.437.3037L} {437, 3037}

\bibitem[\protect\citeauthoryear{{Laibe} \& {Price}}{{Laibe} \&
  {Price}}{2012a}]{laibeprice2012a}
{Laibe} G.,  {Price} D.~J.,  2012a, \mn@doi [\mnras]
  {10.1111/j.1365-2966.2011.20202.x}, \href
  {https://ui.adsabs.harvard.edu/abs/2012MNRAS.420.2345L} {420, 2345}

\bibitem[\protect\citeauthoryear{{Laibe} \& {Price}}{{Laibe} \&
  {Price}}{2012b}]{laibeprice2012b}
{Laibe} G.,  {Price} D.~J.,  2012b, \mn@doi [\mnras]
  {10.1111/j.1365-2966.2011.20201.x}, \href
  {https://ui.adsabs.harvard.edu/abs/2012MNRAS.420.2365L} {420, 2365}

\bibitem[\protect\citeauthoryear{{Laibe} \& {Price}}{{Laibe} \&
  {Price}}{2014a}]{laibeprice14a}
{Laibe} G.,  {Price} D.~J.,  2014a, \mn@doi [\mnras] {10.1093/mnras/stu355},
  \href {https://ui.adsabs.harvard.edu/abs/2014MNRAS.440.2136L} {440, 2136}

\bibitem[\protect\citeauthoryear{{Laibe} \& {Price}}{{Laibe} \&
  {Price}}{2014b}]{laibeprice14b}
{Laibe} G.,  {Price} D.~J.,  2014b, \mn@doi [\mnras] {10.1093/mnras/stu359},
  \href {https://ui.adsabs.harvard.edu/abs/2014MNRAS.440.2147L} {440, 2147}

\bibitem[\protect\citeauthoryear{{Laibe} \& {Price}}{{Laibe} \&
  {Price}}{2014c}]{laibeprice14c}
{Laibe} G.,  {Price} D.~J.,  2014c, \mn@doi [\mnras] {10.1093/mnras/stu1367},
  \href {https://ui.adsabs.harvard.edu/abs/2014MNRAS.444.1940L} {444, 1940}

\bibitem[\protect\citeauthoryear{{Laibe}, {Gonzalez}, {Fouchet}  \&
  {Maddison}}{{Laibe} et~al.}{2008}]{laibe08}
{Laibe} G.,  {Gonzalez} J.-F.,  {Fouchet} L.,   {Maddison} S.~T.,  2008,
  \mn@doi [\aap] {10.1051/0004-6361:200809522}, \href
  {http://cdsads.u-strasbg.fr/abs/2008A%26A...487..265L} {487, 265}

\bibitem[\protect\citeauthoryear{{Laune}, {Li}, {Li}, {Li}, {Walls},
  {Birnstiel}, {Dr{\k{a}}{\.z}kowska}  \& {Stammler}}{{Laune}
  et~al.}{2020}]{laune20}
{Laune} J.,  {Li} H.,  {Li} S.,  {Li} Y.-P.,  {Walls} L.~G.,  {Birnstiel} T.,
  {Dr{\k{a}}{\.z}kowska} J.,   {Stammler} S.,  2020, \mn@doi [\apjl]
  {10.3847/2041-8213/ab65c6}, \href
  {https://ui.adsabs.harvard.edu/abs/2020ApJ...889L...8L} {889, L8}

\bibitem[\protect\citeauthoryear{{Li}, {Li}, {Koller}, {Wendroff}, {Liska},
  {Orban}, {Liang}  \& {Lin}}{{Li} et~al.}{2005}]{li05}
{Li} H.,  {Li} S.,  {Koller} J.,  {Wendroff} B.~B.,  {Liska} R.,  {Orban}
  C.~M.,  {Liang} E. P.~T.,   {Lin} D. N.~C.,  2005, \mn@doi [\apj]
  {10.1086/429367}, \href
  {https://ui.adsabs.harvard.edu/abs/2005ApJ...624.1003L} {624, 1003}

\bibitem[\protect\citeauthoryear{{Li} et~al.,}{{Li} et~al.}{2019a}]{li19}
{Li} Y.-P.,  et~al., 2019a, \mn@doi [\apj] {10.3847/1538-4357/ab1f64}, \href
  {https://ui.adsabs.harvard.edu/abs/2019ApJ...878...39L} {878, 39}

\bibitem[\protect\citeauthoryear{{Li}, {Youdin}  \& {Simon}}{{Li}
  et~al.}{2019b}]{listreaming19}
{Li} R.,  {Youdin} A.~N.,   {Simon} J.~B.,  2019b, \mn@doi [\apj]
  {10.3847/1538-4357/ab480d}, \href
  {https://ui.adsabs.harvard.edu/abs/2019ApJ...885...69L} {885, 69}

\bibitem[\protect\citeauthoryear{{Lissauer} \& {Stewart}}{{Lissauer} \&
  {Stewart}}{1993}]{lisstev93}
{Lissauer} J.~J.,  {Stewart} G.~R.,  1993, in {Levy} E.~H.,  {Lunine} J.~I.,
  eds, Protostars and Planets III. pp 1061--1088

\bibitem[\protect\citeauthoryear{{Lodato} \& {Price}}{{Lodato} \&
  {Price}}{2010}]{lodatoprice10}
{Lodato} G.,  {Price} D.~J.,  2010, \mn@doi [\mnras]
  {10.1111/j.1365-2966.2010.16526.x}, \href
  {https://ui.adsabs.harvard.edu/abs/2010MNRAS.405.1212L} {405, 1212}

\bibitem[\protect\citeauthoryear{{Lombart} \& {Laibe}}{{Lombart} \&
  {Laibe}}{2021}]{lombart21}
{Lombart} M.,  {Laibe} G.,  2021, \mn@doi [\mnras] {10.1093/mnras/staa3682},
  \href {https://ui.adsabs.harvard.edu/abs/2021MNRAS.501.4298L} {501, 4298}

\bibitem[\protect\citeauthoryear{{Lynden-Bell} \& {Pringle}}{{Lynden-Bell} \&
  {Pringle}}{1974}]{lynden74}
{Lynden-Bell} D.,  {Pringle} J.~E.,  1974, \mn@doi [\mnras]
  {10.1093/mnras/168.3.603}, \href
  {https://ui.adsabs.harvard.edu/abs/1974MNRAS.168..603L} {168, 603}

\bibitem[\protect\citeauthoryear{{Mentiplay}, {Price}  \& {Pinte}}{{Mentiplay}
  et~al.}{2019}]{mentiplay19}
{Mentiplay} D.,  {Price} D.~J.,   {Pinte} C.,  2019, \mn@doi [\mnras]
  {10.1093/mnrasl/sly209}, \href
  {https://ui.adsabs.harvard.edu/abs/2019MNRAS.484L.130M} {484, L130}

\bibitem[\protect\citeauthoryear{{Misener}, {Krijt}  \& {Ciesla}}{{Misener}
  et~al.}{2019}]{misener19}
{Misener} W.,  {Krijt} S.,   {Ciesla} F.~J.,  2019, \mn@doi [\apj]
  {10.3847/1538-4357/ab4a13}, \href
  {https://ui.adsabs.harvard.edu/abs/2019ApJ...885..118M} {885, 118}

\bibitem[\protect\citeauthoryear{{Monaghan}}{{Monaghan}}{1992}]{mona92}
{Monaghan} J.~J.,  1992, \mn@doi [\araa] {10.1146/annurev.aa.30.090192.002551},
  \href {https://ui.adsabs.harvard.edu/abs/1992ARA&A..30..543M} {30, 543}

\bibitem[\protect\citeauthoryear{{Monaghan}}{{Monaghan}}{1997}]{mona97}
{Monaghan} J.~J.,  1997, \mn@doi [Journal of Computational Physics]
  {10.1006/jcph.1997.5846}, \href
  {http://cdsads.u-strasbg.fr/abs/1997JCoPh.138..801M} {138, 801}

\bibitem[\protect\citeauthoryear{{Mu{\~n}oz} \& {Lai}}{{Mu{\~n}oz} \&
  {Lai}}{2016}]{munoz16}
{Mu{\~n}oz} D.~J.,  {Lai} D.,  2016, \mn@doi [\apj]
  {10.3847/0004-637X/827/1/43}, \href
  {https://ui.adsabs.harvard.edu/abs/2016ApJ...827...43M} {827, 43}

\bibitem[\protect\citeauthoryear{{Nakagawa}, {Sekiya}  \& {Hayashi}}{{Nakagawa}
  et~al.}{1986}]{naka86}
{Nakagawa} Y.,  {Sekiya} M.,   {Hayashi} C.,  1986, \mn@doi [\icarus]
  {10.1016/0019-1035(86)90121-1}, \href
  {https://ui.adsabs.harvard.edu/abs/1986Icar...67..375N} {67, 375}

\bibitem[\protect\citeauthoryear{{Nealon}, {Cuello}  \& {Alexander}}{{Nealon}
  et~al.}{2020}]{nealon20}
{Nealon} R.,  {Cuello} N.,   {Alexander} R.,  2020, \mn@doi [\mnras]
  {10.1093/mnras/stz3186}, \href
  {https://ui.adsabs.harvard.edu/abs/2020MNRAS.491.4108N} {491, 4108}

\bibitem[\protect\citeauthoryear{{Okuzumi}, {Momose}, {Sirono}, {Kobayashi}  \&
  {Tanaka}}{{Okuzumi} et~al.}{2016}]{oku16}
{Okuzumi} S.,  {Momose} M.,  {Sirono} S.-i.,  {Kobayashi} H.,   {Tanaka} H.,
  2016, \mn@doi [\apj] {10.3847/0004-637X/821/2/82}, \href
  {https://ui.adsabs.harvard.edu/abs/2016ApJ...821...82O} {821, 82}

\bibitem[\protect\citeauthoryear{{Ormel} \& {Spaans}}{{Ormel} \&
  {Spaans}}{2008}]{ormel08}
{Ormel} C.~W.,  {Spaans} M.,  2008, \mn@doi [\apj] {10.1086/590052}, \href
  {https://ui.adsabs.harvard.edu/abs/2008ApJ...684.1291O} {684, 1291}

\bibitem[\protect\citeauthoryear{{Paardekooper} \& {Mellema}}{{Paardekooper} \&
  {Mellema}}{2004}]{parmel04}
{Paardekooper} S.-J.,  {Mellema} G.,  2004, \mn@doi [\aap]
  {10.1051/0004-6361:200400053}, \href
  {http://adsabs.harvard.edu/abs/2004A%26A...425L...9P} {425, L9}

\bibitem[\protect\citeauthoryear{{P{\'e}rez}, {Casassus}, {Baruteau}, {Dong},
  {Hales}  \& {Cieza}}{{P{\'e}rez} et~al.}{2019}]{perez19}
{P{\'e}rez} S.,  {Casassus} S.,  {Baruteau} C.,  {Dong} R.,  {Hales} A.,
  {Cieza} L.,  2019, \mn@doi [\aj] {10.3847/1538-3881/ab1f88}, \href
  {https://ui.adsabs.harvard.edu/abs/2019AJ....158...15P} {158, 15}

\bibitem[\protect\citeauthoryear{{Pignatale}, {Gonzalez}, {Bourdon}  \&
  {Fitoussi}}{{Pignatale} et~al.}{2019}]{pignatale19}
{Pignatale} F.~C.,  {Gonzalez} J.~F.,  {Bourdon} B.,   {Fitoussi} C.,  2019,
  \mn@doi [\mnras] {10.1093/mnras/stz2883}, \href
  {https://ui.adsabs.harvard.edu/abs/2019MNRAS.490.4428P} {490, 4428}

\bibitem[\protect\citeauthoryear{{Pinilla}, {Benisty}  \&
  {Birnstiel}}{{Pinilla} et~al.}{2012}]{pinilla12}
{Pinilla} P.,  {Benisty} M.,   {Birnstiel} T.,  2012, \mn@doi [\aap]
  {10.1051/0004-6361/201219315}, \href
  {https://ui.adsabs.harvard.edu/abs/2012A&A...545A..81P} {545, A81}

\bibitem[\protect\citeauthoryear{{Pinte}, {M{\'e}nard}, {Duch{\^e}ne}  \&
  {Bastien}}{{Pinte} et~al.}{2006}]{mcfost06}
{Pinte} C.,  {M{\'e}nard} F.,  {Duch{\^e}ne} G.,   {Bastien} P.,  2006, \mn@doi
  [\aap] {10.1051/0004-6361:20053275}, \href
  {http://adsabs.harvard.edu/abs/2006A%26A...459..797P} {459, 797}

\bibitem[\protect\citeauthoryear{{Pinte}, {Harries}, {Min}, {Watson},
  {Dullemond}, {Woitke}, {M{\'e}nard}  \& {Dur{\'a}n-Rojas}}{{Pinte}
  et~al.}{2009}]{pinte09}
{Pinte} C.,  {Harries} T.~J.,  {Min} M.,  {Watson} A.~M.,  {Dullemond} C.~P.,
  {Woitke} P.,  {M{\'e}nard} F.,   {Dur{\'a}n-Rojas} M.~C.,  2009, \mn@doi
  [\aap] {10.1051/0004-6361/200811555}, \href
  {https://ui.adsabs.harvard.edu/abs/2009A&A...498..967P} {498, 967}

\bibitem[\protect\citeauthoryear{{Pinte} et~al.,}{{Pinte}
  et~al.}{2018}]{pinte18}
{Pinte} C.,  et~al., 2018, \mn@doi [\apjl] {10.3847/2041-8213/aac6dc}, \href
  {https://ui.adsabs.harvard.edu/abs/2018ApJ...860L..13P} {860, L13}

\bibitem[\protect\citeauthoryear{{Pinte} et~al.,}{{Pinte}
  et~al.}{2019}]{pinte19}
{Pinte} C.,  et~al., 2019, \mn@doi [Nature Astronomy]
  {10.1038/s41550-019-0852-6}, \href
  {https://ui.adsabs.harvard.edu/abs/2019NatAs...3.1109P} {3, 1109}

\bibitem[\protect\citeauthoryear{{Pinte} et~al.,}{{Pinte}
  et~al.}{2020}]{pinte20}
{Pinte} C.,  et~al., 2020, \mn@doi [\apjl] {10.3847/2041-8213/ab6dda}, \href
  {https://ui.adsabs.harvard.edu/abs/2020ApJ...890L...9P} {890, L9}

\bibitem[\protect\citeauthoryear{{Poblete}, {Cuello}  \& {Cuadra}}{{Poblete}
  et~al.}{2019}]{poblete19}
{Poblete} P.~P.,  {Cuello} N.,   {Cuadra} J.,  2019, \mn@doi [\mnras]
  {10.1093/mnras/stz2297}, \href
  {https://ui.adsabs.harvard.edu/abs/2019MNRAS.489.2204P} {489, 2204}

\bibitem[\protect\citeauthoryear{{Price}}{{Price}}{2007}]{splash07}
{Price} D.~J.,  2007, \mn@doi [\pasa] {10.1071/AS07022}, \href
  {http://adsabs.harvard.edu/abs/2007PASA...24..159P} {24, 159}

\bibitem[\protect\citeauthoryear{{Price}}{{Price}}{2011}]{splash11}
{Price} D.~J.,  2011, {SPLASH: An Interactive Visualization Tool for Smoothed
  Particle Hydrodynamics Simulations}, Astrophysics Source Code Library
  (\mn@eprint {ascl} {1103.004})

\bibitem[\protect\citeauthoryear{{Price} \& {Laibe}}{{Price} \&
  {Laibe}}{2015}]{price15}
{Price} D.~J.,  {Laibe} G.,  2015, \mn@doi [\mnras] {10.1093/mnras/stv996},
  \href {https://ui.adsabs.harvard.edu/abs/2015MNRAS.451..813P} {451, 813}

\bibitem[\protect\citeauthoryear{{Price} \& {Laibe}}{{Price} \&
  {Laibe}}{2020}]{price20}
{Price} D.~J.,  {Laibe} G.,  2020, \mn@doi [\mnras] {10.1093/mnras/staa1366},
  \href {https://ui.adsabs.harvard.edu/abs/2020MNRAS.495.3929P} {495, 3929}

\bibitem[\protect\citeauthoryear{{Price} et~al.,}{{Price}
  et~al.}{2017}]{phantom17ascl}
{Price} D.~J.,  et~al., 2017, {PHANTOM: Smoothed particle hydrodynamics and
  magnetohydrodynamics code} (\mn@eprint {ascl} {1709.002})

\bibitem[\protect\citeauthoryear{{Price} et~al.,}{{Price}
  et~al.}{2018a}]{phantom18}
{Price} D.~J.,  et~al., 2018a, \mn@doi [\pasa] {10.1017/pasa.2018.25}, \href
  {https://ui.adsabs.harvard.edu/abs/2018PASA...35...31P} {35, e031}

\bibitem[\protect\citeauthoryear{{Price} et~al.,}{{Price}
  et~al.}{2018b}]{price18}
{Price} D.~J.,  et~al., 2018b, \mn@doi [\mnras] {10.1093/mnras/sty647}, \href
  {https://ui.adsabs.harvard.edu/abs/2018MNRAS.477.1270P} {477, 1270}

\bibitem[\protect\citeauthoryear{{Ragusa}, {Dipierro}, {Lodato}, {Laibe}  \&
  {Price}}{{Ragusa} et~al.}{2017}]{ragusa17}
{Ragusa} E.,  {Dipierro} G.,  {Lodato} G.,  {Laibe} G.,   {Price} D.~J.,  2017,
  \mn@doi [\mnras] {10.1093/mnras/stw2456}, \href
  {https://ui.adsabs.harvard.edu/abs/2017MNRAS.464.1449R} {464, 1449}

\bibitem[\protect\citeauthoryear{{Rice}, {Armitage}, {Wood}  \&
  {Lodato}}{{Rice} et~al.}{2006}]{rice06}
{Rice} W.~K.~M.,  {Armitage} P.~J.,  {Wood} K.,   {Lodato} G.,  2006, \mn@doi
  [\mnras] {10.1111/j.1365-2966.2006.11113.x}, \href
  {https://ui.adsabs.harvard.edu/abs/2006MNRAS.373.1619R} {373, 1619}

\bibitem[\protect\citeauthoryear{{Rubenstein} \& {Bailyn}}{{Rubenstein} \&
  {Bailyn}}{1997}]{ruben97}
{Rubenstein} E.~P.,  {Bailyn} C.~D.,  1997, \mn@doi [\apj] {10.1086/303498},
  \href {https://ui.adsabs.harvard.edu/abs/1997ApJ...474..701R} {474, 701}

\bibitem[\protect\citeauthoryear{{Safronov}}{{Safronov}}{1969}]{safro69}
{Safronov} V.~S.,  1969, {Evoliutsiia doplanetnogo oblaka.}

\bibitem[\protect\citeauthoryear{{Sch{\"a}fer}, {Yang}  \&
  {Johansen}}{{Sch{\"a}fer} et~al.}{2017}]{schafer16}
{Sch{\"a}fer} U.,  {Yang} C.-C.,   {Johansen} A.,  2017, \mn@doi [\aap]
  {10.1051/0004-6361/201629561}, \href
  {https://ui.adsabs.harvard.edu/abs/2017A&A...597A..69S} {597, A69}

\bibitem[\protect\citeauthoryear{{Schoonenberg}, {Ormel}  \&
  {Krijt}}{{Schoonenberg} et~al.}{2018}]{schoo18}
{Schoonenberg} D.,  {Ormel} C.~W.,   {Krijt} S.,  2018, \mn@doi [\aap]
  {10.1051/0004-6361/201834047}, \href
  {https://ui.adsabs.harvard.edu/abs/2018A&A...620A.134S} {620, A134}

\bibitem[\protect\citeauthoryear{{Shakura} \& {Sunyaev}}{{Shakura} \&
  {Sunyaev}}{1973}]{shaksuny73}
{Shakura} N.~I.,  {Sunyaev} R.~A.,  1973, \aap, \href
  {http://cdsads.u-strasbg.fr/abs/1973A%26A....24..337S} {24, 337}

\bibitem[\protect\citeauthoryear{{Smoluchowski}}{{Smoluchowski}}{1916}]{smolu16}
{Smoluchowski} M.~V.,  1916, Zeitschrift fur Physik, \href
  {http://adsabs.harvard.edu/abs/1916ZPhy...17..557S} {17, 557}

\bibitem[\protect\citeauthoryear{{Stepinski} \& {Valageas}}{{Stepinski} \&
  {Valageas}}{1996}]{stepvala96}
{Stepinski} T.~F.,  {Valageas} P.,  1996, \aap, \href
  {https://ui.adsabs.harvard.edu/abs/1996A&A...309..301S} {309, 301}

\bibitem[\protect\citeauthoryear{{Stepinski} \& {Valageas}}{{Stepinski} \&
  {Valageas}}{1997}]{stepvala97}
{Stepinski} T.~F.,  {Valageas} P.,  1997, \aap, \href
  {http://cdsads.u-strasbg.fr/abs/1997A%26A...319.1007S} {319, 1007}

\bibitem[\protect\citeauthoryear{{Toci}, {Lodato}, {Fedele}, {Testi}  \&
  {Pinte}}{{Toci} et~al.}{2020}]{toci20}
{Toci} C.,  {Lodato} G.,  {Fedele} D.,  {Testi} L.,   {Pinte} C.,  2020,
  \mn@doi [\apjl] {10.3847/2041-8213/ab5c87}, \href
  {https://ui.adsabs.harvard.edu/abs/2020ApJ...888L...4T} {888, L4}

\bibitem[\protect\citeauthoryear{{Vericel} \& {Gonzalez}}{{Vericel} \&
  {Gonzalez}}{2020}]{vericelgonzalez20}
{Vericel} A.,  {Gonzalez} J.-F.,  2020, \mn@doi [\mnras]
  {10.1093/mnras/stz3444}, \href
  {https://ui.adsabs.harvard.edu/abs/2020MNRAS.492..210V} {492, 210}

\bibitem[\protect\citeauthoryear{{Veronesi} et~al.,}{{Veronesi}
  et~al.}{2020}]{benni20}
{Veronesi} B.,  et~al., 2020, \mn@doi [\mnras] {10.1093/mnras/staa1278}, \href
  {https://ui.adsabs.harvard.edu/abs/2020MNRAS.495.1913V} {495, 1913}

\bibitem[\protect\citeauthoryear{{Wada}, {Tanaka}, {Suyama}, {Kimura}  \&
  {Yamamoto}}{{Wada} et~al.}{2009}]{wada09}
{Wada} K.,  {Tanaka} H.,  {Suyama} T.,  {Kimura} H.,   {Yamamoto} T.,  2009, in
  {Henning} T.,  {Gr{\"u}n} E.,   {Steinacker} J.,  eds,  Astronomical Society
  of the Pacific Conference Series Vol. 414, Cosmic Dust - Near and Far. p.~347

\bibitem[\protect\citeauthoryear{{Weidenschilling}}{{Weidenschilling}}{1977}]{weiden77}
{Weidenschilling} S.~J.,  1977, \mn@doi [\mnras] {10.1093/mnras/180.1.57},
  \href {http://cdsads.u-strasbg.fr/abs/1977MNRAS.180...57W} {180, 57}

\bibitem[\protect\citeauthoryear{{Weidenschilling}}{{Weidenschilling}}{1997}]{weiden97}
{Weidenschilling} S.~J.,  1997, \mn@doi [\icarus] {10.1006/icar.1997.5712},
  \href {https://ui.adsabs.harvard.edu/abs/1997Icar..127..290W} {127, 290}

\bibitem[\protect\citeauthoryear{{Weingartner} \& {Draine}}{{Weingartner} \&
  {Draine}}{2001}]{Weingartner01}
{Weingartner} J.~C.,  {Draine} B.~T.,  2001, \mn@doi [\apj] {10.1086/318651},
  \href
  {http://adsabs.harvard.edu/cgi-bin/nph-bib_query?bibcode=2001ApJ...548..296W&db_key=AST}
  {548, 296}

\bibitem[\protect\citeauthoryear{{Whipple}}{{Whipple}}{1973}]{whipple73}
{Whipple} F.~L.,  1973, NASA Special Publication, \href
  {http://adsabs.harvard.edu/abs/1973NASSP.319..355W} {319, 355}

\bibitem[\protect\citeauthoryear{{Yang}, {Johansen}  \& {Carrera}}{{Yang}
  et~al.}{2017}]{yang17}
{Yang} C.~C.,  {Johansen} A.,   {Carrera} D.,  2017, \mn@doi [\aap]
  {10.1051/0004-6361/201630106}, \href
  {https://ui.adsabs.harvard.edu/abs/2017A&A...606A..80Y} {606, A80}

\bibitem[\protect\citeauthoryear{{Youdin} \& {Goodman}}{{Youdin} \&
  {Goodman}}{2005}]{streaming05}
{Youdin} A.~N.,  {Goodman} J.,  2005, \mn@doi [\apj] {10.1086/426895}, \href
  {http://adsabs.harvard.edu/abs/2005ApJ...620..459Y} {620, 459}

\bibitem[\protect\citeauthoryear{{Zhu}, {Dong}, {Stone}  \& {Rafikov}}{{Zhu}
  et~al.}{2015}]{zhu15}
{Zhu} Z.,  {Dong} R.,  {Stone} J.~M.,   {Rafikov} R.~R.,  2015, \mn@doi [\apj]
  {10.1088/0004-637X/813/2/88}, \href
  {https://ui.adsabs.harvard.edu/abs/2015ApJ...813...88Z} {813, 88}

\bibitem[\protect\citeauthoryear{{Zsom} \& {Dullemond}}{{Zsom} \&
  {Dullemond}}{2008}]{zsom08}
{Zsom} A.,  {Dullemond} C.~P.,  2008, \mn@doi [\aap]
  {10.1051/0004-6361:200809921}, \href
  {https://ui.adsabs.harvard.edu/abs/2008A&A...489..931Z} {489, 931}

\makeatother
\end{thebibliography}



\appendix

\section{Pure growth: influence of the initial grain size}
\label{app:s0}

The effects of the initial size has been discussed by \citetalias{laibe08}, where they found that dust growth was fast between 1 and $10$~$\mu$m and therefore the choice of the initial size between those values was inconsequential on dust evolution.
The initial size is numerically very important to control, since small grains dictate the timestepping in the two-fluid dust method. As a matter of fact, slightly increasing the initial size can lead to a simulation being a few times faster.

\begin{figure*}
\centering
\resizebox{\hsize}{!}{
\includegraphics[]{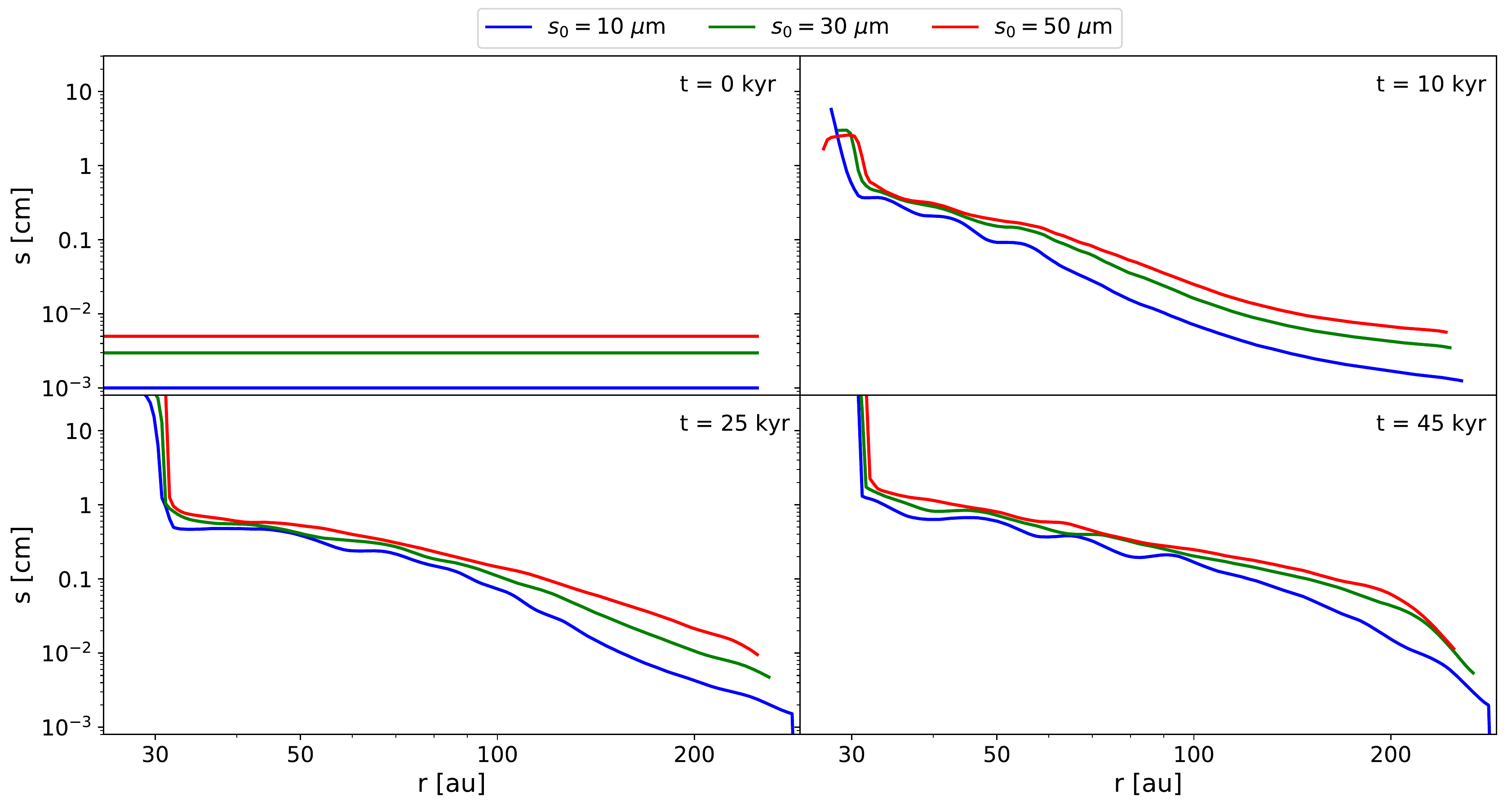}
}
\caption{Dust size radial profiles in pure growth simulations for $s_0 =$ 10 (blue), 30 (green) and 50 (red)~$\mu$m at 0 (top left), 10 (top right), 25 (bottom left) and 45 (bottom right)~kyr.}
\label{AppA-sizecomp}
\end{figure*}

\begin{figure*}
\centering
\resizebox{\hsize}{!}{
\includegraphics[]{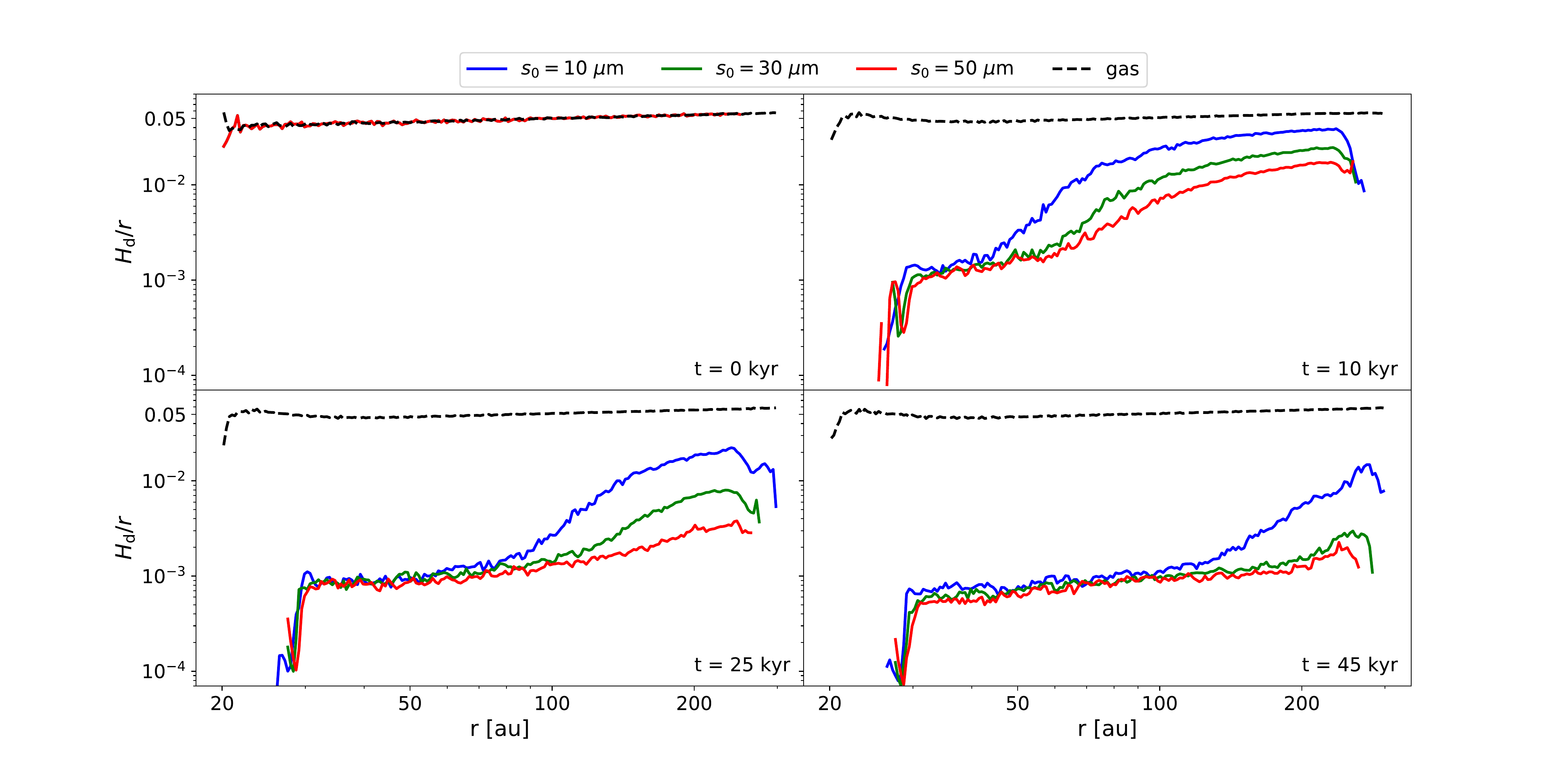}
}
\caption{Same as Fig.~\ref{AppA-sizecomp} but for the gas (black dashed line) and dust (solid coloured lines) disc scale heights.}
\label{AppA-aspratio}
\end{figure*}

Here, we tested 3 initial sizes with 10, 30 (reference value used throughout the paper) and 50~$\mu$m. We can see their effects on grain growth and settling in Figs.~\ref{AppA-sizecomp} and~\ref{AppA-aspratio}.
Grains in the innermost parts of the disc reach similar sizes very rapidly (a few thousand years), since the growth timescale there is the shortest. This allows the grains to quickly forget their initial size. In terms of vertical settling, the inner parts of the discs also match exactly as quickly.
At later times, the distance to the star for which simulations have converged to the same state increases. At 45~kyr for instance, we can notice that the simulations are similar up to $\sim 100$~au.

It seems clear that two simulations with different initial sizes can reach a similar state if enough time has passed so that most of the disc can forget its initial size, that is respecting the condition $\Delta s \gg s_0$, where $\Delta s$ is the size gained through collisions. In practice, the initial size is therefore constrained upward by this limitation, and downward by the timestepping constraint we already mentioned. Our choice of $30$~$\mu$m seems adequate for most cases but can be adapted to the user's needs.


\section{Influence of the disc model on self-induced dust trap formation}
\label{app:disc_model}

\subsection{Inner boundary}
\label{app:rin}

\begin{figure*}
\centering
\resizebox{\hsize}{!}{
\includegraphics[]{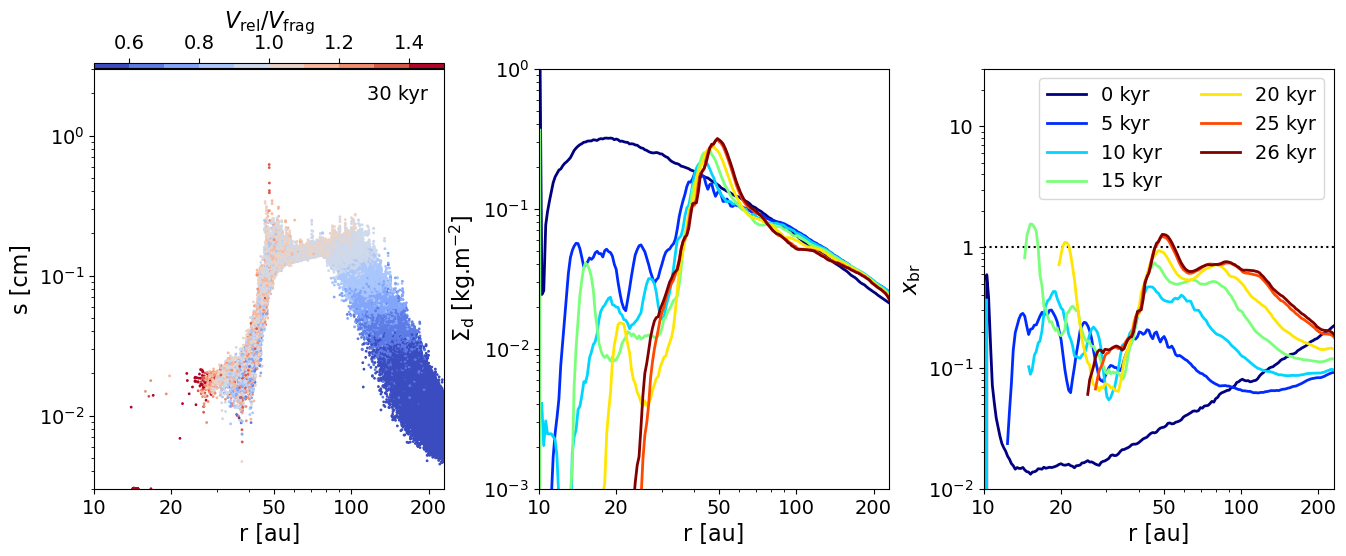}
}
\caption{Simulation for the disc with $r_\mathrm{in}=10$~au. Dust size radial distribution at the end of the simulation where particles are coloured with their ratio $V_\mathrm{rel}/V_\mathrm{frag}$ (left) and radial profiles of the dust surface density (center) and of $x_\mathrm{br}$ (right) at different times.}
\label{fig:racc10}
\end{figure*}

One may wonder whether the location of the disc inner boundary can affect the formation of self-induced dust traps. We ran a new simulation of our standard disc with a smaller inner radius of $r_\mathrm{in}=10$~au, up to the trap formation and before the onset of runaway growth. Figure~\ref{fig:racc10} shows that the self-induced dust trap forms at the same location of $\sim50$~au, where both the grain size (left) and dust surface density (centre) peak, and where $x_\mathrm{br}$ becomes larger than 1. Due to the smaller inner radius, the disc regions interior to the trap can supply more material outwards, helping the trap to from sooner than in the simulation with $r_\mathrm{in}=20$~au presented in Section~\ref{subsec:G+FRAG}.

\subsection{Surface density and temperature profiles}
\label{app:pq}

\begin{figure*}
\centering
\resizebox{\hsize}{!}{
\includegraphics[]{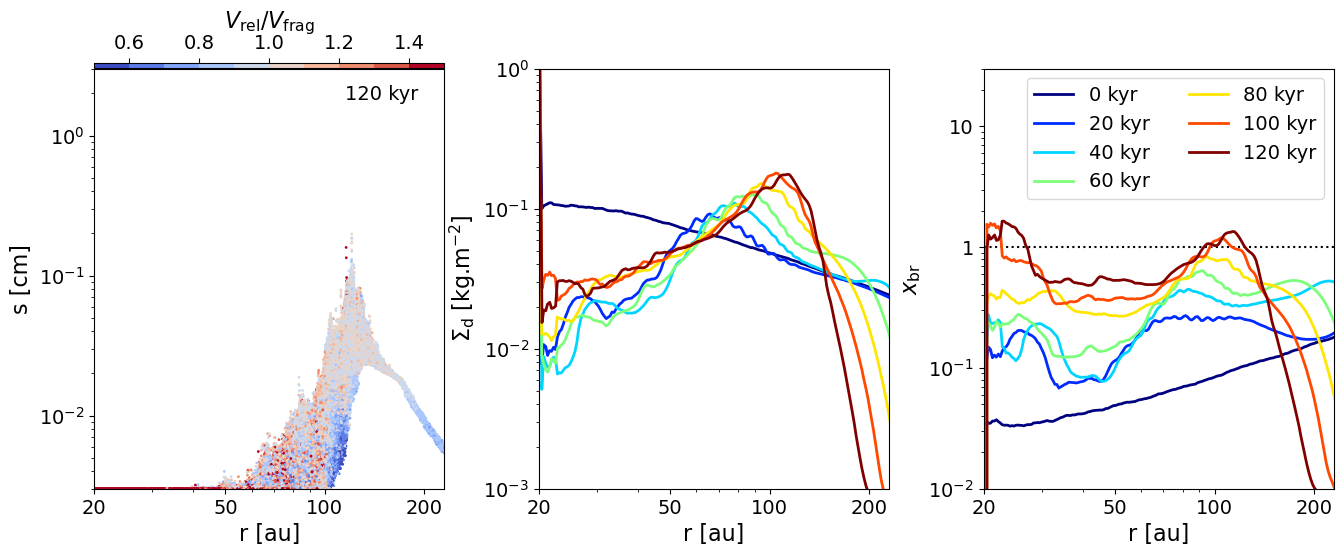}
}
\caption{Same as Fig.~\ref{fig:racc10} for the disc with $p=1$ and $q=1/2$.}
\label{fig:p1q05}
\end{figure*}

We present the case of a disc with $p=1$ and $q=1/2$, all other parameters being equal to those of our standard disc. This configuration also results in self-induced dust trap formation, as can be seen in Fig.~\ref{fig:p1q05}. The different profiles, and in particular the smaller value of $q$, lead to the trap forming at larger radii, as reported in \citet{gonzalez17}, here at $\sim120$~au. Since evolution timescales increase with radius, the trap also forms later than in our standard disc.

\bsp	
\label{lastpage}
\end{document}